\documentclass[a4paper]{article}

\usepackage[utf8]{inputenc}
\usepackage[T1]{fontenc}
\usepackage{amsmath,amssymb}
\usepackage{graphicx}
\usepackage{geometry}
\usepackage{hyperref}
\usepackage{url}
\usepackage{listings}
\usepackage[conor]{agda}
\usepackage{ebproof}
\usepackage{xspace}
\usepackage{soul}
\usepackage{cancel}
\usepackage{stmaryrd}

\newcommand{\lambdabar}{{\mkern0.75mu\mathchar'26\mkern-9.75mu\lambda}}

\newcommand\infer[3]{\ensuremath{{\dfrac{#3}{#2}}{\scriptstyle #1}}}
\newcommand{\hastypei}[2]{\ensuremath{\Gamma \vdash #1 : #2}}

\newcommand\pair[2]{\ensuremath{\langle #1,#2 \rangle}}

\newcommand{\app}[2]{\ensuremath{#1 #2}}
\newcommand{\abs}[2]{\ensuremath{\lambda #1 . #2}}

\newcommand{\timpl}[2]{\ensuremath{#1 \rightarrow #2}}
\newcommand{\tconj}[2]{\ensuremath{#1 \times #2}}

\newcommand{\sep}{\ |\ }
\newcommand{\re}{\ensuremath{\hookrightarrow}}

\newcommand\eq{\ensuremath{\rightleftarrows}}

\newcommand{\unit}{\ensuremath{\star}}
\newcommand{\com}[1]{} 
\newcommand{\ttop}{\AgdaInductiveConstructor{⊤}}
\newcommand{\juicio}{\AgdaOperator{\AgdaDatatype{⊢}}}
\newcommand{\tpi}{\AgdaInductiveConstructor{π} }
\newcommand{\reduces}{\AgdaSpace \AgdaDatatype{↪}\AgdaSpace}
\newcommand{\isoterm}{\AgdaSpace\AgdaDatatype{⇄}\AgdaSpace}
\newcommand{\rediso}{\AgdaSpace \AgdaOperator{\AgdaFunction{⇝}}\AgdaSpace}
\newcommand{\const}[1]{\AgdaInductiveConstructor{#1}}
\newcommand{\bound}[1]{\AgdaBound{#1}}
\newcommand{\func}[1]{\AgdaFunction{#1}}
\newcommand{\type}[1]{\AgdaDatatype{#1}}
\newcommand{\snstar}{\type{SN*}~\func{⟦\_⟧}~}

\newcommand{\parens}[1]{\AgdaSymbol{(}#1\AgdaSymbol{)}}
\newcommand{\substi}[2]{\func{⟪}~#1~\func{⟫}~#2}
\newcommand{\cons}[2]{#1~\func{•}~#2}
\newcommand{\ids}{\func{ids}}

\newcommand{\comp}[2]{#1~\func{∘}~#2}
\newcommand{\ppair}[2]{\AgdaOperator{\AgdaInductiveConstructor{⟨}}
\AgdaBound{#1}\AgdaOperator{\AgdaInductiveConstructor{,}}\AgdaSpace{}
\AgdaBound{#2} \AgdaOperator{\AgdaInductiveConstructor{⟩}}}

\newcommand{\ie}{\emph{i.e.}\xspace}

\newcommand{\teq}[2][\rho]{\ensuremath{[#1]\!\!\equiv\!#2}}
\newcommand{\Vars}{\ensuremath{\mathcal{V}}\xspace}
\newcommand{\ruleEquiv}{\ensuremath{(\equiv)}\xspace}

\newcommand{\new}[1]{#1}
\newcommand{\cris}[1]{#1}

\begin{document}

\title{A formalization of System I with type Top in Agda}
\author{Agustín Séttimo\thanks{Dpto.~de Ciencias de la Computación, Universidad Nacional de Rosario, Rosario, Argentina. \texttt{agustinsettimo.ips@gmail.com}}
\and Cristian Sottile\thanks{Dpto.~de Ciencia y Tecnología, Universidad Nacional de Quilmes, Bernal, Buenos Aires, Argentina. Instituto de Investigación en Ciencias de la Computación (ICC), CONICET / Universidad de Buenos Aires, Buenos Aires, Argentina. \texttt{csottile@dc.uba.ar}}
\and Cecilia Manzino\thanks{Dpto.~de Ciencias de la Computación, Universidad Nacional de Rosario, Rosario, Argentina. \texttt{ceciliam@fceia.unr.edu.ar}}}
\date{}

\maketitle

\begin{abstract}
   System I is a recently introduced simply-typed lambda calculus with pairs
   where isomorphic types are considered equal.
   In this work we propose a variant of System I with the type Top, and
   present a complete formalization of this calculus in Agda, which includes
   the proofs of progress and strong normalization.
\end{abstract}


\section{Introduction}
\label{sec:intro}
The study of type isomorphisms is a recent field of research with multiple
applications. From the perspective of programming languages, if we identify
isomorphic types, we can identify programs of the same type with different
syntax but equivalent semantics. This allows a novel form of programming, for
example, a function could take its arguments in any order, since the types
$A \to B \to C$ and $B \to A \to C$ are isomorphic. On the other hand, from the
perspective of proof systems, since types can be thought of as propositions and
programs correspond to proofs, considering different proofs of isomorphic
propositions as the same proof means that we have a form of proof-irrelevance.
For a deeper discussion on such applications, consult the introduction of the
foundational paper on $\lambda$-calculus \emph{modulo isomorphisms} by Díaz-Caro
and Dowek~\cite{DiazcaroDowekFSCD19}.

Most of these systems modulo isomorphisms use simply typed lambda calculus as a framework, and extend it so that isomorphic types are considered identical.
The first of them, which is also the most relevant for this work, was System I~\cite{DiazcaroDowekFSCD19}: a simply typed lambda calculus with pairs modulo isomorphisms. An interpreter of a preliminary version of this calculus has been implemented in Haskell~\cite{DiazcaroMartinezlopezIFL15}, and variants with $\eta$-expansion rules~\cite{DCD23}, fewer isomorphisms with no type system~\cite{ADC20}, and polymorphism~\cite{PSI} have been introduced.

The main contributions of this work are:
\begin{enumerate}
 \item A variant of System I with the type Top as a base type, which includes adaptations to make it
 more suitable to formalization. Adding new type constructors to a calculus modulo isomorphisms requires extending the equivalence relation on types, considering the isomorphisms involving the new constructor and also the equivalences between terms of isomorphic types. The choice of these term isomorphisms is also motivated by the goal
 of having a normalizing calculus.
\item The formalization in Agda of this calculus, defining its syntax, semantics, and typing rules. The approach we follow in the representation of typed terms and the notation is similar to that used in \cite{plfa}.
\item The proofs in Agda of progress and strong normalization. In our setting, strong normalization means that every reduction sequence starting from a typed term is finite.

\end{enumerate}

The complete Agda code is available in the GitHub repository \url{https://github.com/AgusSett/thesis}.

The paper is organized as follows. Section \ref{sec:calculus} presents an overview
of some calculi modulo isomorphisms.
Section \ref{sec:si-top} extends System I with the type Top.
Section \ref{sec:formalization} presents the formalization in Agda of the extended calculus and proves Progress.
Section \ref{sec:normalization} proves the strong normalization property.
Section \ref{sec:eval} implements the evaluation function.
Section \ref{sec:conclusions} concludes the paper. Section \ref{sec:futurework} presents possible future work.

\newcommand{\stlc}{STLC\xspace}

\section{\stlc modulo isomorphisms}
\label{sec:calculus}
In this section we introduce System I~\cite{DiazcaroDowekFSCD19}, which is the base system that we later extend and formalize.
Two types $A$ and $B$ are isomorphic (denoted by $\equiv$) iff there exist two
functions $f : A \to B$ and $g : B \to A$ such that $f \circ g = id_B$ and $g \circ f = id_A$.

All the type isomorphisms that occur in these systems were characterized and grouped into axiomatic sets by Di Cosmo~\cite{DiCosmo95}. For example, the set that corresponds
to simply typed lambda calculus is formed by the isomorphism called swap
($A \to (B \to C) \equiv B \to (A \to C)$).
In the simply typed lambda calculus with cartesian product (denoted as $\times$), all isomorphisms can be derived from these four isomorphisms and the congruence rules:

\begin{table}[h]
\centering

\begin{align*}
 A \times B & \equiv B \times A \tag{\textsc{\small{comm}}} \label{si:iso:comm} \\
 A \times (B \times C) & \equiv (A \times B) \times C \tag{\textsc{\small{asso}}} \label{si:iso:asso} \\
 A \rightarrow (B \times C) & \equiv (A \rightarrow B) \times (A \rightarrow C)
 \tag{\textsc{\small{dist}}}\label{si:iso:dist} \\
    (A \times B) \rightarrow C & \equiv A \rightarrow B \rightarrow C \tag{\textsc{\small{curry}}} \label{si:iso:curry}
\end{align*}

\vspace{0.3cm}
\caption{Types equivalence in System I}
\label{tab:EqTypes}
\end{table}

We note that the swap isomorphism is not included since it can be constructed
using \ref{si:iso:curry} and \ref{si:iso:comm}.

The grammar of System I for types is the same as in traditional lambda calculus,
 while its grammar for terms differs only in projection, which
is parametrized by a type rather than a position:

\vspace{0.1cm}
\begin{tabular}{l}
$A \quad:=\quad \tau \sep \timpl{A}{A} \sep \tconj{A}{A} $\\
$r\quad:=\quad x\mid \lambda x^A.r\mid rr\mid \pair{r}{r}\mid \pi_A(r)$
\end{tabular}

\vspace{0.1cm}

The type system of this calculus is extended, with respect to typed lambda
calculus, with a typing rule that asserts that if a term $t$ has type $A$ and
$A \equiv B$, then $t$ also has type $B$:

\[
\infer{{(\equiv)}}{\Gamma\vdash r:B}{\Gamma\vdash r:A \qquad A\equiv B}
\]

This rule with the four isomorphisms that characterize the calculus induces
some equivalences between terms. For example, since the types
$A\times B$ and $B\times A$ are equivalent, the terms $\pair rs$ and $\pair sr$
where $r : A$ and $s : B$, must represent the same term.
Therefore, it is necessary to incorporate into the reduction relation the consequences of isomorphisms at the term level. To do so, an equivalence relation between terms is defined. It is denoted as $\rightleftarrows$ and presented in Table~\ref{tab:EqTerms}.
The relation $\rightleftarrows^*$ is its reflexive and transitive
closure.

\begin{table}[t]
  \centering
  \begin{align*}
    \pair{r}{s} & \rightleftarrows \pair{s}{r}
                  \tag{\text{\tiny{COMM}}} \label{si:eq:comm}\\
    \pair{r}{\pair{s}{t}} & \rightleftarrows \pair{\pair{r}{s}}{t}
                            \tag{\text{\tiny{ASSO}}} \label{si:eq:asso}\\
    \abs{x^{A}}{\pair{r}{s}} & \rightleftarrows \pair{\abs{x^{A}}{r}}{\abs{x^{A}}{s}}
                               \tag{\text{\tiny{DIST$_\lambda$}}} \label{si:eq:dist-abs}\\
    \app{\pair{r}{s}}{t} & \rightleftarrows \pair{\app{r}{t}}{\app{s}{t}}
                           \tag{\text{\tiny{DIST$_{\text{\tiny{app}}}$}}} \label{si:eq:dist-app}\\
    \app{r}{\pair{s}{t}} & \rightleftarrows \app{\app{r}{s}}{t}
                           \tag{\text{\tiny{CURRY}}} \label{si:eq:curry} \\
    \end{align*}
  \[
    \infer{}{\lambda x^A.r\rightleftarrows\lambda x^A.s}{r\rightleftarrows s}
    \quad
    \infer{}{rt\rightleftarrows st}{r\rightleftarrows s}
    \quad
    \infer{}{tr\rightleftarrows ts}{r\rightleftarrows s}
  \]
  \[
    \infer{}{\pair{t}{r}\rightleftarrows \pair{t}{s}}{r\rightleftarrows s}
    \quad
    \infer{}{\pair{r}{t}\rightleftarrows \pair{s}{t}}{r\rightleftarrows s}
    \quad
    \infer{}{\pi_A(r)\rightleftarrows\pi_A(s)}{r\rightleftarrows s}
  \]
  \vspace{0.2cm}
  \caption{Terms equivalence in System I}
  \label{tab:EqTerms}
\end{table}

The operational semantics of System I was defined by two relations: the equivalence
relation between terms and a reduction relation which is similar to the classical
$\beta$-reduction. We will show here why $\beta$-reduction cannot be used directly with the rules of Table~\ref{tab:EqTerms}.

The usual projection rules access a pair through the position of the
elements, but the \eqref{si:eq:comm} isomorphism allows the order of a pair to be changed, thus allowing either of the two elements to be projected:

\vspace{0.5em}
\begin{tabular}{l}
$\pi_1 \pair{r}{s} \re r$ \\
$\pi_1 \pair{r}{s} \rightleftarrows  \pi_1 \pair{s}{r} \re s$
\end{tabular}
\vspace{0.5em}

This poses a problem for type preservation and also introduces non-determinism.
The solution is to access the element of a pair through its type, so a new rule was defined: if $\hastypei{r}{A}$, then $\pi^A \pair{r}{s} \re r$.
This rule resolves the problem of type preservation but maintains the non-determinism in the calculus. However, it is possible to encode a deterministic projection even when both
terms are of the same type. Then, non-determinism of System I is considered a feature and not a problem.

Another conflict with classical $\beta$-reduction and the equality rules of
terms occurs as a consequence of the isomorphism: $\timpl{A}{\timpl{B}{C}} \equiv
\timpl{B}{\timpl{A}{C}}$, which allows us to change the order in which we pass arguments to a function.

To solve the problem of type preservation in this case, $\beta$-reduction is modified by requiring the argument having the same type as the parameter: if $\Gamma\vdash s:A$, then $(\lambda x^A.r) s \re r[s/x]$.

Finally, the reduction relation $\rightsquigarrow$ is defined as the relation
$\re$ modulo $\rightleftarrows$ (i.e. $\rightsquigarrow \, := \, \rightleftarrows^* \circ
\re \circ \rightleftarrows^*$), and $\rightsquigarrow^*$ its reflexive and transitive closure.

As an example, consider the following reduction for a term that
is typed in this calculus:

\vspace{0.3cm}

\begin{tabular}{l}
$(\lambda x^A. \lambda y^B.r) z^B w^A   \rightleftarrows^* (\lambda x^A. \lambda y^B.r) w^A z^B   \hookrightarrow_{\beta} $ \\
$(\lambda y^B.r [w^A / x^A]) z^B \hookrightarrow_{\beta} r [ w^A / x^A, z^B /y^B]$
\end{tabular}

\section{Adding Top and transitioning to explicitness}
\label{sec:si-top}

In this section we extend System I with the type Top and we adapt it to make it
more suitable for formalization, by directing equivalence relations and thus
obtaining rewriting relations. We also introduce some modifications regarding
the implicit witnesses chosen for the corresponding terms of isomorphic types,
in order to ensure progress and preserve strong normalization.

The syntax of types is modified using $\top$ as the base type:
$$A \quad:=\quad \top \sep \timpl{A}{A} \sep \tconj{A}{A} $$
The set of type isomorphisms that allow to derive any isomorphism in this
calculus is the following, consisting of adding the three related to $\top$ to
those from System I.

\begin{align*}
  A \times B & \equiv B \times A \tag{\textsc{\small{comm}}} \label{iso:comm} \\
  A \times (B \times C) & \equiv (A \times B) \times C \tag{\textsc{\small{asso}}} \label{iso:asso} \\
  A \rightarrow (B \times C) & \equiv (A \rightarrow B) \times (A \rightarrow C)
                               \tag{\textsc{\small{dist}}}\label{iso:dist} \\
  (A \times B) \rightarrow C & \equiv A \rightarrow B \rightarrow C \tag{\textsc{\small{curry}}} \label{iso:curry} \\
  A \times \top & \equiv  A  \tag{\textsc{id-$\times$}} \label{iso:prodtop} \\ 
  \top \to A & \equiv  A  \tag{\textsc{id-$\rightarrow$}} \label{iso:funtop} \\
  A \to \top & \equiv  \top \tag{\textsc{abs}} \label{iso:funtop2}
\end{align*}

\noindent
Due to the adaptations towards formalization, which will involve a syntactic tracking of each applied isomorphism in the construction of terms, we also explicitly include the following congruence rules to the type isomorphisms relation:

 $$\infer{{(\textsc{refl})}}{A \equiv A}{\;} \quad
 \infer{{(\textsc{sym})}}{B \equiv A}{A \equiv B} \quad
 \infer{{(\textsc{trans})}}{A \equiv C}{A \equiv B \quad B \equiv C}$$

 $$\infer{{(\textsc{cong$\times_1$})}}{A \times C \equiv B \times C}{A \equiv B} \quad
 \infer{{(\textsc{cong$\times_2$})}}{C \times A \equiv C \times B}{A \equiv B}$$

 $$\infer{{(\textsc{cong$\rightarrow_1$})}}{A \rightarrow C \equiv B \rightarrow C}{A \equiv B} \quad
 \infer{{(\textsc{cong$\rightarrow_2$})}}{C \rightarrow A \equiv C \rightarrow B}{A \equiv B}$$

\vspace{0.2cm}

\newcommand\true{\texttt{True}\xspace}
In logic, the type $\top$ corresponds to the proposition \true.
Isomorphism \textsc{(id-$\times$)} corresponds to the neutrality of \true w.r.t. conjunction, while
\textsc{(id-$\rightarrow$)} and \textsc{(abs)} correspond to the neutral and absorbing nature of \true w.r.t. implication.

The grammar of terms is extended with two constructors: the value $\star$ of the
type $\top$, and a syntactic constructor to explicitly witness the application
of an isomorphism. This involves that, in order to apply the conversion rule \ruleEquiv,
\ie to give type $B$ to a term $t$ of type $A$ when $A\equiv B$, a witness of
the isomorphism establishing the type equivalence will be required. Witnesses are based on the equivalence relation, so they follow the grammar:

\bigskip
\begin{tabular}{l}
  $\rho \quad:=\quad
  \textsc{\small{comm}} \mid
  \textsc{\small{asso}} \mid
  \textsc{\small{dist}} \mid
  \textsc{\small{curry}} \mid$
  \textsc{\small{id-}$\times$} $\mid$
  \textsc{\small{id-}$\rightarrow$} $\mid
  \textsc{\small{abs}} $ \\
$\quad \quad \mid$ \small{\textsc{cong$\times_1$}} $\, \rho \mid \; $\small{\textsc{cong$\times_2$}} $\, \rho \mid $ \small{\textsc{cong$\rightarrow_1$}} $\, \rho \mid \, $ \small{\textsc{cong$\rightarrow_2$}}
$\, \rho$ \\
  $\quad \quad \mid \textsc{\small{sym}} \; \rho \mid \textsc{\small{trans}} \; \rho \; \rho $
\end{tabular}

\bigskip
\noindent The syntax of terms is then as follows:
$$t\quad:=\quad \unit \mid x \mid \lambda x^A.t\mid tt\mid \pair{t}{t}\mid \pi_A(t) \mid \teq{t}$$
\noindent where $x \in \Vars$, a set of infinite variables.

The type system has one additional rule than System I, the one that gives type $\top$ to the term $\unit$, and modifies the rule for isomorphic types \ruleEquiv to type the constructor \teq{t}. The typing rules of this calculus are given in Figure~\ref{TS}.

\begin{figure}[t]
	\centering
	\begin{prooftree}
		\infer0[($ax$)]{ \Gamma, x:A \vdash x:A }
	\end{prooftree}
	\quad
	\begin{prooftree}
		\hypo{\Gamma, x:A \vdash t:B}
		\infer1[($\rightarrow_i$)]{ \Gamma \vdash \lambda x^A.t : A \rightarrow B }
	\end{prooftree}
	\vspace{1em}
  \\
	\begin{prooftree}
		\hypo{\Gamma \vdash t : A \rightarrow B}
		\hypo{\Gamma \vdash s:A}
		\infer2[($\rightarrow_e$)]{ \Gamma \vdash ts : B }
	\end{prooftree}
	\vspace{1em}
	\\
	\begin{prooftree}
		\hypo{\Gamma \vdash t:A}
		\hypo{\Gamma \vdash s:B}
		\infer2[($\times_i$)]{ \Gamma \vdash \langle t, s \rangle : A \times B }
	\end{prooftree}
	\quad
	\begin{prooftree}
		\hypo{\Gamma \vdash t : A \times B}
		\infer1[($\times_e$)]{ \Gamma \vdash \pi_A(t) : A }
	\end{prooftree}
  \\[2ex]
 \begin{prooftree}
		\infer0[($\top_i$)]{ \Gamma \vdash \star : \top }
	\end{prooftree}
	\vspace{1em}
	\quad
\begin{prooftree}
		\hypo{A \equiv_\rho B}
		\hypo{\Gamma \vdash t:A}
		\infer2[($\equiv$)]{ \Gamma \vdash \teq[\rho]t:B }
	\end{prooftree}
	\caption{Typing rules}
\label{TS}
\end{figure}

\noindent Note that we label the equivalence rule applied in \ruleEquiv to relate $A$ and $B$ as a subscript $\rho$ in $\equiv$.

An important observation is that, due to the isomorphism
$\top \rightarrow \top \equiv \top$, the term
$\Omega = (\lambda x^\top.xx)(\lambda x^\top.xx)$ has type in this calculus.
In Figure \ref{Omega} we show a possible typing derivation for it.

\begin{figure}[h]
$\boldsymbol{\delta} = $ \\
\[
\begin{prooftree}
	\hypo{\top \equiv_{\textsc{\tiny{sym abs}}} \top \rightarrow \top}
	\infer0[($ax$)]{ x:\top \vdash x:\top }
	\infer2[($\equiv$)]{ x:\top \vdash \teq[\tiny{\textsc{sym abs}}]x: \top \rightarrow \top }
	\infer0[($ax$)]{ x:\top \vdash x:\top }
	\infer2[($\rightarrow_e$)]{ x:\top\vdash x x: \top }
	\infer1[($\rightarrow_i$)]{ \vdash \lambda x.x x: \top \rightarrow \top }
\end{prooftree}
\]
\vspace{1em}
\begin{prooftree*}
	\hypo{\boldsymbol{\delta}}

	\hypo{\top \rightarrow \top \equiv_{\textsc{\tiny{sym}}} \top}
	\hypo{\boldsymbol{\delta}}
	\infer2[($\equiv$)]{ \vdash \teq[\tiny{\textsc{sym}}] \lambda x.x x: \top }

	\infer2[($\rightarrow_e$)]{ \vdash (\lambda x.xx) (\lambda x.xx): \top }
\end{prooftree*}
\caption{Typing derivation of $\Omega$}
\label{Omega}
\end{figure}

The question of whether this calculus can be kept strongly normalizing under
such a term naturally arises. Our approach of carrying witnesses for type
isomorphisms applied to terms allows us to preserve termination. Other
approaches can be followed. The key observation is that we are working with a
special case of $\delta$ that takes a $\top$. For instance, in a variant without
witnesses, one could add a reduction rule
$(\lambda x^A . t)s \hookrightarrow t[\unit/x]$ if $A \equiv \top$. In Section 6
we show how $\Omega$ reduces in our system, as an example of the evaluation
function.

The next step is to define the relation between terms based on type
  isomorphisms. We will call these corresponding terms of isomorphic types
  ``term isomorphisms'' for simplicity. The definition comes with two important
  changes. The first one is extensional, \ie including the terms related
  by the isomorphisms involving $\top$. The second one is structural, related
  to the explicitization we adopt to bring the calculus closer to the
  formalization, and consists of duplicating and directing the rules, in order
  to decompose the isomorphism witnesses. This poses a restriction on term
  isomorphisms: in System I, term equivalences can be applied arbitrarily,
  whereas here, term isomorphisms can only be applied following the labels
  accompanying the constructor \teq[]{}\ in a term. Note that, although we have
  a rewriting relation rather than an equivalence one, we keep the double arrows
  for uniformity.

Table~\ref{tab:EqTermsT} outlines  the selected term isomorphisms, 
where their exact symmetricals, and the congruence rules that are shared with System I
(Table~\ref{tab:EqTerms}), were omitted.

The choice of term isomorphisms were made following two main targets: preserving strong
normalization, and ensuring progress. There are several ways of picking the
equivalences, according to the possible combinations of introductions and
eliminations of the type constructors; for instance,
$\pi_B(rs) \eq (\pi_{A\to B}r)s$ involves the elimination of both type
constructors, while $\pair{\abs x r}{\abs x s} \eq \abs x {\pair r s}$
involves their introduction. In this case, we followed an all–introductions
strategy. Equivalences involving eliminations were used jointly with
$\eta$–expansion, in combined rules that guarantee that no term gets stuck.
Also, due to the Curry isomorphism now being based on introductions instead of
eliminations, the rearrangement of arguments must be treated separately with the
special rule \ref{eq:cong-arr-1}.

\begin{table}[h]
  \centering
  \begin{align*} 
    \teq[\scriptsize{\textsc{comm}}] \pair{t}{s} & \rightleftarrows \pair{s}{t}
                  \tag{\text{\scriptsize{COMM$_i$}}} \label{eq:comm}\\
    \teq[\scriptsize{\textsc{asso}}] \pair{r}{\pair{s}{t}} & \rightleftarrows \pair{\pair{r}{s}}{t}
                            \tag{\text{\scriptsize{ASSO$_i$}}} \label{eq:asso}\\
    \teq[\scriptsize{\textsc{dist}}] \abs{x^{A}}{\pair{r}{s}} & \rightleftarrows \pair{\abs{x^{A}}{r}}{\abs{x^{A}}{s}}
                               \tag{\text{\scriptsize{DIST$_\lambda$}}} \label{eq:dist-abs}\\
    \teq[\scriptsize{\textsc{curry}}] \abs{x^{A}}{\abs{y^{B}}{t}} & \rightleftarrows \abs{z^{A \times B}}{t[\pi_A z/x, \pi_B z/y]}
     \tag{\text{\scriptsize{CURRY}}} \label{eq:curry}\\
    \teq[\scriptsize{\textsc{sym curry}}] \abs{x^{A \times B}}{t} & \rightleftarrows \abs{y^A}{\abs{z^B}{t[\pair{y}{z}/x]}}
    \tag{\text{\scriptsize{UNCURRY}}} \label{eq:uncurry}\\ \\
   \mbox{If $ \hastypei{t}{\top},$} \hspace{2mm}
   \teq[\scriptsize{\textsc{id-$\times$}}] \pair{s}{t} & \rightleftarrows s
    \tag{\text{\scriptsize{ID$_{\times_e}$}}} \label{eq:idx-e} \\
   \teq[\scriptsize{\textsc{sym id-$\times$}}] t & \rightleftarrows \pair{t}{\star}
    \tag{\text{\scriptsize{ID$_{\times_i}$}}} \label{eq:idx-i} \\
    \mbox{If $ \hastypei{t}{\top}$,} \hspace{2mm}
   \teq[\scriptsize{\textsc{sym abs}}] t & \rightleftarrows \abs{x^{A}}{t}
    \tag{\text{\scriptsize{ABS$_\lambda$}}} \label{eq:abs-intro} \\
    \mbox{If $ \hastypei{t}{\timpl{A}{\top}},$} \hspace{2mm}
    \teq[\scriptsize{\textsc{abs}}] t & \rightleftarrows \unit
    \tag{\text{\scriptsize{ABS$_\top$}}} \label{eq:abs-elim}\\
    \mbox{If $ \hastypei{t}{\timpl{\top}{A}},$} \hspace{2mm}
   \teq[\scriptsize{\textsc{id-$\to$}}] t & \rightleftarrows t \, \unit
    \tag{\text{\scriptsize{ID$_{\rightarrow_e}$}}} \label{eq:id-1} \\
   \teq[\scriptsize{\textsc{sym id-$\to$}}] t & \rightleftarrows \abs{x^{\top}}{t}
    \tag{\text{\scriptsize{ID$_{\rightarrow_i}$}}} \label{eq:id-2} \\ \\
     \teq[\scriptsize{\textsc{asso}}] \pair{t}{s} & \rightleftarrows \pair{\pair{t}{\pi_B (s)}}{\pi_C (s)}
                            \tag{\text{\scriptsize{SPLIT-ASSO}}} \label{eq:asso-split}\\
     \mbox{If $ \hastypei{t}{B \times C},$} \hspace{2mm} \teq[\scriptsize{\textsc{dist}}] \abs{x^{A}}{t} & \rightleftarrows \pair{\abs{x^{A}}{\pi_B (t)}}{\abs{x^{A}}{\pi_C (t)}}
        \tag{\text{\scriptsize{SPLIT-DIST$_\lambda$}}} \label{eq:dist-split}\\
     \mbox{If $ \hastypei{t}{\timpl{A}{B}}, \hastypei{s}{\timpl{A}{C}}, $} \\
     \teq[\scriptsize{\textsc{sym dist}}] \pair{t}{s} & \rightleftarrows \abs{x^{A}}{\pair{t \,x}{s \,x}}
        \tag{\text{\scriptsize{$\eta$-DIST$_{\text{\scriptsize{app}}}$}}} \label{eq:dist-le}\\
    \mbox{If $ \hastypei{t}{\timpl{B}{C}},$} \\ 
    \teq[\scriptsize{\textsc{sym curry}}] \abs{x^{A}}{t} & \rightleftarrows \abs{z^{A \times B}}{t[\pi_A (z)/x]\pi_B (z)}
                     \tag{\text{\scriptsize{$\eta$-CURRY}}} \label{eq:curry-eta}\\
   \mbox{If $A \equiv_{\rho} B$ and $y \notin FV(t)$,} \\ 
    \teq[\scriptsize{\textsc{cong-$\rightarrow_1 \; \rho$}}] \abs{x^{A}}{t} & \rightleftarrows
    \abs{x^{B}}{ t[\teq[\scriptsize{\textsc{sym $\rho$}}] x^B/x^A]}
    \tag{\text{\scriptsize{T-SUBST}}} \label{eq:cong-arr-1}
     \end{align*}
  \caption{Corresponding terms of isomorphic types in System I with Top.}
  \label{tab:EqTermsT}
\end{table}

We have then that the equivalences included are not always a direct consequence
of type isomorphisms. The first five in the table are consequence of
\cris{the type isomorphisms related with $\to$ and $\times$, \ie \ref{iso:comm}, \ref{iso:asso}, \ref{iso:dist}, and \ref{iso:curry}}. The next five are consequence of the \cris{newly added type isomorphisms related to $\top$, \ie \ref{iso:prodtop}, \ref{iso:funtop}, and \ref{iso:funtop2}}. The last five are required to unstuck terms.

To further explain the selection, consider $\eta$-expansion and the split rule ($r \rightleftarrows \pair{\pi_A r}{\pi_B r}$ where $r : A \times B$), which are needed ---in some form--- to avoid some terms getting stuck.
But, if they were included directly, the normalization property would be lost,
since, for instance, $\eta$-expansion can be applied an infinite number of times.
\cris{The solution we adopt is similar to the one used in \cite{DiazcaroMartinezlopezIFL15},
we define new constructors that embed these rules,
such as~\ref{eq:asso-split} and \ref{eq:dist-split} that embed the split rule:}

\vspace{0.2cm}

\noindent $ \langle r,s \rangle \rightleftarrows_{\textsc{split}}  \langle r, \boldsymbol{\langle} \pi_B s, \pi_C s \boldsymbol{\rangle} \rangle \rightleftarrows_{\textsc{asso}_i} \langle \boldsymbol{\langle} r, \pi_B s   \boldsymbol{\rangle}, \pi_C s \rangle $

\noindent $\lambda x^A . r \rightleftarrows_{\textsc{split}}
\lambda x^A . \langle  \pi_B r, \pi_C r \rangle \rightleftarrows_{\textsc{dist$_{\lambda}$}} \langle \lambda x^A . \pi_B r,
\lambda x^A . \pi_C r \rangle $
\vspace{0.2cm}

\noindent
The first one is used when trying to apply the association to a pair where the second component term does not have the shape of a pair.
And the second one, for example, when trying to apply a projection to an abstraction whose body does not have the form of a pair.
Similarly, the rules \ref{eq:dist-le} and \ref{eq:curry-eta} embed $\eta$-expansion.

The operational semantics of this calculus is defined as
$\rightsquigarrow \, := \, \rightleftarrows \cup \re$. We write
$\rightsquigarrow^*$ for its reflexive and transitive closure.
As stated at the beginning of this section, $\rightsquigarrow$ is not
defined exactly like in System I, \ie $\re$ modulo $\rightleftarrows$.
Instead, we define it as an extension of $\re$ with the relation
$\rightleftarrows$, so corresponding terms of isomorphic types now
constitute a reduction step.

\section{Formalization}
\label{sec:formalization}
In this section we present a formalization in Agda of a simply typed lambda calculus
with pairs and Top extended with type isomorphisms.

Much of the code presented in this work is based on \cite{plfa}. The adaptation
consists mainly of the addition of type isomorphisms and the semantics of the
calculus, where $\beta$-reduction is extended so that it includes term
equivalences (\ie $\rightleftarrows$) as possible reduction steps.

There are two approaches to formalizing a typed $\lambda$-calculus:
using extrinsically typed terms, where terms and types are defined independently (so a term can be typed or not); or using intrinsically typed terms, where types are defined first and terms are formed with a given type. 
There are also two ways of representing variable names: using named variables,
which are easier to read, or de Bruijn indices, which make the formalization more compact.
In this work we decided to use an intrinsic formulation and de Bruijn indices.

\subsection{Types, contexts and variables}

The types consist of the type Top, function types, and pairs:

\begin{code}%
	\>[0]\AgdaKeyword{data}\AgdaSpace{}%
	\AgdaDatatype{Type}\AgdaSpace{}%
	\AgdaSymbol{:}\AgdaSpace{}%
	\AgdaPrimitive{Set}\AgdaSpace{}%
	\AgdaKeyword{where}\<%
	\\
	\>[0][@{}l@{\AgdaIndent{0}}]%
	\>[2]\AgdaInductiveConstructor{⊤}%
	\>[7]\AgdaSymbol{:}\AgdaSpace{}%
	\AgdaDatatype{Type}\<%
	\\
	\>[2]\AgdaOperator{\AgdaInductiveConstructor{\AgdaUnderscore{}⇒\AgdaUnderscore{}}}\AgdaSpace{}%
	\>[7]\AgdaSymbol{:}\AgdaSpace{}%
	\AgdaDatatype{Type}\AgdaSpace{}%
	\AgdaSymbol{→}\AgdaSpace{}%
	\AgdaDatatype{Type}\AgdaSpace{}%
	\AgdaSymbol{→}\AgdaSpace{}%
	\AgdaDatatype{Type}\<%
	\\
	\>[2]\AgdaOperator{\AgdaInductiveConstructor{\AgdaUnderscore{}×\AgdaUnderscore{}}}%
	\>[7]\AgdaSymbol{:}\AgdaSpace{}%
	\AgdaDatatype{Type}\AgdaSpace{}%
	\AgdaSymbol{→}\AgdaSpace{}%
	\AgdaDatatype{Type}\AgdaSpace{}%
	\AgdaSymbol{→}\AgdaSpace{}%
	\AgdaDatatype{Type}\<%
\end{code}

Since we use natural numbers to represent variables, type contexts are
formalized as lists of types. Unlike classical lists, contexts are read from
right to left.

\begin{code}
\>[0]\AgdaKeyword{data}\AgdaSpace{}%
\AgdaDatatype{Context}\AgdaSpace{}%
\AgdaSymbol{:}\AgdaSpace{}%
\AgdaPrimitive{Set}\AgdaSpace{}%
\AgdaKeyword{where}\<%
\\
\>[0][@{}l@{\AgdaIndent{0}}]%
\>[2]\AgdaInductiveConstructor{∅}%
\>[6]\AgdaSymbol{:}\AgdaSpace{}%
\AgdaDatatype{Context}\<%
\\
\>[2]\AgdaOperator{\AgdaInductiveConstructor{\AgdaUnderscore{},\AgdaUnderscore{}}}\AgdaSpace{}%
\AgdaSymbol{:}\AgdaSpace{}%
\AgdaDatatype{Context}\AgdaSpace{}%
\AgdaSymbol{→}\AgdaSpace{}%
\AgdaDatatype{Type}\AgdaSpace{}%
\AgdaSymbol{→}\AgdaSpace{}%
\AgdaDatatype{Context}\<%
\\[\AgdaEmptyExtraSkip]%
\end{code}

Intrinsically typed variables are represented by
de Bruijn indices. Each variable is indexed by its type and a context
in which the variable is typed.

\begin{code}
\>[0]\AgdaKeyword{data}\AgdaSpace{}%
\AgdaOperator{\AgdaDatatype{\AgdaUnderscore{}∋\AgdaUnderscore{}}}\AgdaSpace{}%
\AgdaSymbol{:}\AgdaSpace{}%
\AgdaDatatype{Context}\AgdaSpace{}%
\AgdaSymbol{→}\AgdaSpace{}%
\AgdaDatatype{Type}\AgdaSpace{}%
\AgdaSymbol{→}\AgdaSpace{}%
\AgdaPrimitive{Set}\AgdaSpace{}%
\AgdaKeyword{where}\<%
\\
\>[0][@{}l@{\AgdaIndent{0}}]%
\>[2]\AgdaInductiveConstructor{Z}%
\>[44I]\AgdaSymbol{:}%
\>[45I]\AgdaSymbol{∀}\AgdaSpace{}%
\AgdaSymbol{\{}\AgdaBound{Γ}\AgdaSpace{}%
\AgdaBound{A}\AgdaSymbol{\}}
\AgdaSymbol{→}\AgdaSpace{}%
\AgdaBound{Γ}\AgdaSpace{}%
\AgdaOperator{\AgdaInductiveConstructor{,}}\AgdaSpace{}%
\AgdaBound{A}\AgdaSpace{}%
\AgdaOperator{\AgdaDatatype{∋}}\AgdaSpace{}%
\AgdaBound{A}\<%
\\
\\[\AgdaEmptyExtraSkip]%
\>[2]\AgdaOperator{\AgdaInductiveConstructor{S\AgdaUnderscore{}}}\AgdaSpace{}%
\AgdaSymbol{:}\AgdaSpace{}%
\AgdaSymbol{∀}\AgdaSpace{}%
\AgdaSymbol{\{}\AgdaBound{Γ}\AgdaSpace{}%
\AgdaBound{A}\AgdaSpace{}%
\AgdaBound{B}\AgdaSymbol{\}} \AgdaSpace{}
\AgdaSymbol{→}%
\>[58I]\AgdaBound{Γ}\AgdaSpace{}%
\AgdaOperator{\AgdaDatatype{∋}}\AgdaSpace{}%
\AgdaBound{B} \AgdaSpace{}
\AgdaSymbol{→}\AgdaSpace{}%
\AgdaBound{Γ}\AgdaSpace{}%
\AgdaOperator{\AgdaInductiveConstructor{,}}\AgdaSpace{}%
\AgdaBound{A}\AgdaSpace{}%
\AgdaOperator{\AgdaDatatype{∋}}\AgdaSpace{}%
\AgdaBound{B}\<%
\end{code}

Then, \AgdaBound{Γ} \AgdaDatatype{∋} \AgdaBound{A} is the type of variables that have type \AgdaBound{A} in the context \AgdaBound{Γ}.

For example, the following variables with types \AgdaInductiveConstructor{⊤} and
\AgdaInductiveConstructor{⊤} \AgdaInductiveConstructor{⇒} \AgdaInductiveConstructor{⊤} are typable in the context \AgdaInductiveConstructor{∅}, \ttop \AgdaInductiveConstructor{⇒} \ttop, \ttop and also represent proofs of this.

\begin{code}
\>[0]\AgdaFunction{\AgdaUnderscore{}}\AgdaSpace{}%
\AgdaSymbol{:}\AgdaSpace{}%
\AgdaInductiveConstructor{∅}\AgdaSpace{}%
\AgdaOperator{\AgdaInductiveConstructor{,}}\AgdaSpace{}%
\AgdaInductiveConstructor{⊤}\AgdaSpace{}%
\AgdaOperator{\AgdaInductiveConstructor{⇒}}\AgdaSpace{}%
\AgdaInductiveConstructor{⊤}\AgdaSpace{}%
\AgdaOperator{\AgdaInductiveConstructor{,}}\AgdaSpace{}%
\AgdaInductiveConstructor{⊤}\AgdaSpace{}%
\AgdaOperator{\AgdaDatatype{∋}}\AgdaSpace{}%
\AgdaInductiveConstructor{⊤}\<%
\\
\>[0]\AgdaSymbol{\AgdaUnderscore{}}\AgdaSpace{}%
\AgdaSymbol{=}\AgdaSpace{}%
\AgdaInductiveConstructor{Z}\<%
\\
\\[\AgdaEmptyExtraSkip]%
\>[0]\AgdaFunction{\AgdaUnderscore{}}\AgdaSpace{}%
\AgdaSymbol{:}\AgdaSpace{}%
\AgdaInductiveConstructor{∅}\AgdaSpace{}%
\AgdaOperator{\AgdaInductiveConstructor{,}}\AgdaSpace{}%
\AgdaInductiveConstructor{⊤}\AgdaSpace{}%
\AgdaOperator{\AgdaInductiveConstructor{⇒}}\AgdaSpace{}%
\AgdaInductiveConstructor{⊤}\AgdaSpace{}%
\AgdaOperator{\AgdaInductiveConstructor{,}}\AgdaSpace{}%
\AgdaInductiveConstructor{⊤}\AgdaSpace{}%
\AgdaOperator{\AgdaDatatype{∋}}\AgdaSpace{}%
\AgdaInductiveConstructor{⊤}\AgdaSpace{}%
\AgdaOperator{\AgdaInductiveConstructor{⇒}}\AgdaSpace{}%
\AgdaInductiveConstructor{⊤}\<%
\\
\>[0]\AgdaSymbol{\AgdaUnderscore{}}\AgdaSpace{}%
\AgdaSymbol{=}\AgdaSpace{}%
\AgdaOperator{\AgdaInductiveConstructor{S}}\AgdaSpace{}%
\AgdaInductiveConstructor{Z}\<%
\end{code}

As we can see in the examples, the proof that a variable has a type in a context
is a de Bruijn index.

\subsection{Terms}
Here we present the typing rules of the calculus. Each constructor of this data
type, which encodes a typing rule since we use an intrinsically typed
representation of terms, represents a term of the calculus, with the
exception of the constructor
\AgdaInductiveConstructor{[\AgdaUnderscore{}]≡\AgdaUnderscore{}}.

Taking into account this exception, \AgdaBound{Γ} \juicio \AgdaBound{A} is the type of terms that have type \AgdaBound{A} in the context \AgdaBound{Γ}, and each term of this type is a proof of that, so terms are actually typing derivations.
The constructor \AgdaInductiveConstructor{[\AgdaUnderscore{}]≡\AgdaUnderscore{}} that
does not represent a term is used in the typing derivation of some terms.

\begin{code}
\>[0]\AgdaKeyword{data}\AgdaSpace{}%
\AgdaOperator{\AgdaDatatype{\AgdaUnderscore{}⊢\AgdaUnderscore{}}}\AgdaSpace{}%
\AgdaSymbol{:}\AgdaSpace{}%
\AgdaDatatype{Context}\AgdaSpace{}%
\AgdaSymbol{→}\AgdaSpace{}%
\AgdaDatatype{Type}\AgdaSpace{}%
\AgdaSymbol{→}\AgdaSpace{}%
\AgdaPrimitive{Set}\AgdaSpace{}%
\AgdaKeyword{where}\<%
\\
\\[\AgdaEmptyExtraSkip]%
\>[0][@{}l@{\AgdaIndent{0}}]%
\>[2]\AgdaOperator{\AgdaInductiveConstructor{`\AgdaUnderscore{}}}\AgdaSpace{}%
\AgdaSymbol{:}\AgdaSpace{}%
\AgdaSymbol{∀}\AgdaSpace{}%
\AgdaSymbol{\{}\AgdaBound{Γ}\AgdaSpace{}%
\AgdaBound{A}\AgdaSymbol{\}}\AgdaSpace{}%
\>[3]\AgdaComment{--\ (ax)}\<%
\\
\>[2][@{}l@{\AgdaIndent{0}}]%
\>[4]\AgdaSymbol{→}%
\>[79I]\AgdaBound{Γ}\AgdaSpace{}%
\AgdaOperator{\AgdaDatatype{∋}}\AgdaSpace{}%
\AgdaBound{A}\<%
\\
%
\>[4]\AgdaSymbol{→}\AgdaSpace{}%
\AgdaBound{Γ}\AgdaSpace{}%
\AgdaOperator{\AgdaDatatype{⊢}}\AgdaSpace{}%
\AgdaBound{A}\<%
\\
\\[\AgdaEmptyExtraSkip]%
\>[2]\AgdaInductiveConstructor{⋆}\AgdaSpace{}%
\AgdaSymbol{:}\AgdaSpace{}%
\AgdaSymbol{∀}\AgdaSpace{}%
\AgdaSymbol{\{}\AgdaBound{Γ}\AgdaSymbol{\}}\AgdaSpace{}%
\AgdaSymbol{→}\AgdaSpace{}%
\AgdaBound{Γ}\AgdaSpace{}%
\AgdaOperator{\AgdaDatatype{⊢}}\AgdaSpace{}%
\AgdaInductiveConstructor{⊤}\AgdaSpace{}%
\>[3]\AgdaComment{--\ (⊤ᵢ)}\<%
\\
\\[\AgdaEmptyExtraSkip]%
\>[2]\AgdaOperator{\AgdaInductiveConstructor{[\AgdaUnderscore{}]≡\AgdaUnderscore{}}}\AgdaSpace{}%
\AgdaSymbol{:}\AgdaSpace{}%
\AgdaSymbol{∀}\AgdaSpace{}%
\AgdaSymbol{\{}\AgdaBound{Γ}\AgdaSpace{}%
\AgdaBound{A}\AgdaSpace{}%
\AgdaBound{B}\AgdaSymbol{\}}\AgdaSpace{}%
\>[3]\AgdaComment{--\ (≡)}\<%
\\
\>[2][@{}l@{\AgdaIndent{0}}]%
\>[4]\AgdaSymbol{→}\AgdaSpace{}%
\AgdaBound{A}\AgdaSpace{}%
\AgdaOperator{\AgdaDatatype{≡}}\AgdaSpace{}%
\AgdaBound{B}\<%
\\
\>[4]\AgdaSymbol{→}%
\>[102I]\AgdaBound{Γ}\AgdaSpace{}%
\AgdaOperator{\AgdaDatatype{⊢}}\AgdaSpace{}%
\AgdaBound{A}\<%
\\
%
\>[4]\AgdaSymbol{→}\AgdaSpace{}%
\AgdaBound{Γ}\AgdaSpace{}%
\AgdaOperator{\AgdaDatatype{⊢}}\AgdaSpace{}%
\AgdaBound{B}\<%
\\
\\[\AgdaEmptyExtraSkip]%
\>[2]\AgdaOperator{\AgdaInductiveConstructor{ƛ\AgdaUnderscore{}}}%
\>[6]\AgdaSymbol{:}%
\>[9]\AgdaSymbol{∀}\AgdaSpace{}%
\AgdaSymbol{\{}\AgdaBound{Γ}\AgdaSpace{}%
\AgdaBound{A}\AgdaSpace{}%
\AgdaBound{B}\AgdaSymbol{\}}\AgdaSpace{}%
\>[3]\AgdaComment{--\ (⇒ᵢ)}\<%
\\
\>[2][@{}l@{\AgdaIndent{0}}]%
\>[4]\AgdaSymbol{→}%
\>[112I]\AgdaBound{Γ}\AgdaSpace{}%
\AgdaOperator{\AgdaInductiveConstructor{,}}\AgdaSpace{}%
\AgdaBound{A}\AgdaSpace{}%
\AgdaOperator{\AgdaDatatype{⊢}}\AgdaSpace{}%
\AgdaBound{B}\<%
\\
%
\>[4]\AgdaSymbol{→}\AgdaSpace{}%
\AgdaBound{Γ}\AgdaSpace{}%
\AgdaOperator{\AgdaDatatype{⊢}}\AgdaSpace{}%
\AgdaBound{A}\AgdaSpace{}%
\AgdaOperator{\AgdaInductiveConstructor{⇒}}\AgdaSpace{}%
\AgdaBound{B}\<%
\\
\\[\AgdaEmptyExtraSkip]%
\>[2]\AgdaOperator{\AgdaInductiveConstructor{\AgdaUnderscore{}·\AgdaUnderscore{}}}\AgdaSpace{}%
\AgdaSymbol{:}\AgdaSpace{}%
\AgdaSymbol{∀}\AgdaSpace{}%
\AgdaSymbol{\{}\AgdaBound{Γ}\AgdaSpace{}%
\AgdaBound{A}\AgdaSpace{}%
\AgdaBound{B}\AgdaSymbol{\}}%
\>[3]\AgdaComment{--\ (⇒ₑ)}\<%
\\
\>[2][@{}l@{\AgdaIndent{0}}]%
\>[4]\AgdaSymbol{→}\AgdaSpace{}%
\AgdaBound{Γ}\AgdaSpace{}%
\AgdaOperator{\AgdaDatatype{⊢}}\AgdaSpace{}%
\AgdaBound{A}\AgdaSpace{}%
\AgdaOperator{\AgdaInductiveConstructor{⇒}}\AgdaSpace{}%
\AgdaBound{B}\<%
\\
\>[4]\AgdaSymbol{→}%
\>[132I]\AgdaBound{Γ}\AgdaSpace{}%
\AgdaOperator{\AgdaDatatype{⊢}}\AgdaSpace{}%
\AgdaBound{A}\<%
\\
%
\>[4]\AgdaSymbol{→}\AgdaSpace{}%
\AgdaBound{Γ}\AgdaSpace{}%
\AgdaOperator{\AgdaDatatype{⊢}}\AgdaSpace{}%
\AgdaBound{B}\<%
\\
\\[\AgdaEmptyExtraSkip]%
\>[2]\AgdaOperator{\AgdaInductiveConstructor{⟨\AgdaUnderscore{},\AgdaUnderscore{}⟩}}\AgdaSpace{}%
\AgdaSymbol{:}\AgdaSpace{}%
\AgdaSymbol{∀}\AgdaSpace{}%
\AgdaSymbol{\{}\AgdaBound{Γ}\AgdaSpace{}%
\AgdaBound{A}\AgdaSpace{}%
\AgdaBound{B}\AgdaSymbol{\}}\AgdaSpace{}%
\>[3]\AgdaComment{--\ (×ᵢ)}\<%
\\
\>[2][@{}l@{\AgdaIndent{0}}]%
\>[4]\AgdaSymbol{→}\AgdaSpace{}%
\AgdaBound{Γ}\AgdaSpace{}%
\AgdaOperator{\AgdaDatatype{⊢}}\AgdaSpace{}%
\AgdaBound{A}\<%
\\
\>[4]\AgdaSymbol{→}%
\>[147I]\AgdaBound{Γ}\AgdaSpace{}%
\AgdaOperator{\AgdaDatatype{⊢}}\AgdaSpace{}%
\AgdaBound{B}\<%
\\
%
\>[4]\AgdaSymbol{→}\AgdaSpace{}%
\AgdaBound{Γ}\AgdaSpace{}%
\AgdaOperator{\AgdaDatatype{⊢}}\AgdaSpace{}%
\AgdaBound{A}\AgdaSpace{}%
\AgdaOperator{\AgdaInductiveConstructor{×}}\AgdaSpace{}%
\AgdaBound{B}\<%
\\
\\[\AgdaEmptyExtraSkip]%
\>[2]\AgdaInductiveConstructor{π}%
\>[155I]\AgdaSymbol{:}\AgdaSpace{}%
\AgdaSymbol{∀}\AgdaSpace{}%
\AgdaSymbol{\{}\AgdaBound{Γ}\AgdaSpace{}%
\AgdaBound{A}\AgdaSpace{}%
\AgdaBound{B}\AgdaSymbol{\}}%
\>[3]\AgdaComment{--\ (×ₑ)}\<%
\\
\>[.][@{}l@{}]\<[155I]%
\>[4]\AgdaSymbol{→}\AgdaSpace{}%
\AgdaSymbol{(}\AgdaBound{C}\AgdaSpace{}%
\AgdaSymbol{:}\AgdaSpace{}%
\AgdaDatatype{Type}\AgdaSymbol{)}\<%
\\
\>[4]\AgdaSymbol{→}\AgdaSpace{}%
\AgdaSymbol{\{}\AgdaBound{proof}\AgdaSpace{}%
\AgdaSymbol{:}\AgdaSpace{}%
\AgdaSymbol{(}\AgdaBound{C}\AgdaSpace{}%
\AgdaOperator{\AgdaDatatype{≋}}\AgdaSpace{}%
\AgdaBound{A}\AgdaSymbol{)}\AgdaSpace{}%
\AgdaOperator{\AgdaDatatype{⊎}}\AgdaSpace{}%
\AgdaSymbol{(}\AgdaBound{C}\AgdaSpace{}%
\AgdaOperator{\AgdaDatatype{≋}}\AgdaSpace{}%
\AgdaBound{B}\AgdaSymbol{)\}}\<%
\\
\>[4]\AgdaSymbol{→}%
\>[172I]\AgdaBound{Γ}\AgdaSpace{}%
\AgdaOperator{\AgdaDatatype{⊢}}\AgdaSpace{}%
\AgdaBound{A}\AgdaSpace{}%
\AgdaOperator{\AgdaInductiveConstructor{×}}\AgdaSpace{}%
\AgdaBound{B}\<%
\\
%
\>[4]\AgdaSymbol{→}\AgdaSpace{}%
\AgdaBound{Γ}\AgdaSpace{}%
\AgdaOperator{\AgdaDatatype{⊢}}\AgdaSpace{}%
\AgdaBound{C}\<%
\end{code}

We note that the constructor \tpi takes as an argument the type \AgdaBound{C},
which is the type that this function uses to carry out the projection,
and an implicit argument that serves as proof that \AgdaBound{C} is either equal
to type \AgdaBound{A} or type \AgdaBound{B}. In the type of this argument we use
propositional equality denoted as \AgdaOperator{\AgdaDatatype{≋}} in this work.

The following are some examples of terms. Each of them is a proof that it types in the given context.

\begin{code}%
\\[\AgdaEmptyExtraSkip]%
\>[0]\AgdaFunction{T₁}\AgdaSpace{}%
\AgdaSymbol{:}\AgdaSpace{}%
\AgdaInductiveConstructor{∅}\AgdaSpace{}%
\AgdaOperator{\AgdaInductiveConstructor{,}}\AgdaSpace{}%
\AgdaInductiveConstructor{⊤}\AgdaSpace{}%
\AgdaOperator{\AgdaDatatype{⊢}}\AgdaSpace{}%
\AgdaInductiveConstructor{⊤}\AgdaSpace{}%
\AgdaOperator{\AgdaInductiveConstructor{⇒}}\AgdaSpace{}%
\AgdaInductiveConstructor{⊤}\<%
\\
\>[0]\AgdaFunction{T₁}\AgdaSpace{}%
\AgdaSymbol{=}\AgdaSpace{}%
\AgdaOperator{\AgdaInductiveConstructor{ƛ}}\AgdaSpace{}%
\AgdaOperator{\AgdaInductiveConstructor{`}}\AgdaSpace{}%
\AgdaInductiveConstructor{Z}\<%
\\
\\[\AgdaEmptyExtraSkip]%
\>[0]\AgdaFunction{T₂}\AgdaSpace{}%
\AgdaSymbol{:}\AgdaSpace{}%
\AgdaInductiveConstructor{∅}\AgdaSpace{}%
\AgdaOperator{\AgdaInductiveConstructor{,}}\AgdaSpace{}%
\AgdaInductiveConstructor{⊤}\AgdaSpace{}%
\AgdaOperator{\AgdaDatatype{⊢}}\AgdaSpace{}%
\AgdaSymbol{(}\AgdaInductiveConstructor{⊤}\AgdaSpace{}%
\AgdaOperator{\AgdaInductiveConstructor{⇒}}\AgdaSpace{}%
\AgdaInductiveConstructor{⊤}\AgdaSymbol{)}\AgdaSpace{}
\AgdaOperator{\AgdaInductiveConstructor{×}}\AgdaSpace{}%
\AgdaInductiveConstructor{⊤}\<%
\\
\>[0]\AgdaFunction{T₂}\AgdaSpace{}%
\AgdaSymbol{=}\AgdaSpace{}%
\AgdaOperator{\AgdaInductiveConstructor{⟨}}\AgdaSpace{}%
\AgdaFunction{T₁}\AgdaSpace{}%
\AgdaOperator{\AgdaInductiveConstructor{,}}\AgdaSpace{}%
\AgdaInductiveConstructor{⋆}\AgdaSpace{}%
\AgdaOperator{\AgdaInductiveConstructor{⟩}}\<%
\\
\\[\AgdaEmptyExtraSkip]%
\>[0]\AgdaFunction{T₃}\AgdaSpace{}%
\AgdaSymbol{:}\AgdaSpace{}%
\AgdaInductiveConstructor{∅}\AgdaSpace{}%
\AgdaOperator{\AgdaInductiveConstructor{,}}\AgdaSpace{}%
\AgdaInductiveConstructor{⊤}\AgdaSpace{}%
\AgdaOperator{\AgdaDatatype{⊢}}\AgdaSpace{}%
\AgdaInductiveConstructor{⊤}\<%
\\
\>[0]\AgdaFunction{T₃}\AgdaSpace{}%
\AgdaSymbol{=}\AgdaSpace{}%
\AgdaSymbol{(}\AgdaInductiveConstructor{π}\AgdaSpace{}%
\AgdaFunction{fun}\AgdaSpace{}%
\AgdaSymbol{\{}\AgdaInductiveConstructor{inj₁}\AgdaSpace{}%
\AgdaInductiveConstructor{refl}\AgdaSymbol{\}}\AgdaSpace{}%
\AgdaFunction{T₂}\AgdaSymbol{)}\AgdaSpace{}%
\AgdaOperator{\AgdaInductiveConstructor{·}}\AgdaSpace{}%
\AgdaOperator{\AgdaInductiveConstructor{`}}\AgdaSpace{}%
\AgdaInductiveConstructor{Z}\<%
\end{code}

\subsection{Substitutions and reduction}
Using de Bruijn representation, substitutions are simply mappings of natural numbers to terms, so they can be interpreted as infinite sequences of terms.
These sequences can be constructed using some operators~\cite{explicit_subs}:
\begin{itemize}
 \item $id$: the identity substitution: $\{i \mapsto i\}$
 \item $\uparrow$: the shift operator: $\{i \mapsto i +1 \}$
 \item $a \bullet s$: the concatenation of the term $a$ with the substitution $s$:
 $\{0 \mapsto a, i+1 \mapsto s(i)\}$
 \item $\circ$: the composition of substitutions
\end{itemize}

With these operations we can give an inductive definition of the application of a substitution $s$ on a term $t$, denoted as $\llangle s \rrangle t$.
Then, $\beta$-reduction can be defined as follows:

\[ (\lambda t)r \hookrightarrow_{\beta} \llangle r \bullet id \rrangle t \]

To implement substitutions, we use an approach given by
Altenkirch and Reus \cite{Thorsten} that was formalized
by McBride in~\cite{ren-sub}. Substitutions are implemented using renamings, which
are functions from variables in one context to variables
in another that preserve typing.
\new{In Agda, the types for renaming and substitutions are these:}

\begin{code}%
	\>[0]\AgdaFunction{Rename}\AgdaSpace{}%
	\AgdaSymbol{:}\AgdaSpace{}%
	\AgdaDatatype{Context}\AgdaSpace{}%
	\AgdaSymbol{→}\AgdaSpace{}%
	\AgdaDatatype{Context}\AgdaSpace{}%
	\AgdaSymbol{→}\AgdaSpace{}%
	\AgdaPrimitive{Set}\<%
	\\
	\>[0]\AgdaFunction{Rename}\AgdaSpace{}%
	\AgdaBound{Γ}\AgdaSpace{}%
	\AgdaBound{Δ}\AgdaSpace{}%
	\AgdaSymbol{=}\AgdaSpace{}%
	\AgdaSymbol{∀\{}\AgdaBound{A}\AgdaSymbol{\}}\AgdaSpace{}%
	\AgdaSymbol{→}\AgdaSpace{}%
	\AgdaBound{Γ}\AgdaSpace{}%
	\AgdaOperator{\AgdaDatatype{∋}}\AgdaSpace{}%
	\AgdaBound{A}\AgdaSpace{}%
	\AgdaSymbol{→}\AgdaSpace{}%
	\AgdaBound{Δ}\AgdaSpace{}%
	\AgdaOperator{\AgdaDatatype{∋}}\AgdaSpace{}%
	\AgdaBound{A}\<%
	\\
	\\[\AgdaEmptyExtraSkip]%
	\>[0]\AgdaFunction{Subst}\AgdaSpace{}%
	\AgdaSymbol{:}\AgdaSpace{}%
	\AgdaDatatype{Context}\AgdaSpace{}%
	\AgdaSymbol{→}\AgdaSpace{}%
	\AgdaDatatype{Context}\AgdaSpace{}%
	\AgdaSymbol{→}\AgdaSpace{}%
	\AgdaPrimitive{Set}\<%
	\\
	\>[0]\AgdaFunction{Subst}\AgdaSpace{}%
	\AgdaBound{Γ}\AgdaSpace{}%
	\AgdaBound{Δ}\AgdaSpace{}%
	\AgdaSymbol{=}\AgdaSpace{}%
	\AgdaSymbol{∀\{}\AgdaBound{A}\AgdaSymbol{\}}\AgdaSpace{}%
	\AgdaSymbol{→}\AgdaSpace{}%
	\AgdaBound{Γ}\AgdaSpace{}%
	\AgdaOperator{\AgdaDatatype{∋}}\AgdaSpace{}%
	\AgdaBound{A}\AgdaSpace{}%
	\AgdaSymbol{→}\AgdaSpace{}%
	\AgdaBound{Δ}\AgdaSpace{}%
	\AgdaOperator{\AgdaDatatype{⊢}}\AgdaSpace{}%
	\AgdaBound{A}\<%
\end{code}

We call the function that applies a renaming of variables in a term \AgdaFunction{rename} \AgdaSpace{}\AgdaSymbol{:}\AgdaSpace{}
\AgdaSymbol{∀}\AgdaSpace{} \AgdaSymbol{\{}\AgdaBound{Γ}\AgdaSpace{}
\AgdaBound{Δ}\AgdaSymbol{\}} \AgdaSpace{}\AgdaSymbol{:}\AgdaSpace{}\AgdaFunction{Rename}\AgdaSpace{}\AgdaBound{Γ}\AgdaSpace{}\AgdaBound{Δ}\AgdaSpace{} \AgdaSymbol{→}
\AgdaSymbol{(∀}\AgdaSpace{} \AgdaSymbol{\{}\AgdaBound{A}\AgdaSymbol{\}}
\AgdaSymbol{→} \AgdaBound{Γ} 
\AgdaOperator{\AgdaDatatype{⊢}} 
\AgdaBound{A}\AgdaSpace{} \AgdaSymbol{→}\AgdaSpace{}
\AgdaBound{Δ}\AgdaSpace{} \AgdaOperator{\AgdaDatatype{⊢}}\AgdaSpace{}
\AgdaBound{A}\AgdaSymbol{)}, and substitution and simple substitution are implemented with the functions
\AgdaFunction{⟪\AgdaUnderscore{}⟫} \AgdaSymbol{:}
\AgdaSymbol{∀}\AgdaSpace{}
	\AgdaSymbol{\{}\AgdaBound{Γ}\AgdaSpace{}
	\AgdaBound{Δ}\AgdaSymbol{\}} \AgdaSymbol{→}
	\AgdaFunction{Subst}\AgdaSpace{} \AgdaBound{Γ}\AgdaSpace{} \AgdaBound{Δ}
	\AgdaSymbol{(∀}\AgdaSpace{}
	\AgdaSymbol{\{}\AgdaBound{C}\AgdaSymbol{\}} \AgdaSymbol{→}
	\AgdaBound{Γ} 
	\AgdaOperator{\AgdaDatatype{⊢}} 
	\AgdaBound{C} \AgdaSymbol{→} \AgdaBound{Δ}\AgdaSpace{}
	\AgdaOperator{\AgdaDatatype{⊢}} \AgdaBound{C}\AgdaSymbol{)} and
\AgdaOperator{\AgdaFunction{\AgdaUnderscore{}[\AgdaUnderscore{}]}} \AgdaSymbol{:}
\AgdaSymbol{∀} \AgdaSymbol{\{}\AgdaBound{Γ}\AgdaSpace{} \AgdaBound{A}\AgdaSpace{}
	\AgdaBound{B}\AgdaSymbol{\}} \AgdaSymbol{→} \AgdaSpace{} \AgdaBound{Γ}\AgdaSpace{}
	\AgdaOperator{\AgdaInductiveConstructor{,}}\AgdaSpace{} \AgdaBound{B} 
	\AgdaOperator{\AgdaDatatype{⊢}}
\AgdaBound{A} \AgdaSymbol{→} \AgdaBound{Γ} \AgdaOperator{\AgdaDatatype{⊢}} 
	\AgdaBound{B} \AgdaSymbol{→} \AgdaBound{Γ}
	\AgdaOperator{\AgdaDatatype{⊢}} \AgdaBound{A}, respectively.
Since the implementation of these functions is fairly standard, we omit their definitions here.

Then, we present the reduction relation which is defined as the following data type:

\begin{code}%
	\>[0]\AgdaKeyword{data}\AgdaSpace{}%
	\AgdaOperator{\AgdaDatatype{\AgdaUnderscore{}↪\AgdaUnderscore{}}}\AgdaSpace{}%
	\AgdaSymbol{:}\AgdaSpace{}%
	\AgdaSymbol{(}\AgdaBound{Γ}\AgdaSpace{}%
	\AgdaOperator{\AgdaDatatype{⊢}}\AgdaSpace{}%
	\AgdaBound{A}\AgdaSymbol{)}\AgdaSpace{}%
	\AgdaSymbol{→}\AgdaSpace{}%
	\AgdaSymbol{(}\AgdaBound{Γ}\AgdaSpace{}%
	\AgdaOperator{\AgdaDatatype{⊢}}\AgdaSpace{}%
	\AgdaBound{A}\AgdaSymbol{)}\AgdaSpace{}%
	\AgdaSymbol{→}\AgdaSpace{}%
	\AgdaPrimitive{Set}\AgdaSpace{}%
	\AgdaKeyword{where}\<%
	\\
	\>[0][@{}l@{\AgdaIndent{0}}]%
	\>[2]\AgdaInductiveConstructor{β-ƛ}\AgdaSpace{}%
	\AgdaSymbol{:}\AgdaSpace{}%
	\AgdaSymbol{∀}\AgdaSpace{}%
	\AgdaSymbol{\{}\AgdaBound{t}\AgdaSpace{}%
	\AgdaSymbol{:}\AgdaSpace{}%
	\AgdaBound{Γ}\AgdaSpace{}%
	\AgdaOperator{\AgdaInductiveConstructor{,}}\AgdaSpace{}%
	\AgdaBound{A}\AgdaSpace{}%
	\AgdaOperator{\AgdaDatatype{⊢}}\AgdaSpace{}%
	\AgdaBound{B}\AgdaSymbol{\}}\AgdaSpace{}%
	\AgdaSymbol{\{}\AgdaBound{s}\AgdaSpace{}%
	\AgdaSymbol{:}\AgdaSpace{}%
	\AgdaBound{Γ}\AgdaSpace{}%
	\AgdaOperator{\AgdaDatatype{⊢}}\AgdaSpace{}%
	\AgdaBound{A}\AgdaSymbol{\}}\<%
	\\
	\>[0][@{}l@{\AgdaIndent{0}}]%
	\>[4]\AgdaSymbol{→}\AgdaSpace{}%
	\AgdaSymbol{(}\AgdaOperator{\AgdaInductiveConstructor{ƛ}}\AgdaSpace{}%
	\AgdaBound{t}\AgdaSymbol{)}\AgdaSpace{}%
	\AgdaOperator{\AgdaInductiveConstructor{·}}\AgdaSpace{}%
	\AgdaBound{s}\AgdaSpace{}%
	\AgdaOperator{\AgdaDatatype{↪}}\AgdaSpace{}%
	\AgdaBound{t}\AgdaSpace{}%
	\AgdaOperator{\AgdaFunction{[}}\AgdaSpace{}%
	\AgdaBound{s}\AgdaSpace{}%
	\AgdaOperator{\AgdaFunction{]}}\<%
	\\
	\\[\AgdaEmptyExtraSkip]%
	\>[2]\AgdaInductiveConstructor{β-π₁}\AgdaSpace{}%
	\AgdaSymbol{:}\AgdaSpace{}%
	\AgdaSymbol{∀}\AgdaSpace{}%
	\AgdaSymbol{\{}\AgdaBound{r}\AgdaSpace{}%
	\AgdaSymbol{:}\AgdaSpace{}%
	\AgdaBound{Γ}\AgdaSpace{}%
	\AgdaOperator{\AgdaDatatype{⊢}}\AgdaSpace{}%
	\AgdaBound{A}\AgdaSymbol{\}}\AgdaSpace{}%
	\AgdaSymbol{\{}\AgdaBound{s}\AgdaSpace{}%
	\AgdaSymbol{:}\AgdaSpace{}%
	\AgdaBound{Γ}\AgdaSpace{}%
	\AgdaOperator{\AgdaDatatype{⊢}}\AgdaSpace{}%
	\AgdaBound{B}\AgdaSymbol{\}}\<%
	\\
	\>[0][@{}l@{\AgdaIndent{0}}]%
	\>[4]\AgdaSymbol{→}\AgdaSpace{}%
	\AgdaInductiveConstructor{π}\AgdaSpace{}%
	\AgdaBound{A}\AgdaSpace{}%
	\AgdaSymbol{\{}\AgdaInductiveConstructor{inj₁}\AgdaSpace{}%
	\AgdaInductiveConstructor{refl}\AgdaSymbol{\}}\AgdaSpace{}%
	\AgdaOperator{\AgdaInductiveConstructor{⟨}}\AgdaSpace{}%
	\AgdaBound{r}\AgdaSpace{}%
	\AgdaOperator{\AgdaInductiveConstructor{,}}\AgdaSpace{}%
	\AgdaBound{s}\AgdaSpace{}%
	\AgdaOperator{\AgdaInductiveConstructor{⟩}}\AgdaSpace{}%
	\AgdaOperator{\AgdaDatatype{↪}}\AgdaSpace{}%
	\AgdaBound{r}\<%
	\\
	\\[\AgdaEmptyExtraSkip]%
	\>[0][@{}l@{\AgdaIndent{0}}]%
	\>[2]\AgdaInductiveConstructor{β-π₂}\AgdaSpace{}%
	\AgdaSymbol{:}\AgdaSpace{}%
	\AgdaSymbol{∀}\AgdaSpace{}%
	\AgdaSymbol{\{}\AgdaBound{r}\AgdaSpace{}%
	\AgdaSymbol{:}\AgdaSpace{}%
	\AgdaBound{Γ}\AgdaSpace{}%
	\AgdaOperator{\AgdaDatatype{⊢}}\AgdaSpace{}%
	\AgdaBound{A}\AgdaSymbol{\}}\AgdaSpace{}%
	\AgdaSymbol{\{}\AgdaBound{s}\AgdaSpace{}%
	\AgdaSymbol{:}\AgdaSpace{}%
	\AgdaBound{Γ}\AgdaSpace{}%
	\AgdaOperator{\AgdaDatatype{⊢}}\AgdaSpace{}%
	\AgdaBound{B}\AgdaSymbol{\}}\<%
	\\
	\>[0][@{}l@{\AgdaIndent{0}}]%
	\>[4]\AgdaSymbol{→}\AgdaSpace{}%
	\AgdaInductiveConstructor{π}\AgdaSpace{}%
	\AgdaBound{B}\AgdaSpace{}%
	\AgdaSymbol{\{}\AgdaInductiveConstructor{inj₂}\AgdaSpace{}%
	\AgdaInductiveConstructor{refl}\AgdaSymbol{\}}\AgdaSpace{}%
	\AgdaOperator{\AgdaInductiveConstructor{⟨}}\AgdaSpace{}%
	\AgdaBound{r}\AgdaSpace{}%
	\AgdaOperator{\AgdaInductiveConstructor{,}}\AgdaSpace{}%
	\AgdaBound{s}\AgdaSpace{}%
	\AgdaOperator{\AgdaInductiveConstructor{⟩}}\AgdaSpace{}%
	\AgdaOperator{\AgdaDatatype{↪}}\AgdaSpace{}%
	\AgdaBound{s}\<%
\end{code}

In this definition we omit the constructors \AgdaInductiveConstructor{ξ-·₁},
\AgdaInductiveConstructor{ξ-·₂}, \AgdaInductiveConstructor{ξ-⟨,⟩₁},
\AgdaInductiveConstructor{ξ-⟨,⟩₂}, \AgdaInductiveConstructor{ξ-π},
\AgdaInductiveConstructor{ξ-≡} and \AgdaInductiveConstructor{ζ} that represent the
congruence rules. The constructor \AgdaInductiveConstructor{β-λ} corresponds to beta reduction, and the constructors \AgdaInductiveConstructor{β-π1} and \AgdaInductiveConstructor{β-π2} correspond to the applications of the
projections.

Since the relation \AgdaOperator{\AgdaDatatype{\AgdaUnderscore{}↪\AgdaUnderscore{}}} is indexed by two terms of the same type,
it is not necessary to prove that it preserves types. This is one of
the advantages of using intrinsically typed terms.

\subsection{Type isomorphisms}

The type isomorphisms included in this formalization correspond to the axiomatic set
given in Section 2, the isomorphism that represents the
symmetry of $\equiv$, and some isomorphisms that represent congruence rules.

\begin{code}%
\>[0]\AgdaKeyword{data}\AgdaSpace{}%
\AgdaOperator{\AgdaDatatype{\AgdaUnderscore{}≡\AgdaUnderscore{}}}\AgdaSpace{}%
\AgdaSymbol{:}\AgdaSpace{}%
\AgdaDatatype{Type}\AgdaSpace{}%
\AgdaSymbol{→}\AgdaSpace{}%
\AgdaDatatype{Type}\AgdaSpace{}%
\AgdaSymbol{→}\AgdaSpace{}%
\AgdaPrimitive{Set}\AgdaSpace{}%
\AgdaKeyword{where}\<%
\\
\>[0][@{}l@{\AgdaIndent{0}}]%
\>[2]\AgdaInductiveConstructor{comm}%
\>[9]\AgdaSymbol{:}\AgdaSpace{}%
\AgdaSymbol{∀}\AgdaSpace{}%
\AgdaSymbol{\{}\AgdaBound{A}\AgdaSpace{}%
\AgdaBound{B}\AgdaSymbol{\}}\AgdaSpace{}%
\AgdaSymbol{→}\AgdaSpace{}%
\AgdaBound{A}\AgdaSpace{}%
\AgdaOperator{\AgdaInductiveConstructor{×}}\AgdaSpace{}%
\AgdaBound{B}\AgdaSpace{}%
\AgdaOperator{\AgdaDatatype{≡}}\AgdaSpace{}%
\AgdaBound{B}\AgdaSpace{}%
\AgdaOperator{\AgdaInductiveConstructor{×}}\AgdaSpace{}%
\AgdaBound{A}\<%
\\
\>[2]\AgdaInductiveConstructor{asso}%
\>[9]\AgdaSymbol{:}\AgdaSpace{}%
\AgdaSymbol{∀}\AgdaSpace{}%
\AgdaSymbol{\{}\AgdaBound{A}\AgdaSpace{}%
\AgdaBound{B}\AgdaSpace{}%
\AgdaBound{C}\AgdaSymbol{\}}\AgdaSpace{}%
\AgdaSymbol{→}\AgdaSpace{}%
\AgdaBound{A}\AgdaSpace{}%
\AgdaOperator{\AgdaInductiveConstructor{×}}\AgdaSpace{}%
\AgdaSymbol{(}\AgdaBound{B}\AgdaSpace{}%
\AgdaOperator{\AgdaInductiveConstructor{×}}\AgdaSpace{}%
\AgdaBound{C}\AgdaSymbol{)}\AgdaSpace{}%
\AgdaOperator{\AgdaDatatype{≡}}\AgdaSpace{}%
\AgdaSymbol{(}\AgdaBound{A}\AgdaSpace{}%
\AgdaOperator{\AgdaInductiveConstructor{×}}\AgdaSpace{}%
\AgdaBound{B}\AgdaSymbol{)}\AgdaSpace{}%
\AgdaOperator{\AgdaInductiveConstructor{×}}\AgdaSpace{}%
\AgdaBound{C}\<%
\\
\>[2]\AgdaInductiveConstructor{dist}%
\>[9]\AgdaSymbol{:}\AgdaSpace{}%
\AgdaSymbol{∀}\AgdaSpace{}%
\AgdaSymbol{\{}\AgdaBound{A}\AgdaSpace{}%
\AgdaBound{B}\AgdaSpace{}%
\AgdaBound{C}\AgdaSymbol{\}}\AgdaSpace{}%
\AgdaSymbol{→}\AgdaSpace{}%
\AgdaSymbol{(}\AgdaBound{A}\AgdaSpace{}%
\AgdaOperator{\AgdaInductiveConstructor{⇒}}\AgdaSpace{}%
\AgdaBound{B}\AgdaSymbol{)}\AgdaSpace{}%
\AgdaOperator{\AgdaInductiveConstructor{×}}\AgdaSpace{}%
\AgdaSymbol{(}\AgdaBound{A}\AgdaSpace{}%
\AgdaOperator{\AgdaInductiveConstructor{⇒}}\AgdaSpace{}%
\AgdaBound{C}\AgdaSymbol{)}\AgdaSpace{}%
\AgdaOperator{\AgdaDatatype{≡}}\AgdaSpace{}%
\AgdaBound{A}\AgdaSpace{}%
\AgdaOperator{\AgdaInductiveConstructor{⇒}}\AgdaSpace{}%
\AgdaBound{B}\AgdaSpace{}%
\AgdaOperator{\AgdaInductiveConstructor{×}}\AgdaSpace{}%
\AgdaBound{C}\<%
\\
\>[2]\AgdaInductiveConstructor{curry}%
\>[9]\AgdaSymbol{:}\AgdaSpace{}%
\AgdaSymbol{∀}\AgdaSpace{}%
\AgdaSymbol{\{}\AgdaBound{A}\AgdaSpace{}%
\AgdaBound{B}\AgdaSpace{}%
\AgdaBound{C}\AgdaSymbol{\}}\AgdaSpace{}%
\AgdaSymbol{→}\AgdaSpace{}%
\AgdaBound{A}\AgdaSpace{}%
\AgdaOperator{\AgdaInductiveConstructor{⇒}}\AgdaSpace{}%
\AgdaBound{B}\AgdaSpace{}%
\AgdaOperator{\AgdaInductiveConstructor{⇒}}\AgdaSpace{}%
\AgdaBound{C}\AgdaSpace{}%
\AgdaOperator{\AgdaDatatype{≡}}\AgdaSpace{}%
\AgdaSymbol{(}\AgdaBound{A}\AgdaSpace{}%
\AgdaOperator{\AgdaInductiveConstructor{×}}\AgdaSpace{}%
\AgdaBound{B}\AgdaSymbol{)}\AgdaSpace{}%
\AgdaOperator{\AgdaInductiveConstructor{⇒}}\AgdaSpace{}%
\AgdaBound{C}\<%
\\
\>[2]\AgdaInductiveConstructor{id-×}%
\>[9]\AgdaSymbol{:}\AgdaSpace{}%
\AgdaSymbol{∀}\AgdaSpace{}%
\AgdaSymbol{\{}\AgdaBound{A}\AgdaSymbol{\}}\AgdaSpace{}%
\AgdaSymbol{→}\AgdaSpace{}%
\AgdaBound{A}\AgdaSpace{}%
\AgdaOperator{\AgdaInductiveConstructor{×}}\AgdaSpace{}%
\AgdaInductiveConstructor{⊤}\AgdaSpace{}%
\AgdaOperator{\AgdaDatatype{≡}}\AgdaSpace{}%
\AgdaBound{A}\<%
\\
\>[2]\AgdaInductiveConstructor{id-⇒}%
\>[8]\AgdaSymbol{:}\AgdaSpace{}%
\AgdaSymbol{∀}\AgdaSpace{}%
\AgdaSymbol{\{}\AgdaBound{A}\AgdaSymbol{\}}\AgdaSpace{}%
\AgdaSymbol{→}\AgdaSpace{}%
\AgdaInductiveConstructor{⊤}\AgdaSpace{}%
\AgdaOperator{\AgdaInductiveConstructor{⇒}}\AgdaSpace{}%
\AgdaBound{A}\AgdaSpace{}%
\AgdaOperator{\AgdaDatatype{≡}}\AgdaSpace{}%
\AgdaBound{A}\<%
\\
\>[2]\AgdaInductiveConstructor{abs}%
\>[9]\AgdaSymbol{:}\AgdaSpace{}%
\AgdaSymbol{∀}\AgdaSpace{}%
\AgdaSymbol{\{}\AgdaBound{A}\AgdaSymbol{\}}\AgdaSpace{}%
\AgdaSymbol{→}\AgdaSpace{}%
\AgdaBound{A}\AgdaSpace{}%
\AgdaOperator{\AgdaInductiveConstructor{⇒}}\AgdaSpace{}%
\AgdaInductiveConstructor{⊤}\AgdaSpace{}%
\AgdaOperator{\AgdaDatatype{≡}}\AgdaSpace{}%
\AgdaInductiveConstructor{⊤}\<%
\\
\\[\AgdaEmptyExtraSkip]%
\>[2]\AgdaInductiveConstructor{sym}%
\>[11]\AgdaSymbol{:}\AgdaSpace{}%
\AgdaSymbol{∀}\AgdaSpace{}%
\AgdaSymbol{\{}\AgdaBound{A}\AgdaSpace{}%
\AgdaBound{B}\AgdaSymbol{\}}\AgdaSpace{}%
\AgdaSymbol{→}\AgdaSpace{}%
\AgdaBound{A}\AgdaSpace{}%
\AgdaOperator{\AgdaDatatype{≡}}\AgdaSpace{}%
\AgdaBound{B}\AgdaSpace{}%
\AgdaSymbol{→}\AgdaSpace{}%
\AgdaBound{B}\AgdaSpace{}%
\AgdaOperator{\AgdaDatatype{≡}}\AgdaSpace{}%
\AgdaBound{A}\<%
\\
\>[2]\AgdaInductiveConstructor{cong⇒₁}%
\>[11]\AgdaSymbol{:}\AgdaSpace{}%
\AgdaSymbol{∀}\AgdaSpace{}%
\AgdaSymbol{\{}\AgdaBound{A}\AgdaSpace{}%
\AgdaBound{B}\AgdaSpace{}%
\AgdaBound{C}\AgdaSymbol{\}}\AgdaSpace{}%
\AgdaSymbol{→}\AgdaSpace{}%
\AgdaBound{A}\AgdaSpace{}%
\AgdaOperator{\AgdaDatatype{≡}}\AgdaSpace{}%
\AgdaBound{B}\AgdaSpace{}%
\AgdaSymbol{→}\AgdaSpace{}%
\AgdaBound{A}\AgdaSpace{}%
\AgdaOperator{\AgdaInductiveConstructor{⇒}}\AgdaSpace{}%
\AgdaBound{C}\AgdaSpace{}%
\AgdaOperator{\AgdaDatatype{≡}}\AgdaSpace{}%
\AgdaBound{B}\AgdaSpace{}%
\AgdaOperator{\AgdaInductiveConstructor{⇒}}\AgdaSpace{}%
\AgdaBound{C}\<%
\\
\>[2]\AgdaInductiveConstructor{cong⇒₂}%
\>[11]\AgdaSymbol{:}\AgdaSpace{}%
\AgdaSymbol{∀}\AgdaSpace{}%
\AgdaSymbol{\{}\AgdaBound{A}\AgdaSpace{}%
\AgdaBound{B}\AgdaSpace{}%
\AgdaBound{C}\AgdaSymbol{\}}\AgdaSpace{}%
\AgdaSymbol{→}\AgdaSpace{}%
\AgdaBound{A}\AgdaSpace{}%
\AgdaOperator{\AgdaDatatype{≡}}\AgdaSpace{}%
\AgdaBound{B}\AgdaSpace{}%
\AgdaSymbol{→}\AgdaSpace{}%
\AgdaBound{C}\AgdaSpace{}%
\AgdaOperator{\AgdaInductiveConstructor{⇒}}\AgdaSpace{}%
\AgdaBound{A}\AgdaSpace{}%
\AgdaOperator{\AgdaDatatype{≡}}\AgdaSpace{}%
\AgdaBound{C}\AgdaSpace{}%
\AgdaOperator{\AgdaInductiveConstructor{⇒}}\AgdaSpace{}%
\AgdaBound{B}\<%
\\
\>[2]\AgdaInductiveConstructor{cong×₁}%
\>[11]\AgdaSymbol{:}\AgdaSpace{}%
\AgdaSymbol{∀}\AgdaSpace{}%
\AgdaSymbol{\{}\AgdaBound{A}\AgdaSpace{}%
\AgdaBound{B}\AgdaSpace{}%
\AgdaBound{C}\AgdaSymbol{\}}\AgdaSpace{}%
\AgdaSymbol{→}\AgdaSpace{}%
\AgdaBound{A}\AgdaSpace{}%
\AgdaOperator{\AgdaDatatype{≡}}\AgdaSpace{}%
\AgdaBound{B}\AgdaSpace{}%
\AgdaSymbol{→}\AgdaSpace{}%
\AgdaBound{A}\AgdaSpace{}%
\AgdaOperator{\AgdaInductiveConstructor{×}}\AgdaSpace{}%
\AgdaBound{C}\AgdaSpace{}%
\AgdaOperator{\AgdaDatatype{≡}}\AgdaSpace{}%
\AgdaBound{B}\AgdaSpace{}%
\AgdaOperator{\AgdaInductiveConstructor{×}}\AgdaSpace{}%
\AgdaBound{C}\<%
\\
\>[2]\AgdaInductiveConstructor{cong×₂}%
\>[11]\AgdaSymbol{:}\AgdaSpace{}%
\AgdaSymbol{∀}\AgdaSpace{}%
\AgdaSymbol{\{}\AgdaBound{A}\AgdaSpace{}%
\AgdaBound{B}\AgdaSpace{}%
\AgdaBound{C}\AgdaSymbol{\}}\AgdaSpace{}%
\AgdaSymbol{→}\AgdaSpace{}%
\AgdaBound{A}\AgdaSpace{}%
\AgdaOperator{\AgdaDatatype{≡}}\AgdaSpace{}%
\AgdaBound{B}\AgdaSpace{}%
\AgdaSymbol{→}\AgdaSpace{}%
\AgdaBound{C}\AgdaSpace{}%
\AgdaOperator{\AgdaInductiveConstructor{×}}\AgdaSpace{}%
\AgdaBound{A}\AgdaSpace{}%
\AgdaOperator{\AgdaDatatype{≡}}\AgdaSpace{}%
\AgdaBound{C}\AgdaSpace{}%
\AgdaOperator{\AgdaInductiveConstructor{×}}\AgdaSpace{}%
\AgdaBound{B}\<%
\end{code}

We do not include in the formalization the isomorphism corresponding to
the transitivity of $\equiv$, since it can be obtained using the constructor
\AgdaInductiveConstructor{[\_]≡\_} that corresponds to the typing rule $(\equiv)$,
and the reflexivity of $\equiv$, since it does not add expressiveness to the formalization.

\subsection{Equivalence of terms}
The formalization of the equivalence relation between terms, corresponding to isomorphic types ($\rightleftarrows$) is presented below:

\begin{code}
	\>[0]\AgdaKeyword{data}\AgdaSpace{}%
	\AgdaOperator{\AgdaDatatype{\AgdaUnderscore{}⇄\AgdaUnderscore{}}}\AgdaSpace{}%
	\AgdaSymbol{:}\AgdaSpace{}%
	\AgdaSymbol{(}\AgdaBound{Γ}\AgdaSpace{}%
	\AgdaOperator{\AgdaDatatype{⊢}}\AgdaSpace{}%
	\AgdaBound{A}\AgdaSymbol{)}\AgdaSpace{}%
	\AgdaSymbol{→}\AgdaSpace{}%
	\AgdaSymbol{(}\AgdaBound{Γ}\AgdaSpace{}%
	\AgdaOperator{\AgdaDatatype{⊢}}\AgdaSpace{}%
	\AgdaBound{A}\AgdaSymbol{)}\AgdaSpace{}%
	\AgdaSymbol{→}\AgdaSpace{}%
	\AgdaPrimitive{Set}\AgdaSpace{}%
	\AgdaKeyword{where}\<%
	\\
	\>[0][@{}l@{\AgdaIndent{0}}]%
	\>[2]\AgdaInductiveConstructor{comm}\AgdaSpace{}%
	\AgdaSymbol{:}\AgdaSpace{}%
	\AgdaSymbol{∀}\AgdaSpace{}%
	\AgdaSymbol{\{}\AgdaBound{r}\AgdaSpace{}%
	\AgdaSymbol{:}\AgdaSpace{}%
	\AgdaBound{Γ}\AgdaSpace{}%
	\AgdaOperator{\AgdaDatatype{⊢}}\AgdaSpace{}%
	\AgdaBound{A}\AgdaSymbol{\}}\AgdaSpace{}%
	\AgdaSymbol{→}\AgdaSpace{}%
	\AgdaSymbol{\{}\AgdaBound{s}\AgdaSpace{}%
	\AgdaSymbol{:}\AgdaSpace{}%
	\AgdaBound{Γ}\AgdaSpace{}%
	\AgdaOperator{\AgdaDatatype{⊢}}\AgdaSpace{}%
	\AgdaBound{B}\AgdaSymbol{\}}\<%
	\\
	\>[2][@{}l@{\AgdaIndent{0}}]%
	\>[4]\AgdaSymbol{→}\AgdaSpace{}%
	\AgdaOperator{\AgdaInductiveConstructor{[}}\AgdaSpace{}%
	\AgdaInductiveConstructor{comm}\AgdaSpace{}%
	\AgdaOperator{\AgdaInductiveConstructor{]≡}}\AgdaSpace{}%
	\AgdaOperator{\AgdaInductiveConstructor{⟨}}\AgdaSpace{}%
	\AgdaBound{r}\AgdaSpace{}%
	\AgdaOperator{\AgdaInductiveConstructor{,}}\AgdaSpace{}%
	\AgdaBound{s}\AgdaSpace{}%
	\AgdaOperator{\AgdaInductiveConstructor{⟩}}\AgdaSpace{}%
	\AgdaOperator{\AgdaDatatype{⇄}}\AgdaSpace{}%
	\AgdaOperator{\AgdaInductiveConstructor{⟨}}\AgdaSpace{}%
	\AgdaBound{s}\AgdaSpace{}%
	\AgdaOperator{\AgdaInductiveConstructor{,}}\AgdaSpace{}%
	\AgdaBound{r}\AgdaSpace{}%
	\AgdaOperator{\AgdaInductiveConstructor{⟩}}\<%
	\\
	\\[\AgdaEmptyExtraSkip]%
	\>[2]\AgdaInductiveConstructor{asso}\AgdaSpace{}%
	\AgdaSymbol{:}\AgdaSpace{}%
	\AgdaSymbol{∀}\AgdaSpace{}%
	\AgdaSymbol{\{}\AgdaBound{r}\AgdaSpace{}%
	\AgdaSymbol{:}\AgdaSpace{}%
	\AgdaBound{Γ}\AgdaSpace{}%
	\AgdaOperator{\AgdaDatatype{⊢}}\AgdaSpace{}%
	\AgdaBound{A}\AgdaSymbol{\}}\AgdaSpace{}%
	\AgdaSymbol{→}\AgdaSpace{}%
	\AgdaSymbol{\{}\AgdaBound{s}\AgdaSpace{}%
	\AgdaSymbol{:}\AgdaSpace{}%
	\AgdaBound{Γ}\AgdaSpace{}%
	\AgdaOperator{\AgdaDatatype{⊢}}\AgdaSpace{}%
	\AgdaBound{B}\AgdaSymbol{\}}\AgdaSpace{}%
	\AgdaSymbol{→}\AgdaSpace{}%
	\AgdaSymbol{\{}\AgdaBound{t}\AgdaSpace{}%
	\AgdaSymbol{:}\AgdaSpace{}%
	\AgdaBound{Γ}\AgdaSpace{}%
	\AgdaOperator{\AgdaDatatype{⊢}}\AgdaSpace{}%
	\AgdaBound{C}\AgdaSymbol{\}}\<%
	\\
	\>[2][@{}l@{\AgdaIndent{0}}]%
	\>[4]\AgdaSymbol{→}\AgdaSpace{}%
	\AgdaOperator{\AgdaInductiveConstructor{[}}\AgdaSpace{}%
	\AgdaInductiveConstructor{asso}\AgdaSpace{}%
	\AgdaOperator{\AgdaInductiveConstructor{]≡}}\AgdaSpace{}%
	\AgdaOperator{\AgdaInductiveConstructor{⟨}}\AgdaSpace{}%
	\AgdaBound{r}\AgdaSpace{}%
	\AgdaOperator{\AgdaInductiveConstructor{,}}\AgdaSpace{}%
	\AgdaOperator{\AgdaInductiveConstructor{⟨}}\AgdaSpace{}%
	\AgdaBound{s}\AgdaSpace{}%
	\AgdaOperator{\AgdaInductiveConstructor{,}}\AgdaSpace{}%
	\AgdaBound{t}\AgdaSpace{}%
	\AgdaOperator{\AgdaInductiveConstructor{⟩}}\AgdaSpace{}%
	\AgdaOperator{\AgdaInductiveConstructor{⟩}}\AgdaSpace{}%
	\AgdaOperator{\AgdaDatatype{⇄}}\AgdaSpace{}%
	\AgdaOperator{\AgdaInductiveConstructor{⟨}}\AgdaSpace{}%
	\AgdaOperator{\AgdaInductiveConstructor{⟨}}\AgdaSpace{}%
	\AgdaBound{r}\AgdaSpace{}%
	\AgdaOperator{\AgdaInductiveConstructor{,}}\AgdaSpace{}%
	\AgdaBound{s}\AgdaSpace{}%
	\AgdaOperator{\AgdaInductiveConstructor{⟩}}\AgdaSpace{}%
	\AgdaOperator{\AgdaInductiveConstructor{,}}\AgdaSpace{}%
	\AgdaBound{t}\AgdaSpace{}%
	\AgdaOperator{\AgdaInductiveConstructor{⟩}}\<%
	\\
	\\[\AgdaEmptyExtraSkip]%
	\>[2]\AgdaInductiveConstructor{dist-ƛ}\AgdaSpace{}%
	\AgdaSymbol{:}\AgdaSpace{}%
	\AgdaSymbol{∀}\AgdaSpace{}%
	\AgdaSymbol{\{}\AgdaBound{r}\AgdaSpace{}%
	\AgdaSymbol{:}\AgdaSpace{}%
	\AgdaBound{Γ}\AgdaSpace{}%
	\AgdaOperator{\AgdaInductiveConstructor{,}}\AgdaSpace{}%
	\AgdaBound{C}\AgdaSpace{}%
	\AgdaOperator{\AgdaDatatype{⊢}}\AgdaSpace{}%
	\AgdaBound{A}\AgdaSymbol{\}}\AgdaSpace{}%
	\AgdaSymbol{→}\AgdaSpace{}%
	\AgdaSymbol{\{}\AgdaBound{s}\AgdaSpace{}%
	\AgdaSymbol{:}\AgdaSpace{}%
	\AgdaBound{Γ}\AgdaSpace{}%
	\AgdaOperator{\AgdaInductiveConstructor{,}}\AgdaSpace{}%
	\AgdaBound{C}\AgdaSpace{}%
	\AgdaOperator{\AgdaDatatype{⊢}}\AgdaSpace{}%
	\AgdaBound{B}\AgdaSymbol{\}}\<%
	\\
	\>[2][@{}l@{\AgdaIndent{0}}]%
	\>[4]\AgdaSymbol{→}\AgdaSpace{}%
	\AgdaOperator{\AgdaInductiveConstructor{[}}\AgdaSpace{}%
	\AgdaInductiveConstructor{dist}\AgdaSpace{}%
	\AgdaOperator{\AgdaInductiveConstructor{]≡}}\AgdaSpace{}%
	\AgdaOperator{\AgdaInductiveConstructor{⟨}}\AgdaSpace{}%
	\AgdaOperator{\AgdaInductiveConstructor{ƛ}}\AgdaSpace{}%
	\AgdaBound{r}\AgdaSpace{}%
	\AgdaOperator{\AgdaInductiveConstructor{,}}\AgdaSpace{}%
	\AgdaOperator{\AgdaInductiveConstructor{ƛ}}\AgdaSpace{}%
	\AgdaBound{s}\AgdaSpace{}%
	\AgdaOperator{\AgdaInductiveConstructor{⟩}}\AgdaSpace{}%
	\AgdaOperator{\AgdaDatatype{⇄}}\AgdaSpace{}%
	\AgdaOperator{\AgdaInductiveConstructor{ƛ}}\AgdaSpace{}%
	\AgdaOperator{\AgdaInductiveConstructor{⟨}}\AgdaSpace{}%
	\AgdaBound{r}\AgdaSpace{}%
	\AgdaOperator{\AgdaInductiveConstructor{,}}\AgdaSpace{}%
	\AgdaBound{s}\AgdaSpace{}%
	\AgdaOperator{\AgdaInductiveConstructor{⟩}}\<%
	\\
	\\
	\>[2]\AgdaInductiveConstructor{curry}\AgdaSpace{}%
	\AgdaSymbol{:}\AgdaSpace{}%
	\AgdaSymbol{∀}\AgdaSpace{}%
	\AgdaSymbol{\{}\AgdaBound{r}\AgdaSpace{}%
	\AgdaSymbol{:}\AgdaSpace{}%
	\AgdaBound{Γ}\AgdaSpace{}%
	\AgdaOperator{\AgdaInductiveConstructor{,}}\AgdaSpace{}%
	\AgdaBound{A}\AgdaSpace{}%
	\AgdaOperator{\AgdaInductiveConstructor{,}}\AgdaSpace{}%
	\AgdaBound{B}\AgdaSpace{}%
	\AgdaOperator{\AgdaDatatype{⊢}}\AgdaSpace{}%
	\AgdaBound{C}\AgdaSymbol{\}}\<%
	\\
	\>[2][@{}l@{\AgdaIndent{0}}]%
	\>[4]\AgdaSymbol{→}\AgdaSpace{}%
	\AgdaOperator{\AgdaInductiveConstructor{[}}\AgdaSpace{}%
	\AgdaInductiveConstructor{curry}\AgdaSpace{}%
	\AgdaOperator{\AgdaInductiveConstructor{]≡}}\AgdaSpace{}%
	\AgdaSymbol{(}\AgdaOperator{\AgdaInductiveConstructor{ƛ}}\AgdaSpace{}%
	\AgdaOperator{\AgdaInductiveConstructor{ƛ}}\AgdaSpace{}%
	\AgdaBound{r}\AgdaSymbol{)}\AgdaSpace{}%
	\AgdaOperator{\AgdaDatatype{⇄}}\AgdaSpace{}%
	\AgdaOperator{\AgdaInductiveConstructor{ƛ}}\AgdaSpace{}%
	\AgdaFunction{subst}\AgdaSpace{}%
	\AgdaFunction{σ-curry}\AgdaSpace{}%
	\AgdaBound{r}\<%
	\\
	\>[2]\AgdaInductiveConstructor{uncurry}\AgdaSpace{}%
	\AgdaSymbol{:}\AgdaSpace{}%
	\AgdaSymbol{∀}\AgdaSpace{}%
	\AgdaSymbol{\{}\AgdaBound{r}\AgdaSpace{}%
	\AgdaSymbol{:}\AgdaSpace{}%
	\AgdaBound{Γ}\AgdaSpace{}%
	\AgdaOperator{\AgdaInductiveConstructor{,}}\AgdaSpace{}%
	\AgdaBound{A}\AgdaSpace{}%
	\AgdaOperator{\AgdaInductiveConstructor{×}}\AgdaSpace{}%
	\AgdaBound{B}\AgdaSpace{}%
	\AgdaOperator{\AgdaDatatype{⊢}}\AgdaSpace{}%
	\AgdaBound{C}\AgdaSymbol{\}}\<%
	\\
	\>[2][@{}l@{\AgdaIndent{0}}]%
	\>[4]\AgdaSymbol{→}\AgdaSpace{}%
	\AgdaOperator{\AgdaInductiveConstructor{[}}\AgdaSpace{}%
	\AgdaInductiveConstructor{sym}\AgdaSpace{}%
	\AgdaInductiveConstructor{curry}\AgdaSpace{}%
	\AgdaOperator{\AgdaInductiveConstructor{]≡}}\AgdaSpace{}%
	\AgdaSymbol{(}\AgdaOperator{\AgdaInductiveConstructor{ƛ}}\AgdaSpace{}%
	\AgdaBound{r}\AgdaSymbol{)}\AgdaSpace{}%
	\AgdaOperator{\AgdaDatatype{⇄}}\AgdaSpace{}%
	\AgdaOperator{\AgdaInductiveConstructor{ƛ}}\AgdaSpace{}%
	\AgdaOperator{\AgdaInductiveConstructor{ƛ}}\AgdaSpace{}%
	\AgdaFunction{subst}\AgdaSpace{}%
	\AgdaFunction{σ-uncurry}\AgdaSpace{}%
	\AgdaBound{r}\<%
	\\ \\
	\>[2]\AgdaInductiveConstructor{split-asso}\AgdaSpace{}%
	\AgdaSymbol{:}\AgdaSpace{}%
	\AgdaSymbol{∀}\AgdaSpace{}%
	\AgdaSymbol{\{}\AgdaBound{r}\AgdaSpace{}%
	\AgdaSymbol{:}\AgdaSpace{}%
	\AgdaBound{Γ}\AgdaSpace{}%
	\AgdaOperator{\AgdaDatatype{⊢}}\AgdaSpace{}%
	\AgdaBound{A}\AgdaSymbol{\}}\AgdaSpace{}%
	\AgdaSymbol{→}\AgdaSpace{}%
	\AgdaSymbol{\{}\AgdaBound{s}\AgdaSpace{}%
	\AgdaSymbol{:}\AgdaSpace{}%
	\AgdaBound{Γ}\AgdaSpace{}%
	\AgdaOperator{\AgdaDatatype{⊢}}\AgdaSpace{}%
	\AgdaBound{B}\AgdaSpace{}%
	\AgdaOperator{\AgdaInductiveConstructor{×}}\AgdaSpace{}%
	\AgdaBound{C}\AgdaSymbol{\}}\<%
	\\
	\>[2][@{}l@{\AgdaIndent{0}}]%
	\>[4]\AgdaSymbol{→}\AgdaSpace{}%
	\AgdaOperator{\AgdaInductiveConstructor{[}}\AgdaSpace{}%
	\AgdaInductiveConstructor{asso}\AgdaSpace{}%
	\AgdaOperator{\AgdaInductiveConstructor{]≡}}\AgdaSpace{}%
	\AgdaOperator{\AgdaInductiveConstructor{⟨}}\AgdaSpace{}%
	\AgdaBound{r}\AgdaSpace{}%
	\AgdaOperator{\AgdaInductiveConstructor{,}}\AgdaSpace{}%
	\AgdaBound{s}\AgdaSpace{}%
	\AgdaOperator{\AgdaInductiveConstructor{⟩}}\AgdaSpace{}%
	\AgdaOperator{\AgdaDatatype{⇄}}
	\\
	\>[2][@{}l@{\AgdaIndent{0}}]%
	\>[4]\AgdaOperator{\AgdaInductiveConstructor{⟨}}\AgdaSpace{}%
	\AgdaOperator{\AgdaInductiveConstructor{⟨}}\AgdaSpace{}%
	\AgdaBound{r}\AgdaSpace{}%
	\AgdaOperator{\AgdaInductiveConstructor{,}}\AgdaSpace{}%
	\AgdaInductiveConstructor{π}\AgdaSpace{}%
	\AgdaBound{B}\AgdaSpace{}%
	\AgdaSymbol{\{}\AgdaInductiveConstructor{inj₁}\AgdaSpace{}%
	\AgdaInductiveConstructor{refl}\AgdaSymbol{\}}\AgdaSpace{}%
	\AgdaBound{s}\AgdaSpace{}%
	\AgdaOperator{\AgdaInductiveConstructor{⟩}}\AgdaSpace{}%
	\AgdaOperator{\AgdaInductiveConstructor{,}}\AgdaSpace{}%
	\AgdaInductiveConstructor{π}\AgdaSpace{}%
	\AgdaBound{C}\AgdaSpace{}%
	\AgdaSymbol{\{}\AgdaInductiveConstructor{inj₂}\AgdaSpace{}%
	\AgdaInductiveConstructor{refl}\AgdaSymbol{\}}\AgdaSpace{}%
	\AgdaBound{s}\AgdaSpace{}%
	\AgdaOperator{\AgdaInductiveConstructor{⟩}}\<%
	\\ \\
	\>[2]\AgdaInductiveConstructor{split-dist-ƛ}\AgdaSpace{}%
	\AgdaSymbol{:}\AgdaSpace{}%
	\AgdaSymbol{∀}\AgdaSpace{}%
	\AgdaSymbol{\{}\AgdaBound{r}\AgdaSpace{}%
	\AgdaSymbol{:}\AgdaSpace{}%
	\AgdaBound{Γ}\AgdaSpace{}%
	\AgdaOperator{\AgdaInductiveConstructor{,}}\AgdaSpace{}%
	\AgdaBound{C}\AgdaSpace{}%
	\AgdaOperator{\AgdaDatatype{⊢}}\AgdaSpace{}%
	\AgdaBound{A}\AgdaSymbol{\}}\AgdaSpace{}%
	\AgdaSymbol{→}\AgdaSpace{}%
	\AgdaSymbol{\{}\AgdaBound{s}\AgdaSpace{}%
	\AgdaSymbol{:}\AgdaSpace{}%
	\AgdaBound{Γ}\AgdaSpace{}%
	\AgdaOperator{\AgdaInductiveConstructor{,}}\AgdaSpace{}%
	\AgdaBound{C}\AgdaSpace{}%
	\AgdaOperator{\AgdaDatatype{⊢}}\AgdaSpace{}%
	\AgdaBound{B}\AgdaSymbol{\}}\<%
	\\
	\>[2][@{}l@{\AgdaIndent{0}}]%
	\>[4]\AgdaSymbol{→}\AgdaSpace{}%
	\AgdaOperator{\AgdaInductiveConstructor{[}}\AgdaSpace{}%
	\AgdaInductiveConstructor{sym}\AgdaSpace{}%
	\AgdaInductiveConstructor{dist}\AgdaSpace{}%
	\AgdaOperator{\AgdaInductiveConstructor{]≡}}\AgdaSpace{}%
	\AgdaSymbol{(}\AgdaOperator{\AgdaInductiveConstructor{ƛ}}\AgdaSpace{}%
	\AgdaOperator{\AgdaInductiveConstructor{⟨}}\AgdaSpace{}%
	\AgdaBound{r}\AgdaSpace{}%
	\AgdaOperator{\AgdaInductiveConstructor{,}}\AgdaSpace{}%
	\AgdaBound{s}\AgdaSpace{}%
	\AgdaOperator{\AgdaInductiveConstructor{⟩}}\AgdaSymbol{)}\AgdaSpace{}%
	\AgdaOperator{\AgdaDatatype{⇄}}\AgdaSpace{}%
	\AgdaOperator{\AgdaInductiveConstructor{⟨}}\AgdaSpace{}%
	\AgdaOperator{\AgdaInductiveConstructor{ƛ}}\AgdaSpace{}%
	\AgdaBound{r}\AgdaSpace{}%
	\AgdaOperator{\AgdaInductiveConstructor{,}}\AgdaSpace{}%
	\AgdaOperator{\AgdaInductiveConstructor{ƛ}}\AgdaSpace{}%
	\AgdaBound{s}\AgdaSpace{}%
	\AgdaOperator{\AgdaInductiveConstructor{⟩}}\<%
	\\ \\
	\>[2]\AgdaInductiveConstructor{η-dist-app}\AgdaSpace{}%
	\AgdaSymbol{:}\AgdaSpace{}%
	\AgdaSymbol{∀}\AgdaSpace{}%
	\AgdaSymbol{\{}\AgdaBound{r}\AgdaSpace{}%
	\AgdaSymbol{:}\AgdaSpace{}%
	\AgdaBound{Γ}\AgdaSpace{}%
	\AgdaOperator{\AgdaDatatype{⊢}}\AgdaSpace{}%
	\AgdaBound{C}\AgdaSpace{}%
	\AgdaOperator{\AgdaInductiveConstructor{⇒}}\AgdaSpace{}%
	\AgdaBound{A}\AgdaSymbol{\}}\AgdaSpace{}%
	\AgdaSymbol{→}\AgdaSpace{}%
	\AgdaSymbol{\{}\AgdaBound{s}\AgdaSpace{}%
	\AgdaSymbol{:}\AgdaSpace{}%
	\AgdaBound{Γ}\AgdaSpace{}%
	\AgdaOperator{\AgdaDatatype{⊢}}\AgdaSpace{}%
	\AgdaBound{C}\AgdaSpace{}%
	\AgdaOperator{\AgdaInductiveConstructor{⇒}}\AgdaSpace{}%
	\AgdaBound{B}\AgdaSymbol{\}}\<%
	\\
	\>[2][@{}l@{\AgdaIndent{0}}]%
	\>[4]\AgdaSymbol{→}\AgdaSpace{}%
	\AgdaOperator{\AgdaInductiveConstructor{[}}\AgdaSpace{}%
	\AgdaInductiveConstructor{dist}\AgdaSpace{}%
	\AgdaOperator{\AgdaInductiveConstructor{]≡}}\AgdaSpace{}%
	\AgdaOperator{\AgdaInductiveConstructor{⟨}}\AgdaSpace{}%
	\AgdaBound{r}\AgdaSpace{}%
	\AgdaOperator{\AgdaInductiveConstructor{,}}\AgdaSpace{}%
	\AgdaBound{s}\AgdaSpace{}%
	\AgdaOperator{\AgdaInductiveConstructor{⟩}}\AgdaSpace{}%
	\AgdaOperator{\AgdaDatatype{⇄}} 
	\\
	\>[2]
	\AgdaOperator{\AgdaInductiveConstructor{ƛ}}\AgdaSpace{}%
	\AgdaOperator{\AgdaInductiveConstructor{⟨}}\AgdaSpace{}%
	\AgdaFunction{rename}\AgdaSpace{}%
	\AgdaOperator{\AgdaInductiveConstructor{S\AgdaUnderscore{}}}\AgdaSpace{}%
	\AgdaBound{r}\AgdaSpace{}%
	\AgdaOperator{\AgdaInductiveConstructor{·}}\AgdaSpace{}%
	\AgdaOperator{\AgdaInductiveConstructor{`}}\AgdaSpace{}%
	\AgdaInductiveConstructor{Z}\AgdaSpace{}%
	\AgdaOperator{\AgdaInductiveConstructor{,}}\AgdaSpace{}%
	\AgdaFunction{rename}\AgdaSpace{}%
	\AgdaOperator{\AgdaInductiveConstructor{S\AgdaUnderscore{}}}\AgdaSpace{}%
	\AgdaBound{s}\AgdaSpace{}%
	\AgdaOperator{\AgdaInductiveConstructor{·}}\AgdaSpace{}%
	\AgdaOperator{\AgdaInductiveConstructor{`}}\AgdaSpace{}%
	\AgdaInductiveConstructor{Z}\AgdaSpace{}%
	\AgdaOperator{\AgdaInductiveConstructor{⟩}}\<%
		\\ \\
	\>[2]\AgdaInductiveConstructor{η-curry}\AgdaSpace{}%
	\AgdaSymbol{:}\AgdaSpace{}%
	\AgdaSymbol{∀}\AgdaSpace{}%
	\AgdaSymbol{\{}\AgdaBound{r}\AgdaSpace{}%
	\AgdaSymbol{:}\AgdaSpace{}%
	\AgdaBound{Γ}\AgdaSpace{}%
	\AgdaOperator{\AgdaInductiveConstructor{,}}\AgdaSpace{}%
	\AgdaBound{A}\AgdaSpace{}%
	\AgdaOperator{\AgdaDatatype{⊢}}\AgdaSpace{}%
	\AgdaBound{B}\AgdaSpace{}%
	\AgdaOperator{\AgdaInductiveConstructor{⇒}}\AgdaSpace{}%
	\AgdaBound{C}\AgdaSymbol{\}}\<%
	\\
	\>[2][@{}l@{\AgdaIndent{0}}]%
	\>[4]\AgdaSymbol{→}\AgdaSpace{}%
	\AgdaOperator{\AgdaInductiveConstructor{[}}\AgdaSpace{}%
	\AgdaInductiveConstructor{curry}\AgdaSpace{}%
	\AgdaOperator{\AgdaInductiveConstructor{]≡}}\AgdaSpace{}%
	\AgdaSymbol{(}\AgdaOperator{\AgdaInductiveConstructor{ƛ}}\AgdaSpace{}%
	\AgdaBound{r}\AgdaSymbol{)}\AgdaSpace{}%
	\AgdaOperator{\AgdaDatatype{⇄}} \<%
	\\
	\>[2][@{}l@{\AgdaIndent{0}}]%
	\>[4]
	\AgdaOperator{\AgdaInductiveConstructor{ƛ}}\AgdaSpace{}%
	\AgdaFunction{subst}\AgdaSpace{}%
	\AgdaFunction{σ-curry}\AgdaSpace{}%
	\AgdaSymbol{(}\AgdaFunction{rename}\AgdaSpace{}%
	\AgdaOperator{\AgdaInductiveConstructor{S\AgdaUnderscore{}}}\AgdaSpace{}%
	\AgdaBound{r}\AgdaSpace{}%
	\AgdaOperator{\AgdaInductiveConstructor{·}}\AgdaSpace{}%
	\AgdaOperator{\AgdaInductiveConstructor{`}}\AgdaSpace{}%
	\AgdaInductiveConstructor{Z}\AgdaSymbol{)}\<%
		\\ \\
	\>[2]\AgdaInductiveConstructor{id-×}\AgdaSpace{}%
	\AgdaSymbol{:}\AgdaSpace{}%
	\AgdaSymbol{∀}\AgdaSpace{}%
	\AgdaSymbol{\{}\AgdaBound{r}\AgdaSpace{}%
	\AgdaSymbol{:}\AgdaSpace{}%
	\AgdaBound{Γ}\AgdaSpace{}%
	\AgdaOperator{\AgdaDatatype{⊢}}\AgdaSpace{}%
	\AgdaBound{A}\AgdaSymbol{\}}\AgdaSpace{}%
	\AgdaSymbol{→}\AgdaSpace{}%
	\AgdaSymbol{\{}\AgdaBound{t}\AgdaSpace{}%
	\AgdaSymbol{:}\AgdaSpace{}%
	\AgdaBound{Γ}\AgdaSpace{}%
	\AgdaOperator{\AgdaDatatype{⊢}}\AgdaSpace{}%
	\AgdaInductiveConstructor{⊤}\AgdaSymbol{\}}\<%
	\\
	\>[2][@{}l@{\AgdaIndent{0}}]%
	\>[4]\AgdaSymbol{→}\AgdaSpace{}%
	\AgdaOperator{\AgdaInductiveConstructor{[}}\AgdaSpace{}%
	\AgdaInductiveConstructor{id-×}\AgdaSpace{}%
	\AgdaOperator{\AgdaInductiveConstructor{]≡}}\AgdaSpace{}%
	\AgdaOperator{\AgdaInductiveConstructor{⟨}}\AgdaSpace{}%
	\AgdaBound{r}\AgdaSpace{}%
	\AgdaOperator{\AgdaInductiveConstructor{,}}\AgdaSpace{}%
	\AgdaBound{t}\AgdaSpace{}%
	\AgdaOperator{\AgdaInductiveConstructor{⟩}}\AgdaSpace{}%
	\AgdaOperator{\AgdaDatatype{⇄}}\AgdaSpace{}%
	\AgdaBound{r}\<%
	\\ \\
	\>[2]\AgdaInductiveConstructor{abs-e}\AgdaSpace{}%
	\AgdaSymbol{:}\AgdaSpace{}%
	\AgdaSymbol{∀}\AgdaSpace{}%
	\AgdaSymbol{\{}\AgdaBound{r}\AgdaSpace{}%
	\AgdaSymbol{:}\AgdaSpace{}%
	\AgdaBound{Γ}\AgdaSpace{}%
	\AgdaOperator{\AgdaDatatype{⊢}}\AgdaSpace{}%
	\AgdaBound{A}\AgdaSpace{}%
	\AgdaOperator{\AgdaInductiveConstructor{⇒}}\AgdaSpace{}%
	\AgdaInductiveConstructor{⊤}\AgdaSymbol{\}} 
	\AgdaSpace{} \AgdaSymbol{→}\AgdaSpace{}%
	\AgdaOperator{\AgdaInductiveConstructor{[}}\AgdaSpace{}%
	\AgdaInductiveConstructor{abs}\AgdaSpace{}%
	\AgdaOperator{\AgdaInductiveConstructor{]≡}}\AgdaSpace{}%
	\AgdaBound{r}\AgdaSpace{}%
	\AgdaOperator{\AgdaDatatype{⇄}}\AgdaSpace{}%
	\AgdaInductiveConstructor{⋆}\<%
	\\ \\
	\>[2]\AgdaInductiveConstructor{abs-i}\AgdaSpace{}%
	\AgdaSymbol{:}\AgdaSpace{}%
	\AgdaSymbol{∀}\AgdaSpace{}%
	\AgdaSymbol{\{}\AgdaBound{t}\AgdaSpace{}%
	\AgdaSymbol{:}\AgdaSpace{}%
	\AgdaBound{Γ}\AgdaSpace{}%
	\AgdaOperator{\AgdaDatatype{⊢}}\AgdaSpace{}%
	\AgdaInductiveConstructor{⊤}\AgdaSymbol{\}}\<%
	\\
	\>[2][@{}l@{\AgdaIndent{0}}]%
	\>[4]\AgdaSymbol{→}\AgdaSpace{}%
	\AgdaOperator{\AgdaInductiveConstructor{[}}\AgdaSpace{}%
	\AgdaInductiveConstructor{sym}\AgdaSpace{}%
	\AgdaInductiveConstructor{abs}\AgdaSpace{}%
	\AgdaOperator{\AgdaInductiveConstructor{]≡}}\AgdaSpace{}%
	\AgdaBound{t}\AgdaSpace{}%
	\AgdaOperator{\AgdaDatatype{⇄}}\AgdaSpace{}%
	\AgdaOperator{\AgdaInductiveConstructor{ƛ}}\AgdaSpace{}%
	\AgdaFunction{rename}\AgdaSpace{}%
	\AgdaSymbol{(}\AgdaOperator{\AgdaInductiveConstructor{S\AgdaUnderscore{}}}\AgdaSpace{}%
	\AgdaSymbol{\{}\AgdaArgument{A}\AgdaSpace{}%
	\AgdaSymbol{=}\AgdaSpace{}%
	\AgdaBound{A}\AgdaSymbol{\})}\AgdaSpace{}%
	\AgdaBound{t}\AgdaSpace{}%
	\\
	\\
	%
 	\>[2]\AgdaInductiveConstructor{id⇒i}\AgdaSpace{}%
	\AgdaSymbol{:}\AgdaSpace{}%
	\AgdaSymbol{∀}\AgdaSpace{}%
	\AgdaSymbol{\{}\AgdaBound{r}\AgdaSpace{}%
	\AgdaSymbol{:}\AgdaSpace{}%
	\AgdaBound{Γ}\AgdaSpace{}%
	\AgdaOperator{\AgdaDatatype{⊢}}\AgdaSpace{}%
	\AgdaInductiveConstructor{⊤}\AgdaSpace{}%
	\AgdaOperator{\AgdaInductiveConstructor{⇒}}\AgdaSpace{}%
	\AgdaBound{A}\AgdaSymbol{\}} 
	\AgdaSymbol{→}\AgdaSpace{}%
	\AgdaOperator{\AgdaInductiveConstructor{[}}\AgdaSpace{}%
	\AgdaInductiveConstructor{id-⇒}\AgdaSpace{}%
	\AgdaOperator{\AgdaInductiveConstructor{]≡}}\AgdaSpace{}%
	\AgdaBound{r}\AgdaSpace{}%
	\AgdaOperator{\AgdaDatatype{⇄}}\AgdaSpace{}%
	\AgdaBound{r}\AgdaSpace{}%
	\AgdaOperator{\AgdaInductiveConstructor{·}}\AgdaSpace{}%
	\AgdaInductiveConstructor{⋆}\<%
	\\ \\
	\>[2]\AgdaInductiveConstructor{id⇒e}\AgdaSpace{}%
	\AgdaSymbol{:}\AgdaSpace{}%
	\AgdaSymbol{∀}\AgdaSpace{}%
	\AgdaSymbol{\{}\AgdaBound{r}\AgdaSpace{}%
	\AgdaSymbol{:}\AgdaSpace{}%
	\AgdaBound{Γ}\AgdaSpace{}%
	\AgdaOperator{\AgdaDatatype{⊢}}\AgdaSpace{}%
	\AgdaBound{A}\AgdaSymbol{\}}
	\<%
	\\
	\>[2][@{}l@{\AgdaIndent{0}}]%
	\>[4]\AgdaSymbol{→}\AgdaSpace{}%
	\AgdaOperator{\AgdaInductiveConstructor{[}}\AgdaSpace{}%
	\AgdaInductiveConstructor{sym}\AgdaSpace{}%
	\AgdaInductiveConstructor{id-⇒}\AgdaSpace{}%
	\AgdaOperator{\AgdaInductiveConstructor{]≡}}\AgdaSpace{}%
	\AgdaBound{r}\AgdaSpace{}%
	\AgdaOperator{\AgdaDatatype{⇄}}\AgdaSpace{}%
	\AgdaOperator{\AgdaInductiveConstructor{ƛ}}\AgdaSpace{}%
	\AgdaFunction{rename}\AgdaSpace{}%
	\AgdaOperator{\AgdaInductiveConstructor{S\AgdaUnderscore{}}}\AgdaSpace{}%
	\AgdaBound{r} 
\end{code}

\noindent where the functions \AgdaFunction{σ-curry} and \AgdaFunction{σ-uncurry} are
substitutions of types \AgdaFunction{Subst}\AgdaSpace{}
\AgdaSymbol{(}\AgdaBound{Γ}\AgdaSpace{}
\AgdaOperator{\AgdaInductiveConstructor{,}}\AgdaSpace{} \AgdaBound{A}\AgdaSpace{}
\AgdaOperator{\AgdaInductiveConstructor{,}}\AgdaSpace{} \AgdaBound{B}\AgdaSymbol{)}
\AgdaSpace{}
\AgdaSymbol{(}\AgdaBound{Γ}\AgdaSpace{} \AgdaOperator{\AgdaInductiveConstructor{,}}\AgdaSpace{} \AgdaBound{A}\AgdaSpace{}
\AgdaOperator{\AgdaInductiveConstructor{×}}\AgdaSpace{}
\AgdaBound{B}\AgdaSymbol{)}\AgdaSpace{}
and
\AgdaFunction{Subst}\AgdaSpace{} \AgdaSymbol{(}\AgdaBound{Γ}\AgdaSpace{} \AgdaOperator{\AgdaInductiveConstructor{,}}\AgdaSpace{} \AgdaBound{A}\AgdaSpace{}
\AgdaOperator{\AgdaInductiveConstructor{×}}\AgdaSpace{}
\AgdaBound{B}\AgdaSymbol{)}\AgdaSpace{} \AgdaSymbol{(}\AgdaBound{Γ}\AgdaSpace{}
\AgdaOperator{\AgdaInductiveConstructor{,}}\AgdaSpace{} \AgdaBound{A}\AgdaSpace{}
\AgdaOperator{\AgdaInductiveConstructor{,}}\AgdaSpace{} \AgdaBound{B}\AgdaSymbol{)}
respectively.

In this definition we omit for simplicity the constructors that
corresponds to congruence rules, given in Table~\ref{tab:EqTermsT}.
We also omit some constructors with the prefix \AgdaInductiveConstructor{sym},
all of which could be obtained from the base equivalences combined with the
\AgdaInductiveConstructor{sym} constructor.
However, they were included in the formalization in order to demonstrate strong normalization.

We note that all constructors have a rule ($\equiv$) on the left side, which
disappears on the right side. For example, the \AgdaInductiveConstructor{comm}
constructor removes the rule \AgdaInductiveConstructor{[comm]≡} from the term.
In the cases of congruence rules, which are not shown here, the applied type
isomorphism will be simplified. For example, the constructor
\AgdaInductiveConstructor{cong⇒$_2$} simplifies the isomorphism
\AgdaInductiveConstructor{[cong⇒$_2$ iso]≡} to
\AgdaInductiveConstructor{[iso]≡}. Therefore, equivalences of terms can only be
applied to eliminate or simplify a constructor ($\equiv$) from the term.

We note that the equivalence relation between terms \isoterm and the reduction relation
\reduces can only relate terms of the same type. The subject reduction property of these relations is a consequence of the intrinsic representation of types.

Then, we define the reduction relation \rediso in Agda, which also satisfies
subject reduction, as follows:

\begin{code}%
	\>[0]\AgdaOperator{\AgdaFunction{\AgdaUnderscore{}⇝\AgdaUnderscore{}}}\AgdaSpace{}%
	\AgdaSymbol{:}\AgdaSpace{}%
	\AgdaSymbol{∀}\AgdaSpace{}%
	\AgdaSymbol{\{}\AgdaBound{Γ}\AgdaSpace{}%
	\AgdaBound{A}\AgdaSymbol{\}}\AgdaSpace{}%
	\AgdaSymbol{→}\AgdaSpace{}%
	\AgdaSymbol{(}\AgdaBound{t}\AgdaSpace{}%
	\AgdaBound{t'}\AgdaSpace{}%
	\AgdaSymbol{:}\AgdaSpace{}%
	\AgdaBound{Γ}\AgdaSpace{}%
	\AgdaOperator{\AgdaDatatype{⊢}}\AgdaSpace{}%
	\AgdaBound{A}\AgdaSymbol{)}\AgdaSpace{}%
	\AgdaSymbol{→}\AgdaSpace{}%
	\AgdaPrimitive{Set}\<%
	\\
	\>[0]\AgdaBound{t}\AgdaSpace{}%
	\AgdaOperator{\AgdaFunction{⇝}}\AgdaSpace{}%
	\AgdaBound{t'}\AgdaSpace{}%
	\AgdaSymbol{=}\AgdaSpace{}%
	\AgdaBound{t}\AgdaSpace{}%
	\AgdaOperator{\AgdaDatatype{↪}}\AgdaSpace{}%
	\AgdaBound{t'}\AgdaSpace{}%
	\AgdaOperator{\AgdaDatatype{⊎}}\AgdaSpace{}%
	\AgdaBound{t}\AgdaSpace{}%
	\AgdaOperator{\AgdaDatatype{⇄}}\AgdaSpace{}%
	\AgdaBound{t'}\<%
\end{code}

\noindent \new{The idea behind this definition is that the constructor \AgdaDatatype{↪} eliminates applications, and \AgdaDatatype{⇄} eliminates isomorphisms.}

\subsection{Progress}

In this section we prove progress for the calculus, which establishes that
any typed and closed term is a value or reduces to some other term.

The first step is to define the notion of value, which in this calculus are:
abstractions, the term $\ttop$, and pairs of values.

\begin{code}
\>[0]\AgdaKeyword{data}\AgdaSpace{}%
\AgdaDatatype{Value}\AgdaSpace{}%
\AgdaSymbol{:}\AgdaSpace{}%
\AgdaSymbol{∀}\AgdaSpace{}%
\AgdaSymbol{\{}\AgdaBound{Γ}\AgdaSpace{}%
\AgdaBound{A}\AgdaSymbol{\}}\AgdaSpace{}%
\AgdaSymbol{→}\AgdaSpace{}%
\AgdaBound{Γ}\AgdaSpace{}%
\AgdaOperator{\AgdaDatatype{⊢}}\AgdaSpace{}%
\AgdaBound{A}\AgdaSpace{}%
\AgdaSymbol{→}\AgdaSpace{}%
\AgdaPrimitive{Set}\AgdaSpace{}%
\AgdaKeyword{where}\<%
\\
\>[0][@{}l@{\AgdaIndent{0}}]%
\>[2]\AgdaInductiveConstructor{V-ƛ}%
\>[191I]\AgdaSymbol{:}\AgdaSpace{}%
\AgdaSymbol{∀}\AgdaSpace{}%
\AgdaSymbol{\{}\AgdaBound{Γ}\AgdaSpace{}%
\AgdaBound{A}\AgdaSpace{}%
\AgdaBound{B}\AgdaSymbol{\}}\AgdaSpace{}%
\AgdaSymbol{\{}\AgdaBound{t}\AgdaSpace{}%
\AgdaSymbol{:}\AgdaSpace{}%
\AgdaBound{Γ}\AgdaSpace{}%
\AgdaOperator{\AgdaInductiveConstructor{,}}\AgdaSpace{}%
\AgdaBound{A}\AgdaSpace{}%
\AgdaOperator{\AgdaDatatype{⊢}}\AgdaSpace{}%
 \AgdaBound{B}\AgdaSymbol{\}} 
\AgdaSymbol{→}\AgdaSpace{}%
\AgdaDatatype{Value}\AgdaSpace{}%
\AgdaSymbol{(}\AgdaOperator{\AgdaInductiveConstructor{ƛ}}\AgdaSpace{}%
\AgdaBound{t}\AgdaSymbol{)}\<%
\\
\>[2]\AgdaInductiveConstructor{V-⋆}%
\>[206I]\AgdaSymbol{:}\AgdaSpace{}%
\AgdaSymbol{∀}\AgdaSpace{}%
\AgdaSymbol{\{}\AgdaBound{Γ}\AgdaSymbol{\}}
\AgdaSymbol{→}\AgdaSpace{}%
\AgdaDatatype{Value}\AgdaSpace{}%
\AgdaSymbol{(}\AgdaInductiveConstructor{⋆}\AgdaSpace{}%
\AgdaSymbol{\{}\AgdaBound{Γ}\AgdaSymbol{\})}\<%
\\
\>[0]\<%
\\[\AgdaEmptyExtraSkip]%
\>[2]\AgdaOperator{\AgdaInductiveConstructor{V-⟨\AgdaUnderscore{},\AgdaUnderscore{}⟩}}\AgdaSpace{}%
\AgdaSymbol{:}\AgdaSpace{}%
\AgdaSymbol{∀}\AgdaSpace{}%
\AgdaSymbol{\{}\AgdaBound{Γ}\AgdaSpace{}%
\AgdaBound{A}\AgdaSpace{}%
\AgdaBound{B}\AgdaSymbol{\}}\AgdaSpace{}%
\AgdaSymbol{\{}\AgdaBound{r}\AgdaSpace{}%
\AgdaSymbol{:}\AgdaSpace{}%
\AgdaBound{Γ}\AgdaSpace{}%
\AgdaOperator{\AgdaDatatype{⊢}}\AgdaSpace{}%
\AgdaBound{A}\AgdaSymbol{\}}\AgdaSpace{}%
\AgdaSymbol{\{}\AgdaBound{s}\AgdaSpace{}%
\AgdaSymbol{:}\AgdaSpace{}%
\AgdaBound{Γ}\AgdaSpace{}%
\AgdaOperator{\AgdaDatatype{⊢}}\AgdaSpace{}%
\AgdaBound{B}\AgdaSymbol{\}}\<%
\\
\>[2][@{}l@{\AgdaIndent{0}}]%
\>[4]\AgdaSymbol{→}\AgdaSpace{}%
\AgdaDatatype{Value}\AgdaSpace{}%
\AgdaBound{r} \AgdaSpace{} \AgdaSymbol{→}\AgdaSpace{}
\AgdaDatatype{Value}\AgdaSpace{}
\AgdaBound{s}\AgdaSpace{} \AgdaSpace{} \AgdaSymbol{→}\AgdaSpace{}
\AgdaDatatype{Value}\AgdaSpace{} \AgdaOperator{\AgdaInductiveConstructor{⟨}}\AgdaSpace{} \AgdaBound{r}\AgdaSpace{} \AgdaOperator{\AgdaInductiveConstructor{,}}\AgdaSpace{} \AgdaBound{s}\AgdaSpace{}%
\AgdaOperator{\AgdaInductiveConstructor{⟩}}
%
\end{code}

In the definition of progress we present below, we use the call-by-value reduction order as a reduction strategy. When reducing under abstractions, it is possible to find free variables. So the final states of reduction to be considered are normal forms instead of values. Then we prove a version of progress for open terms.
To complete the proof of progress for closed terms, we prove that every closed term in normal form is a value.

We start by defining the syntax that characterizes terms in normal form
in this language:

\begin{align*}
	\textbf{norm} &:= \langle \textbf{norm}, \textbf{norm} \rangle \mid \lambda x. \textbf{norm} \mid \star \mid \textbf{neu} \\
	\textbf{neu} &:= var \mid  \textbf{neu} \cdot \textbf{norm} \mid \pi\; \textbf{neu} \mid [\, iso \,]\!\!\equiv \textbf{neu}
\end{align*}

\noindent where $var$ are variables and the syntactic category $\textbf{neu}$ characterizes neutral forms, which are terms that cannot be reduced and are not values. 
In the formalization, the symbols $\Downarrow$ are used for neutral forms and
$\Uparrow$ for normal forms.

\begin{code}
	\>[0]\AgdaKeyword{data}\AgdaSpace{}%
	\AgdaDatatype{⇓}\AgdaSpace{}%
	\AgdaSymbol{:}\AgdaSpace{}%
	\AgdaSymbol{∀}\AgdaSpace{}%
	\AgdaSymbol{\{}\AgdaBound{Γ}\AgdaSpace{}%
	\AgdaBound{A}\AgdaSymbol{\}}\AgdaSpace{}%
	\AgdaSymbol{→}\AgdaSpace{}%
	\AgdaBound{Γ}\AgdaSpace{}%
	\AgdaOperator{\AgdaDatatype{⊢}}\AgdaSpace{}%
	\AgdaBound{A}\AgdaSpace{}%
	\AgdaSymbol{→}\AgdaSpace{}%
	\AgdaPrimitive{Set}\<%
	\\
	\>[0]\AgdaKeyword{data}\AgdaSpace{}%
	\AgdaDatatype{⇑}\AgdaSpace{}%
	\AgdaSymbol{:}\AgdaSpace{}%
	\AgdaSymbol{∀}\AgdaSpace{}%
	\AgdaSymbol{\{}\AgdaBound{Γ}\AgdaSpace{}%
	\AgdaBound{A}\AgdaSymbol{\}}\AgdaSpace{}%
	\AgdaSymbol{→}\AgdaSpace{}%
	\AgdaBound{Γ}\AgdaSpace{}%
	\AgdaOperator{\AgdaDatatype{⊢}}\AgdaSpace{}%
	\AgdaBound{A}\AgdaSpace{}%
	\AgdaSymbol{→}\AgdaSpace{}%
	\AgdaPrimitive{Set}\<%
	\\
	\\[\AgdaEmptyExtraSkip]%
	\>[0]\AgdaKeyword{data}\AgdaSpace{}%
	\AgdaDatatype{⇓}\AgdaSpace{}%
	\AgdaKeyword{where}\<%
	\\
	\>[0][@{}l@{\AgdaIndent{0}}]%
	\>[2]\AgdaOperator{\AgdaInductiveConstructor{`\AgdaUnderscore{}}}%
	\>[5]\AgdaSymbol{:}\AgdaSpace{}%
	\AgdaSymbol{∀}\AgdaSpace{}%
	\AgdaSymbol{\{}\AgdaBound{Γ}\AgdaSpace{}%
	\AgdaBound{A}\AgdaSymbol{\}}\AgdaSpace{}%
	\AgdaSymbol{(}\AgdaBound{x}\AgdaSpace{}%
	\AgdaSymbol{:}\AgdaSpace{}%
	\AgdaBound{Γ}\AgdaSpace{}%
	\AgdaOperator{\AgdaDatatype{∋}}\AgdaSpace{}%
	\AgdaBound{A}\AgdaSymbol{)}
	%
	\AgdaSymbol{→}\AgdaSpace{}%
	\AgdaDatatype{⇓}\AgdaSpace{}%
	\AgdaSymbol{(}\AgdaOperator{\AgdaInductiveConstructor{`}}\AgdaSpace{}%
	\AgdaBound{x}\AgdaSymbol{)}\<%
	\\
	\\[\AgdaEmptyExtraSkip]%
	\>[2]\AgdaOperator{\AgdaInductiveConstructor{\AgdaUnderscore{}·\AgdaUnderscore{}}}%
	\>[7]\AgdaSymbol{:}\AgdaSpace{}%
	\AgdaSymbol{∀}\AgdaSpace{}%
	\AgdaSymbol{\{}\AgdaBound{Γ}\AgdaSpace{}%
	\AgdaBound{A}\AgdaSpace{}%
	\AgdaBound{B}\AgdaSymbol{\}}\AgdaSpace{}%
	\AgdaSymbol{\{}\AgdaBound{r}\AgdaSpace{}%
	\AgdaSymbol{:}\AgdaSpace{}%
	\AgdaBound{Γ}\AgdaSpace{}%
	\AgdaOperator{\AgdaDatatype{⊢}}\AgdaSpace{}%
	\AgdaBound{A}\AgdaSpace{}%
	\AgdaOperator{\AgdaInductiveConstructor{⇒}}\AgdaSpace{}%
	\AgdaBound{B}\AgdaSymbol{\}}\AgdaSpace{}%
	\AgdaSymbol{\{}\AgdaBound{s}\AgdaSpace{}%
	\AgdaSymbol{:}\AgdaSpace{}%
	\AgdaBound{Γ}\AgdaSpace{}%
	\AgdaOperator{\AgdaDatatype{⊢}}\AgdaSpace{}%
	\AgdaBound{A}\AgdaSymbol{\}}\<%
	\\
	\>[2][@{}l@{\AgdaIndent{0}}]%
	\>[4]\AgdaSymbol{→}%
	\>[288I]\AgdaDatatype{⇓}\AgdaSpace{}%
	\AgdaBound{r}\AgdaSpace{}%
	\AgdaSymbol{→}\AgdaSpace{}%
	\AgdaDatatype{⇑}\AgdaSpace{}%
	\AgdaBound{s} \AgdaSpace{}
	%
	\AgdaSymbol{→}\AgdaSpace{}%
	\AgdaDatatype{⇓}\AgdaSpace{}%
	\AgdaSymbol{(}\AgdaBound{r}\AgdaSpace{}%
	\AgdaOperator{\AgdaInductiveConstructor{·}}\AgdaSpace{}
	\AgdaBound{s}\AgdaSymbol{)}
	\<%
	\\
	\>[0]\<%
	\\
	\>[2]\AgdaInductiveConstructor{π}%
	\>[297I]\AgdaSymbol{:}\AgdaSpace{}%
	\AgdaSymbol{∀}\AgdaSpace{}%
	\AgdaSymbol{\{}\AgdaBound{Γ}\AgdaSpace{}%
	\AgdaBound{A}\AgdaSpace{}%
	\AgdaBound{B}\AgdaSpace{}%
	\AgdaBound{C}\AgdaSpace{}%
	\AgdaBound{p}\AgdaSymbol{\}}\AgdaSpace{}%
	\AgdaSymbol{\{}\AgdaBound{t}\AgdaSpace{}%
	\AgdaSymbol{:}\AgdaSpace{}%
	\AgdaBound{Γ}\AgdaSpace{}%
	\AgdaOperator{\AgdaDatatype{⊢}}\AgdaSpace{}%
	\AgdaBound{A}\AgdaSpace{}%
	\AgdaOperator{\AgdaInductiveConstructor{×}}\AgdaSpace{}%
	\AgdaBound{B}\AgdaSymbol{\}}
	\<%
	\\
	\>[.][@{}l@{}]\<[297I]%
	\>[4]\AgdaSymbol{→}%
	\>[311I]\AgdaDatatype{⇓}\AgdaSpace{}%
	\AgdaBound{t} \AgdaSpace{}
	%
	\AgdaSymbol{→}\AgdaSpace{}%
	\AgdaDatatype{⇓}\AgdaSpace{}%
	\AgdaSymbol{(}\AgdaInductiveConstructor{π}\AgdaSpace{}%
	\AgdaBound{C}\AgdaSpace{}%
	\AgdaSymbol{\{}\AgdaBound{p}\AgdaSymbol{\}}\AgdaSpace{}%
	\AgdaBound{t}\AgdaSymbol{)}\<%
	\\
	\>[0]\<%
	\\
	\>[2]\AgdaOperator{\AgdaInductiveConstructor{[\AgdaUnderscore{}]≡\AgdaUnderscore{}}}\AgdaSpace{}%
	\AgdaSymbol{:}\AgdaSpace{}%
	\AgdaSymbol{∀}\AgdaSpace{}%
	\AgdaSymbol{\{}\AgdaBound{Γ}\AgdaSpace{}%
	\AgdaBound{A}\AgdaSpace{}%
	\AgdaBound{B}\AgdaSymbol{\}}\AgdaSpace{}%
	\AgdaSymbol{\{}\AgdaBound{t}\AgdaSpace{}%
	\AgdaSymbol{:}\AgdaSpace{}%
	\AgdaBound{Γ}\AgdaSpace{}%
	\AgdaOperator{\AgdaDatatype{⊢}}\AgdaSpace{}%
	\AgdaBound{A}\AgdaSymbol{\}}
	\<%
	\\
	\>[2][@{}l@{\AgdaIndent{0}}]%
	\>[4]\AgdaSymbol{→}\AgdaSpace{}%
	\AgdaSymbol{(}\AgdaBound{iso}\AgdaSpace{}%
	\AgdaSymbol{:}\AgdaSpace{}%
	\AgdaBound{A}\AgdaSpace{}%
	\AgdaOperator{\AgdaDatatype{≡}}\AgdaSpace{}%
	\AgdaBound{B}\AgdaSymbol{)} \AgdaSpace{}
	%
	\AgdaSymbol{→}\AgdaSpace{}
	\AgdaDatatype{⇓}\AgdaSpace{}%
	\AgdaBound{t}
	%
	\AgdaSpace{} \AgdaSymbol{→}\AgdaSpace{}%
	\AgdaDatatype{⇓}\AgdaSpace{}%
	\AgdaSymbol{(}\AgdaOperator{\AgdaInductiveConstructor{[}}\AgdaSpace{}%
	\AgdaBound{iso}\AgdaSpace{}%
	\AgdaOperator{\AgdaInductiveConstructor{]≡}}\AgdaSpace{}%
	\AgdaBound{t}\AgdaSymbol{)}\<%
\end{code}
\begin{code}
	\>[0]\AgdaKeyword{data}\AgdaSpace{}%
	\AgdaDatatype{⇑}\AgdaSpace{}%
	\AgdaKeyword{where}\<%
	\\
	\>[0][@{}l@{\AgdaIndent{0}}]%
	\>[2]\AgdaOperator{\AgdaInductiveConstructor{\textasciicircum{}\AgdaUnderscore{}}}\AgdaSpace{}%
	\AgdaSymbol{:}\AgdaSpace{}%
	\AgdaSymbol{∀}\AgdaSpace{}%
	\AgdaSymbol{\{}\AgdaBound{Γ}\AgdaSpace{}%
	\AgdaBound{A}\AgdaSymbol{\}}\AgdaSpace{}%
	\AgdaSymbol{\{}\AgdaBound{t}\AgdaSpace{}%
	\AgdaSymbol{:}\AgdaSpace{}%
	\AgdaBound{Γ}\AgdaSpace{}%
	\AgdaOperator{\AgdaDatatype{⊢}}\AgdaSpace{}%
	\AgdaBound{A}\AgdaSymbol{\}} \AgdaSpace{}
	\AgdaSymbol{→}%
	\>[351I]\AgdaDatatype{⇓}\AgdaSpace{}%
	\AgdaBound{t} \AgdaSpace{}
	%
	\AgdaSymbol{→}\AgdaSpace{}%
	\AgdaDatatype{⇑}\AgdaSpace{}%
	\AgdaBound{t}\<%
	\\
	\\[\AgdaEmptyExtraSkip]%
	\>[2]\AgdaInductiveConstructor{N-ƛ}\AgdaSpace{}%
	\AgdaSymbol{:}\AgdaSpace{}%
	\AgdaSymbol{∀}\AgdaSpace{}%
	\AgdaSymbol{\{}\AgdaBound{Γ}\AgdaSpace{}%
	\AgdaBound{A}\AgdaSpace{}%
	\AgdaBound{B}\AgdaSymbol{\}}\AgdaSpace{}%
	\AgdaSymbol{\{}\AgdaBound{t}\AgdaSpace{}%
	\AgdaSymbol{:}\AgdaSpace{}%
	\AgdaBound{Γ}\AgdaSpace{}%
	\AgdaOperator{\AgdaInductiveConstructor{,}}\AgdaSpace{}%
	\AgdaBound{A}\AgdaSpace{}%
	\AgdaOperator{\AgdaDatatype{⊢}}\AgdaSpace{}%
	\AgdaBound{B}\AgdaSymbol{\}}
	%
	\AgdaSpace{} \AgdaSymbol{→}\AgdaSpace{}%
	\AgdaDatatype{⇑}\AgdaSpace{}%
	\AgdaSymbol{(}\AgdaOperator{\AgdaInductiveConstructor{ƛ}}\AgdaSpace{}%
	\AgdaBound{t}\AgdaSymbol{)}\<%
	\\
	\>[0]\<%
	\\
	\>[2]\AgdaOperator{\AgdaInductiveConstructor{N-⟨\AgdaUnderscore{},\AgdaUnderscore{}⟩}}\AgdaSpace{}%
	\AgdaSymbol{:}\AgdaSpace{}%
	\AgdaSymbol{∀}\AgdaSpace{}%
	\AgdaSymbol{\{}\AgdaBound{Γ}\AgdaSpace{}%
	\AgdaBound{A}\AgdaSpace{}%
	\AgdaBound{B}\AgdaSymbol{\}}\AgdaSpace{}%
	\AgdaSymbol{\{}\AgdaBound{r}\AgdaSpace{}%
	\AgdaSymbol{:}\AgdaSpace{}%
	\AgdaBound{Γ}\AgdaSpace{}%
	\AgdaOperator{\AgdaDatatype{⊢}}\AgdaSpace{}%
	\AgdaBound{A}\AgdaSymbol{\}}\AgdaSpace{}%
	\AgdaSymbol{\{}\AgdaBound{s}\AgdaSpace{}%
	\AgdaSymbol{:}\AgdaSpace{}%
	\AgdaBound{Γ}\AgdaSpace{}%
	\AgdaOperator{\AgdaDatatype{⊢}}\AgdaSpace{}%
	\AgdaBound{B}\AgdaSymbol{\}}
	\<%
	\\
	\>[2][@{}l@{\AgdaIndent{0}}]%
	\>[4]\AgdaSymbol{→}\AgdaSpace{}%
	\AgdaDatatype{⇑}\AgdaSpace{}%
	\AgdaBound{r}\AgdaSpace{}
	\AgdaSymbol{→}\AgdaSpace{}
	\AgdaDatatype{⇑}\AgdaSpace{}%
	\AgdaBound{s}
	%
	\AgdaSpace{} \AgdaSymbol{→}\AgdaSpace{}%
	\AgdaDatatype{⇑}\AgdaSpace{}%
	\AgdaOperator{\AgdaInductiveConstructor{⟨}}\AgdaSpace{}%
	\AgdaBound{r}\AgdaSpace{}%
	\AgdaOperator{\AgdaInductiveConstructor{,}}\AgdaSpace{}%
	\AgdaBound{s}\AgdaSpace{}%
	\AgdaOperator{\AgdaInductiveConstructor{⟩}}\<%
	\\
	\>[0]\<%
	\\
	\>[2]\AgdaInductiveConstructor{N-⋆}%
	\>[397I]\AgdaSymbol{:}\AgdaSpace{}%
	\AgdaSymbol{∀}\AgdaSpace{}%
	\AgdaSymbol{\{}\AgdaBound{Γ}\AgdaSymbol{\}}
	\AgdaSpace{} \AgdaSymbol{→}\AgdaSpace{}%
	\AgdaDatatype{⇑}\AgdaSpace{}%
	\AgdaSymbol{(}\AgdaInductiveConstructor{⋆}\AgdaSpace{}%
	\AgdaSymbol{\{}\AgdaBound{Γ}\AgdaSymbol{\})}\<%
	\\
	\>[0]\<%
\end{code}

Now we define a relation that captures the cases in which a term $t$ satisfies progress,
namely: it can be transformed into an equivalent one, it can take a reduction step, or
it is in normal form:

\begin{code}%
	\>[0]\AgdaKeyword{data}\AgdaSpace{}%
	\AgdaDatatype{Progress}\AgdaSpace{}%
	\AgdaSymbol{\{}\AgdaBound{Γ}\AgdaSpace{}%
	\AgdaBound{A}\AgdaSymbol{\}}\AgdaSpace{}%
	\AgdaSymbol{(}\AgdaBound{t}\AgdaSpace{}%
	\AgdaSymbol{:}\AgdaSpace{}%
	\AgdaBound{Γ}\AgdaSpace{}%
	\AgdaOperator{\AgdaDatatype{⊢}}\AgdaSpace{}%
	\AgdaBound{A}\AgdaSymbol{)}\AgdaSpace{}%
	\AgdaSymbol{:}\AgdaSpace{}%
	\AgdaPrimitive{Set}\AgdaSpace{}%
	\AgdaKeyword{where}\<%
	\\
	\>[0][@{}l@{\AgdaIndent{0}}]%
	\>[2]\AgdaInductiveConstructor{step⇄}\AgdaSpace{}%
	\AgdaSymbol{:}\AgdaSpace{}%
	\AgdaSymbol{∀}\AgdaSpace{}%
	\AgdaSymbol{\{}\AgdaBound{t'}\AgdaSpace{}%
	\AgdaSymbol{:}\AgdaSpace{}%
	\AgdaBound{Γ}\AgdaSpace{}%
	\AgdaOperator{\AgdaDatatype{⊢}}\AgdaSpace{}%
	\AgdaBound{A}\AgdaSymbol{\}}
	\AgdaSpace{} \AgdaSymbol{→}%
	\>[41I]\AgdaBound{t}\AgdaSpace{}%
	\AgdaOperator{\AgdaDatatype{⇄}}\AgdaSpace{}%
	\AgdaBound{t'} \AgdaSpace{}
	%
	\AgdaSymbol{→}\AgdaSpace{}%
	\AgdaDatatype{Progress}\AgdaSpace{}%
	\AgdaBound{t}\<%
	\\
	\>[0]\<%
	\\
	\>[2]\AgdaInductiveConstructor{step↪}\AgdaSpace{}%
	\AgdaSymbol{:}\AgdaSpace{}%
	\AgdaSymbol{∀}\AgdaSpace{}%
	\AgdaSymbol{\{}\AgdaBound{t'}\AgdaSpace{}%
	\AgdaSymbol{:}\AgdaSpace{}%
	\AgdaBound{Γ}\AgdaSpace{}%
	\AgdaOperator{\AgdaDatatype{⊢}}\AgdaSpace{}%
	\AgdaBound{A}\AgdaSymbol{\}}
	\AgdaSpace{} \AgdaSymbol{→}%
	\>[53I]\AgdaBound{t}\AgdaSpace{}%
	\AgdaOperator{\AgdaDatatype{↪}}\AgdaSpace{}%
	\AgdaBound{t'}
	\AgdaSpace{} \AgdaSymbol{→}\AgdaSpace{}%
	\AgdaDatatype{Progress}\AgdaSpace{}%
	\AgdaBound{t}\<%
	\\
	\>[0]\<%
	\\
	\>[2]\AgdaInductiveConstructor{done}\AgdaSpace{}%
	\AgdaSymbol{:}
	\AgdaDatatype{⇑}\AgdaSpace{}%
	\AgdaBound{t}
	%
	\AgdaSpace{} \AgdaSymbol{→}\AgdaSpace{}%
	\AgdaDatatype{Progress}\AgdaSpace{}%
	\AgdaBound{t}\<%
\end{code}

We can now prove progress. The variable and unit cases are very simple since both terms are
in normal form:

\begin{code}
\>[0]\AgdaFunction{progress}\AgdaSpace{}%
\AgdaSymbol{:}\AgdaSpace{}%
\AgdaSymbol{∀}\AgdaSpace{}%
\AgdaSymbol{\{}\AgdaBound{Γ}\AgdaSpace{}%
\AgdaBound{A}\AgdaSymbol{\}}\AgdaSpace{}%
\AgdaSymbol{→}\AgdaSpace{}%
\AgdaSymbol{(}\AgdaBound{t}\AgdaSpace{}%
\AgdaSymbol{:}\AgdaSpace{}%
\AgdaBound{Γ}\AgdaSpace{}%
\AgdaOperator{\AgdaDatatype{⊢}}\AgdaSpace{}%
\AgdaBound{A}\AgdaSymbol{)}\AgdaSpace{}%
\AgdaSymbol{→}\AgdaSpace{}%
\AgdaDatatype{Progress}\AgdaSpace{}%
\AgdaBound{t}\<%
\\
\>[0]\AgdaFunction{progress}\AgdaSpace{}%
\AgdaSymbol{(}\AgdaOperator{\AgdaInductiveConstructor{`}}\AgdaSpace{}%
\AgdaBound{x}\AgdaSymbol{)}\AgdaSpace{}%
\>[1]\AgdaSymbol{=}\AgdaSpace{}%
\AgdaInductiveConstructor{done}\AgdaSpace{}%
\AgdaSymbol{(}\AgdaOperator{\AgdaInductiveConstructor{\textasciicircum{}}}\AgdaSpace{}%
\AgdaSymbol{(}\AgdaOperator{\AgdaInductiveConstructor{`}}\AgdaSpace{}%
\AgdaBound{x}\AgdaSymbol{))}\<%
\\
\>[0]\AgdaFunction{progress}\AgdaSpace{}%
\AgdaInductiveConstructor{⋆}\AgdaSpace{}%
\>[1]\AgdaSymbol{=}\AgdaSpace{}%
\AgdaInductiveConstructor{done}\AgdaSpace{}%
\AgdaInductiveConstructor{N-⋆}\<%
\end{code}

If the term is a lambda abstraction, then we call the progress function under the binder;
if this returns a reduction step, the step is extended using $\zeta$ (the congruence rule
for abstractions). Otherwise, the abstraction is in normal form:

\begin{code}
\>[0]\AgdaFunction{progress}\AgdaSpace{}%
\AgdaSymbol{(}\AgdaOperator{\AgdaInductiveConstructor{ƛ}}\AgdaSpace{}%
\AgdaBound{t}\AgdaSymbol{)}\AgdaSpace{}%
\AgdaKeyword{with}\AgdaSpace{}%
\AgdaFunction{progress}\AgdaSpace{}%
\AgdaBound{t}\<%
\\
\>[0]\AgdaSymbol{...}\AgdaSpace{}%
\AgdaSymbol{|}\AgdaSpace{}%
\AgdaInductiveConstructor{step⇄}\AgdaSpace{}%
\AgdaBound{t⇄t'}%
\>[18]\AgdaSymbol{=}\AgdaSpace{}%
\AgdaInductiveConstructor{step⇄}\AgdaSpace{}%
\AgdaSymbol{(}\AgdaInductiveConstructor{ζ}\AgdaSpace{}%
\AgdaBound{t⇄t'}\AgdaSymbol{)}\<%
\\
\>[0]\AgdaSymbol{...}\AgdaSpace{}%
\AgdaSymbol{|}\AgdaSpace{}%
\AgdaInductiveConstructor{step↪}\AgdaSpace{}%
\AgdaBound{t↪t'}%
\>[18]\AgdaSymbol{=}\AgdaSpace{}%
\AgdaInductiveConstructor{step↪}\AgdaSpace{}%
\AgdaSymbol{(}\AgdaInductiveConstructor{ζ}\AgdaSpace{}%
\AgdaBound{t↪t'}\AgdaSymbol{)}\<%
\\
\>[0]\AgdaSymbol{...}\AgdaSpace{}%
\AgdaSymbol{|}\AgdaSpace{}%
\AgdaInductiveConstructor{done}\AgdaSpace{}%
\AgdaBound{⇑t}%
\>[18]\AgdaSymbol{=}\AgdaSpace{}%
\AgdaInductiveConstructor{done}\AgdaSpace{}%
\AgdaInductiveConstructor{N-ƛ}\<%
\end{code}

For pairs the proof is similar.
If the term is a projection or an application, we recursively apply progress
to each subterm; if the result is a reduction or an equivalence case, we use the congruence rule
$\zeta$, and if it is the done case, then we have a normal form or we can apply
$\beta$-reduction.

The most interesting case for this work is when the term is
\AgdaSymbol{[} \AgdaBound{iso} \AgdaSymbol{]≡} \AgdaBound{t}.
In this case we recursively apply progress to the subterm $t$;
if the result is a reduction or an equivalence case, we use the congruence rule
\AgdaInductiveConstructor{ξ-≡}, and if it is the done case, we continue the proof by case analysis on
each normal form and all applicable isomorphisms that can be applied in each case.
We show here just a few cases; the remaining cases are similar.

\begin{code}
\>[0]\AgdaFunction{progress}\AgdaSpace{}%
\AgdaSymbol{(}\AgdaOperator{\AgdaInductiveConstructor{[}}\AgdaSpace{}%
\AgdaBound{iso}\AgdaSpace{}%
\AgdaOperator{\AgdaInductiveConstructor{]≡}}\AgdaSpace{}%
\AgdaBound{t}\AgdaSymbol{)}\AgdaSpace{}%
\AgdaKeyword{with}\AgdaSpace{}%
\AgdaFunction{progress}\AgdaSpace{}%
\AgdaBound{t}\<%
\\
\>[0]\AgdaSymbol{...}\AgdaSpace{}%
\AgdaSymbol{|}\AgdaSpace{}%
\AgdaInductiveConstructor{step⇄}\AgdaSpace{}%
\AgdaBound{t⇄t'}\AgdaSpace{}%
\AgdaSymbol{=}\AgdaSpace{}%
\AgdaInductiveConstructor{step⇄}\AgdaSpace{}%
\AgdaSymbol{(}\AgdaInductiveConstructor{ξ-≡}\AgdaSpace{}%
\AgdaBound{t⇄t'}\AgdaSymbol{)}\<%
\\
\>[0]\AgdaSymbol{...}\AgdaSpace{}%
\AgdaSymbol{|}\AgdaSpace{}%
\AgdaInductiveConstructor{step↪}\AgdaSpace{}%
\AgdaBound{t↪t'}\AgdaSpace{}%
\AgdaSymbol{=}\AgdaSpace{}%
\AgdaInductiveConstructor{step↪}\AgdaSpace{}%
\AgdaSymbol{(}\AgdaInductiveConstructor{ξ-≡}\AgdaSpace{}%
\AgdaBound{t↪t'}\AgdaSymbol{)}\<%
\\ \\
\>[0]\AgdaFunction{progress}\AgdaSpace{}%
\AgdaSymbol{(}\AgdaOperator{\AgdaInductiveConstructor{[}}\AgdaSpace{}%
\AgdaInductiveConstructor{comm}\AgdaSpace{}%
\AgdaOperator{\AgdaInductiveConstructor{]≡}}\AgdaSpace{}%
\AgdaSymbol{\AgdaUnderscore{})}%
\>[28]\AgdaSymbol{|}\AgdaSpace{}%
\AgdaInductiveConstructor{done}\AgdaSpace{}%
\AgdaOperator{\AgdaInductiveConstructor{N-⟨}}\AgdaSpace{}%
\AgdaBound{⇑r}\AgdaSpace{}%
\AgdaOperator{\AgdaInductiveConstructor{,}}\AgdaSpace{}%
\AgdaBound{⇑s}\AgdaSpace{}%
\AgdaOperator{\AgdaInductiveConstructor{⟩}}\AgdaSpace{}%
\AgdaSymbol{=} \\ 
\>[0] \AgdaInductiveConstructor{step⇄}\AgdaSpace{}%
\AgdaInductiveConstructor{comm}\<%
\\
\\
\>[0]\AgdaCatchallClause{\AgdaFunction{progress}}\AgdaSpace{}%
\AgdaCatchallClause{\AgdaSymbol{(}}\AgdaCatchallClause{\AgdaOperator{\AgdaInductiveConstructor{[}}}\AgdaSpace{}%
\AgdaCatchallClause{\AgdaInductiveConstructor{asso}}\AgdaSpace{}%
\AgdaCatchallClause{\AgdaOperator{\AgdaInductiveConstructor{]≡}}}\AgdaSpace{}%
\AgdaCatchallClause{\AgdaSymbol{\AgdaUnderscore{})}}%
\>[28]\AgdaCatchallClause{\AgdaSymbol{|}}\AgdaSpace{}%
\AgdaCatchallClause{\AgdaInductiveConstructor{done}}\AgdaSpace{}%
\AgdaCatchallClause{\AgdaOperator{\AgdaInductiveConstructor{N-⟨}}}\AgdaSpace{}%
\AgdaCatchallClause{\AgdaBound{⇑r}}\AgdaSpace{}%
\AgdaCatchallClause{\AgdaOperator{\AgdaInductiveConstructor{,}}}\AgdaSpace{}%
\AgdaCatchallClause{\AgdaBound{⇑s}}\AgdaSpace{}%
\AgdaCatchallClause{\AgdaOperator{\AgdaInductiveConstructor{⟩}}}\AgdaSpace{}%
\AgdaSymbol{=} \\ 
\>[0]\AgdaInductiveConstructor{step⇄}\AgdaSpace{}%
\AgdaSymbol{(}\AgdaInductiveConstructor{asso-split}\AgdaSymbol{)}\<%
\\
\\
\>[0]\AgdaFunction{progress}\AgdaSpace{}%
\AgdaSymbol{(}\AgdaOperator{\AgdaInductiveConstructor{[}}\AgdaSpace{}%
\AgdaInductiveConstructor{dist}\AgdaSpace{}%
\AgdaOperator{\AgdaInductiveConstructor{]≡}}\AgdaSpace{}%
\AgdaSymbol{\AgdaUnderscore{})}\AgdaSpace{}%
\AgdaSymbol{|}\AgdaSpace{}%
\AgdaInductiveConstructor{done}\AgdaSpace{}%
\AgdaOperator{\AgdaInductiveConstructor{N-⟨}}\AgdaSpace{}%
\AgdaInductiveConstructor{N-ƛ}\AgdaSpace{}%
\AgdaOperator{\AgdaInductiveConstructor{,}}\AgdaSpace{}%
\AgdaInductiveConstructor{N-ƛ}\AgdaSpace{}%
\AgdaOperator{\AgdaInductiveConstructor{⟩}}%
\>[47]\AgdaSymbol{=} \\ 
\>[0]\AgdaInductiveConstructor{step⇄}\AgdaSpace{}%
\AgdaInductiveConstructor{dist-ƛ}\<%
\\
\>[0]\AgdaCatchallClause{\AgdaFunction{progress}}\AgdaSpace{}%
\AgdaCatchallClause{\AgdaSymbol{(}}\AgdaCatchallClause{\AgdaOperator{\AgdaInductiveConstructor{[}}}\AgdaSpace{}%
\AgdaCatchallClause{\AgdaInductiveConstructor{dist}}\AgdaSpace{}%
\AgdaCatchallClause{\AgdaOperator{\AgdaInductiveConstructor{]≡}}}\AgdaSpace{}%
\AgdaCatchallClause{\AgdaSymbol{\AgdaUnderscore{})}}\AgdaSpace{}%
\AgdaCatchallClause{\AgdaSymbol{|}}\AgdaSpace{}%
\AgdaCatchallClause{\AgdaInductiveConstructor{done}}\AgdaSpace{}%
\AgdaCatchallClause{\AgdaOperator{\AgdaInductiveConstructor{N-⟨}}}\AgdaSpace{}%
\AgdaCatchallClause{\AgdaInductiveConstructor{N-ƛ}}\AgdaSpace{}%
\AgdaCatchallClause{\AgdaOperator{\AgdaInductiveConstructor{,}}}\AgdaSpace{}%
\AgdaCatchallClause{\AgdaBound{⇑s}}\AgdaSpace{}%
\AgdaCatchallClause{\AgdaOperator{\AgdaInductiveConstructor{⟩}}}%
\>[47]\AgdaSymbol{=}\\ 
\>[0]\AgdaInductiveConstructor{step⇄}\AgdaSpace{}%
\AgdaSymbol{(}\AgdaInductiveConstructor{dist-ƛηᵣ}\AgdaSymbol{)}\<%
\\
\>[0]\AgdaCatchallClause{\AgdaFunction{progress}}\AgdaSpace{}%
\AgdaCatchallClause{\AgdaSymbol{(}}\AgdaCatchallClause{\AgdaOperator{\AgdaInductiveConstructor{[}}}\AgdaSpace{}%
\AgdaCatchallClause{\AgdaInductiveConstructor{dist}}\AgdaSpace{}%
\AgdaCatchallClause{\AgdaOperator{\AgdaInductiveConstructor{]≡}}}\AgdaSpace{}%
\AgdaCatchallClause{\AgdaSymbol{\AgdaUnderscore{})}}\AgdaSpace{}%
\AgdaCatchallClause{\AgdaSymbol{|}}\AgdaSpace{}%
\AgdaCatchallClause{\AgdaInductiveConstructor{done}}\AgdaSpace{}%
\AgdaCatchallClause{\AgdaOperator{\AgdaInductiveConstructor{N-⟨}}}\AgdaSpace{}%
\AgdaCatchallClause{\AgdaBound{⇑r}}\AgdaSpace{}%
\AgdaCatchallClause{\AgdaOperator{\AgdaInductiveConstructor{,}}}\AgdaSpace{}%
\AgdaCatchallClause{\AgdaInductiveConstructor{N-ƛ}}\AgdaSpace{}%
\AgdaCatchallClause{\AgdaOperator{\AgdaInductiveConstructor{⟩}}}%
\>[47]\AgdaSymbol{=} \\ 
\>[0]\AgdaInductiveConstructor{step⇄}\AgdaSpace{}%
\AgdaSymbol{(}\AgdaInductiveConstructor{dist-ƛηₗ}\AgdaSymbol{)}\<%
\\
\>[0]\AgdaCatchallClause{\AgdaFunction{progress}}\AgdaSpace{}%
\AgdaCatchallClause{\AgdaSymbol{(}}\AgdaCatchallClause{\AgdaOperator{\AgdaInductiveConstructor{[}}}\AgdaSpace{}%
\AgdaCatchallClause{\AgdaInductiveConstructor{dist}}\AgdaSpace{}%
\AgdaCatchallClause{\AgdaOperator{\AgdaInductiveConstructor{]≡}}}\AgdaSpace{}%
\AgdaCatchallClause{\AgdaSymbol{\AgdaUnderscore{})}}\AgdaSpace{}%
\AgdaCatchallClause{\AgdaSymbol{|}}\AgdaSpace{}%
\AgdaCatchallClause{\AgdaInductiveConstructor{done}}\AgdaSpace{}%
\AgdaCatchallClause{\AgdaOperator{\AgdaInductiveConstructor{N-⟨}}}\AgdaSpace{}%
\AgdaCatchallClause{\AgdaBound{⇑r}}\AgdaSpace{}%
\AgdaCatchallClause{\AgdaOperator{\AgdaInductiveConstructor{,}}}\AgdaSpace{}%
\AgdaCatchallClause{\AgdaBound{⇑s}}\AgdaSpace{}%
\AgdaCatchallClause{\AgdaOperator{\AgdaInductiveConstructor{⟩}}}%
\>[47]\AgdaSymbol{=}\\ 
\>[0] \AgdaInductiveConstructor{step⇄}\AgdaSpace{}%
\AgdaSymbol{(}\AgdaInductiveConstructor{dist-ƛηₗᵣ}\AgdaSymbol{)}\<%
\\
\\
\>[0]\AgdaFunction{progress}\AgdaSpace{}%
\AgdaSymbol{(}\AgdaOperator{\AgdaInductiveConstructor{[}}\AgdaSpace{}%
\AgdaInductiveConstructor{curry}\AgdaSpace{}%
\AgdaOperator{\AgdaInductiveConstructor{]≡}}\AgdaSpace{}%
\AgdaSymbol{(}\AgdaOperator{\AgdaInductiveConstructor{ƛ}}\AgdaSpace{}%
\AgdaOperator{\AgdaInductiveConstructor{ƛ}}\AgdaSpace{}%
\AgdaSymbol{\AgdaUnderscore{}))}%
\>[31]\AgdaSymbol{|}\AgdaSpace{}%
\AgdaInductiveConstructor{done}\AgdaSpace{}%
\AgdaInductiveConstructor{N-ƛ}%
\>[43]\AgdaSymbol{=}\\ 
\>[0]\AgdaInductiveConstructor{step⇄}\AgdaSpace{}%
\AgdaInductiveConstructor{curry}\<%
\\
\>[0]\AgdaCatchallClause{\AgdaFunction{progress}}\AgdaSpace{}%
\AgdaCatchallClause{\AgdaSymbol{(}}\AgdaCatchallClause{\AgdaOperator{\AgdaInductiveConstructor{[}}}\AgdaSpace{}%
\AgdaCatchallClause{\AgdaInductiveConstructor{curry}}\AgdaSpace{}%
\AgdaCatchallClause{\AgdaOperator{\AgdaInductiveConstructor{]≡}}}\AgdaSpace{}%
\AgdaCatchallClause{\AgdaSymbol{\AgdaUnderscore{})}}%
\>[31]\AgdaCatchallClause{\AgdaSymbol{|}}\AgdaSpace{}%
\AgdaCatchallClause{\AgdaInductiveConstructor{done}}\AgdaSpace{}%
\AgdaCatchallClause{\AgdaInductiveConstructor{N-ƛ}}%
\>[43]\AgdaSymbol{=} \\ 
\>[0]\AgdaInductiveConstructor{step⇄}\AgdaSpace{}%
\AgdaSymbol{(}\AgdaInductiveConstructor{curry-η}\AgdaSymbol{)}\<%
\\
\>[0]\AgdaFunction{progress}\AgdaSpace{}%
\AgdaSymbol{(}\AgdaOperator{\AgdaInductiveConstructor{[}}\AgdaSpace{}%
\AgdaInductiveConstructor{id-×}\AgdaSpace{}%
\AgdaOperator{\AgdaInductiveConstructor{]≡}}\AgdaSpace{}%
\AgdaSymbol{\AgdaUnderscore{})}%
\>[29]\AgdaSymbol{|}\AgdaSpace{}%
\AgdaInductiveConstructor{done}\AgdaSpace{}%
\AgdaOperator{\AgdaInductiveConstructor{N-⟨}}\AgdaSpace{}%
\AgdaBound{⇑r}\AgdaSpace{}%
\AgdaOperator{\AgdaInductiveConstructor{,}}\AgdaSpace{}%
\AgdaBound{⇑s}\AgdaSpace{}%
\AgdaOperator{\AgdaInductiveConstructor{⟩}}%
\>[51]\AgdaSymbol{=} \\ 
\>[0]\AgdaInductiveConstructor{step⇄}\AgdaSpace{}%
\AgdaInductiveConstructor{id-×}\<%
\\
\>[0]\AgdaCatchallClause{\AgdaFunction{progress}}\AgdaSpace{}%
\AgdaCatchallClause{\AgdaSymbol{(}}\AgdaCatchallClause{\AgdaOperator{\AgdaInductiveConstructor{[}}}\AgdaSpace{}%
\AgdaCatchallClause{\AgdaInductiveConstructor{id-⇒}}\AgdaSpace{}%
\AgdaCatchallClause{\AgdaOperator{\AgdaInductiveConstructor{]≡}}}\AgdaSpace{}%
\AgdaCatchallClause{\AgdaSymbol{\AgdaUnderscore{})}}%
\>[29]\AgdaCatchallClause{\AgdaSymbol{|}}\AgdaSpace{}%
\AgdaCatchallClause{\AgdaInductiveConstructor{done}}\AgdaSpace{}%
\AgdaCatchallClause{\AgdaBound{⇑t}}%
\>[51]\AgdaSymbol{=}\\ 
\>[0]\AgdaInductiveConstructor{step⇄}\AgdaSpace{}%
\AgdaInductiveConstructor{id-⇒}\<%
\\
\>[0]\AgdaCatchallClause{\AgdaFunction{progress}}\AgdaSpace{}%
\AgdaCatchallClause{\AgdaSymbol{(}}\AgdaCatchallClause{\AgdaOperator{\AgdaInductiveConstructor{[}}}\AgdaSpace{}%
\AgdaCatchallClause{\AgdaInductiveConstructor{abs}}\AgdaSpace{}%
\AgdaCatchallClause{\AgdaOperator{\AgdaInductiveConstructor{]≡}}}\AgdaSpace{}%
\AgdaCatchallClause{\AgdaSymbol{\AgdaUnderscore{}}}\AgdaCatchallClause{\AgdaSymbol{)}}%
\>[27]\AgdaCatchallClause{\AgdaSymbol{|}}\AgdaSpace{}%
\AgdaCatchallClause{\AgdaInductiveConstructor{done}}\AgdaSpace{}%
\AgdaCatchallClause{\AgdaBound{⇑t}}\AgdaSpace{}%
\AgdaSymbol{=}\AgdaSpace{}%
\AgdaInductiveConstructor{step⇄}\AgdaSpace{}%
\AgdaInductiveConstructor{abs}\<%
\\
\>[0]\AgdaCatchallClause{\AgdaFunction{progress}}\AgdaSpace{}%
\AgdaCatchallClause{\AgdaSymbol{(}}\AgdaCatchallClause{\AgdaOperator{\AgdaInductiveConstructor{[}}}\AgdaSpace{}%
\AgdaCatchallClause{\AgdaBound{iso}}\AgdaSpace{}%
\AgdaCatchallClause{\AgdaOperator{\AgdaInductiveConstructor{]≡}}}\AgdaSpace{}%
\AgdaCatchallClause{\AgdaSymbol{\AgdaUnderscore{}}}\AgdaCatchallClause{\AgdaSymbol{)}}\AgdaSpace{}%
\AgdaCatchallClause{\AgdaSymbol{|}}\AgdaSpace{}%
\AgdaCatchallClause{\AgdaInductiveConstructor{done}}\AgdaSpace{}%
\AgdaCatchallClause{\AgdaSymbol{(}}\AgdaCatchallClause{\AgdaOperator{\AgdaInductiveConstructor{\textasciicircum{}}}}\AgdaSpace{}%
\AgdaCatchallClause{\AgdaBound{⇓t}}\AgdaCatchallClause{\AgdaSymbol{)}}\AgdaSpace{}%
\AgdaSymbol{=}\AgdaSpace{}%
\AgdaInductiveConstructor{done}\AgdaSpace{}%
\AgdaSymbol{(}\AgdaOperator{\AgdaInductiveConstructor{\textasciicircum{}}}\AgdaSpace{}%
\AgdaOperator{\AgdaInductiveConstructor{[}}\AgdaSpace{}%
\AgdaBound{iso}\AgdaSpace{}%
\AgdaOperator{\AgdaInductiveConstructor{]≡}}\AgdaSpace{}%
\AgdaBound{⇓t}\AgdaSymbol{)}\<%
\end{code}

The definition of progress shows how each term isomorphism corresponds to a case where it is necessary to eliminate a rule (\AgdaInductiveConstructor{≡}) in order to continue with the reduction.

Finally, we complete the proof of progress by proving that any
closed term in normal form is a value; we show just the type of
this function here:

\vspace{0.5em}
\mbox{
\noindent \AgdaFunction{closed⇑→Value} \AgdaSymbol{:}\AgdaSpace{}
\AgdaSymbol{∀}\AgdaSpace{} \AgdaSymbol{\{}\AgdaBound{A}\AgdaSymbol{\}}\AgdaSpace{}
\AgdaSymbol{\{}\AgdaBound{t}\AgdaSpace{}
\AgdaSymbol{:}\AgdaSpace{} \AgdaInductiveConstructor{∅}\AgdaSpace{}
\AgdaOperator{\AgdaDatatype{⊢}}\AgdaSpace{} \AgdaBound{A}\AgdaSymbol{\}}\AgdaSpace{}
\AgdaSymbol{→}\AgdaSpace{} \AgdaDatatype{⇑}\AgdaSpace{}
\AgdaBound{t}\AgdaSpace{} \AgdaSymbol{→}\AgdaSpace{}
\AgdaDatatype{Value}\AgdaSpace{} \AgdaBound{t}
}

\vspace{0.5em}

\noindent The cases where the term is the top value, an
abstraction, or a pair, the definition of this function is very simple; when the term is a neutral form,
we use the function \AgdaFunction{⊥-elim} with a proof that a closed term cannot be neutral. This last proof
can be constructed since a neutral term must contain a free variable, because variables are the only constructor that is not recursive in its definition.

\section{Strong normalization}
\label{sec:normalization}
In this section we prove the strong normalization of the relation
\rediso, which states that every typed term cannot reduce indefinitely, meaning that every
reduction sequence ends in a value.

The most commonly used technique for proving strong normalization is
that of reducibility, introduced by Tait~\cite{tait67} and generalized by Girard~\cite{girard71, girard72}. In this work
we follow Schäfer's approach~\cite{Schafer}, whose changes to the classical formulations of reducibility produces a technique that is more suitable for formalization, which he uses to prove strong normalization of System F in Coq.
Our formalization is based on the adaptation of Schäfer's ideas to Agda for the simply
typed lambda calculus, given by András Kovács\footnote{https://github.com/AndrasKovacs/misc-stuff/blob/master/agda /STLCStrongNorm/StrongNorm2.agda}.

We begin by giving a constructive definition of strongly normalizing terms,
originally presented by Altenkirch.  

The set of strongly normalizing terms with respect to a reduction $\rightsquigarrow$
is inductively defined by the following rule:

\begin{center}
	\begin{prooftree}
		\hypo{ \forall t. s \rightsquigarrow t \implies t \in SN }
		\infer1{ s \in SN }
	\end{prooftree}
\end{center}

This means that if for each term $t$ such that $s \rightsquigarrow t$, it holds that $t$ is
strongly normalizing, then $s$ is also strongly normalizing.

In Agda, we can define the set SN using the following inductive data type:

\begin{code}%
	\>[0]\AgdaKeyword{data}\AgdaSpace{}%
	\AgdaDatatype{SN}\AgdaSpace{}%
	\AgdaSymbol{\{}\AgdaBound{Γ}\AgdaSpace{}%
	\AgdaBound{A}\AgdaSymbol{\}}\AgdaSpace{}%
	\AgdaSymbol{(}\AgdaBound{t}\AgdaSpace{}%
	\AgdaSymbol{:}\AgdaSpace{}%
	\AgdaBound{Γ}\AgdaSpace{}%
	\AgdaOperator{\AgdaDatatype{⊢}}\AgdaSpace{}%
	\AgdaBound{A}\AgdaSymbol{)}\AgdaSpace{}%
	\AgdaSymbol{:}\AgdaSpace{}%
	\AgdaPrimitive{Set}\AgdaSpace{}%
	\AgdaKeyword{where}\<%
	\\
	\>[0][@{}l@{\AgdaIndent{0}}]%
	\>[2]\AgdaInductiveConstructor{sn}\AgdaSpace{}%
	\AgdaSymbol{:}\AgdaSpace{}%
	\AgdaSymbol{(∀}\AgdaSpace{}%
	\AgdaSymbol{\{}\AgdaBound{t'}\AgdaSymbol{\}}\AgdaSpace{}%
	\AgdaSymbol{→}\AgdaSpace{}%
	\AgdaBound{t}\AgdaSpace{}%
	\AgdaOperator{\AgdaDatatype{⇝}}\AgdaSpace{}%
	\AgdaBound{t'}\AgdaSpace{}%
	\AgdaSymbol{→}\AgdaSpace{}%
	\AgdaDatatype{SN}\AgdaSpace{}%
	\AgdaBound{t'}\AgdaSymbol{)}\AgdaSpace{}%
	\AgdaSymbol{→}\AgdaSpace{}%
	\AgdaDatatype{SN}\AgdaSpace{}%
	\AgdaBound{t}\<%
\end{code}

\com{ver si poner los ejemplos}

Our goal is to define a function with this type:

\vspace{0.5em}
\AgdaFunction{strong-norm}
\AgdaSymbol{:}
\AgdaSymbol{∀}
\AgdaSymbol{\{}\AgdaBound{Γ}
\AgdaBound{A}\AgdaSymbol{\}}
\AgdaSymbol{(}\AgdaBound{t}
\AgdaSymbol{:}
\AgdaBound{Γ}
\AgdaOperator{\AgdaDatatype{⊢}}
\AgdaBound{A}\AgdaSymbol{)}
\AgdaSymbol{→}
\AgdaOperator{\AgdaDatatype{SN}}
\AgdaBound{t}
\vspace{0.5em}

We give a definition of this function that represents the proof
of normalization by generalizing the type \AgdaOperator{\AgdaDatatype{SN}} with the addition of a predicate about the term. In the proof, we use this predicate to add information
about the term when constructing introductions, and then we use that information
in the cases of eliminations. 

\begin{code}%
	\>[0]\AgdaKeyword{data}\AgdaSpace{}%
	\AgdaDatatype{SN*}\AgdaSpace{}%
	\AgdaSymbol{\{}\AgdaBound{Γ}\AgdaSpace{}%
	\AgdaBound{A}\AgdaSymbol{\}}\AgdaSpace{}%
	\AgdaSymbol{(}\AgdaBound{P}\AgdaSpace{}%
	\AgdaSymbol{:}\AgdaSpace{}%
	\AgdaBound{Γ}\AgdaSpace{}%
	\AgdaOperator{\AgdaDatatype{⊢}}\AgdaSpace{}%
	\AgdaBound{A}\AgdaSpace{}%
	\AgdaSymbol{→}\AgdaSpace{}%
	\AgdaPrimitive{Set}\AgdaSymbol{)}\AgdaSpace{}%
	\AgdaSymbol{(}\AgdaBound{t}\AgdaSpace{}%
	\AgdaSymbol{:}\AgdaSpace{}%
	\AgdaBound{Γ}\AgdaSpace{}%
	\AgdaOperator{\AgdaDatatype{⊢}}\AgdaSpace{}%
	\AgdaBound{A}\AgdaSymbol{)}\AgdaSpace{}%
	\AgdaSymbol{:}\AgdaSpace{}%
	\AgdaPrimitive{Set}\AgdaSpace{}%
	\AgdaKeyword{where}\<%
	\\
	\>[0][@{}l@{\AgdaIndent{0}}]%
	\>[2]\AgdaInductiveConstructor{sn*}\AgdaSpace{}%
	\AgdaSymbol{:}\AgdaSpace{}%
	\AgdaBound{P}\AgdaSpace{}%
	\AgdaBound{t}\AgdaSpace{}%
	\AgdaSymbol{→}\AgdaSpace{}%
	\AgdaSymbol{(∀}\AgdaSpace{}%
	\AgdaSymbol{\{}\AgdaBound{t'}\AgdaSymbol{\}}\AgdaSpace{}%
	\AgdaSymbol{→}\AgdaSpace{}%
	\AgdaBound{t}\AgdaSpace{}%
	\AgdaOperator{\AgdaDatatype{⇝}}\AgdaSpace{}%
	\AgdaBound{t'}\AgdaSpace{}%
	\AgdaSymbol{→}\AgdaSpace{}%
	\AgdaDatatype{SN*}\AgdaSpace{}%
	\AgdaBound{P}\AgdaSpace{}%
	\AgdaBound{t'}\AgdaSymbol{)}\AgdaSpace{}%
	\AgdaSymbol{→}\AgdaSpace{}%
	\AgdaDatatype{SN*}\AgdaSpace{}%
	\AgdaBound{P}\AgdaSpace{}%
	\AgdaBound{t}\<%
\end{code}

It is easy to see that if a term satisfies \AgdaOperator{\AgdaDatatype{SN*}},
then it also satisfies \AgdaOperator{\AgdaDatatype{SN}}.

To begin the proof, we need to provide some definitions.
First, we define the ``interpretation of the term'', which will be used for adding
some extra hypotheses that allow us to proceed with the proof in the cases of eliminations.
A pair is interpreted as the product of the normalization of each element of the pair,
an abstraction (\AgdaSymbol{λ} \AgdaBound{t}) is interpreted as the normalization of the term
 \AgdaBound{ρ}(\AgdaBound{t})[\AgdaBound{u}], where
\AgdaBound{u} is any normalizing term and \AgdaBound{ρ} is any rename function, and
the other terms are interpreted as the unit type of the module Data.Unit.

\begin{code}%
\>[0]\AgdaOperator{\AgdaFunction{⟦\AgdaUnderscore{}⟧}}\AgdaSpace{}%
\AgdaSymbol{:}\AgdaSpace{}%
\AgdaSymbol{∀}\AgdaSpace{}%
\AgdaSymbol{\{}\AgdaBound{Γ}\AgdaSpace{}%
\AgdaBound{A}\AgdaSymbol{\}}\AgdaSpace{}%
\AgdaSymbol{→}\AgdaSpace{}%
\AgdaBound{Γ}\AgdaSpace{}%
\AgdaOperator{\AgdaDatatype{⊢}}\AgdaSpace{}%
\AgdaBound{A}\AgdaSpace{}%
\AgdaSymbol{→}\AgdaSpace{}%
\AgdaPrimitive{Set}\<%
\\
\>[0]\AgdaOperator{\AgdaFunction{⟦}}\AgdaSpace{}%
\AgdaSymbol{(}\AgdaOperator{\AgdaInductiveConstructor{ƛ}}\AgdaSpace{}%
\AgdaBound{t}\AgdaSymbol{)}\AgdaSpace{}%
\AgdaOperator{\AgdaFunction{⟧}}\AgdaSpace{}%
\>[1]\AgdaSymbol{=}\AgdaSpace{}%
\AgdaSymbol{∀}\AgdaSpace{}%
\AgdaSymbol{\{}\AgdaBound{Δ}\AgdaSymbol{\}\{}\AgdaBound{ρ}\AgdaSpace{}%
\AgdaSymbol{:}\AgdaSpace{}%
\AgdaFunction{Rename}\AgdaSpace{}%
\AgdaSymbol{\AgdaUnderscore{}}\AgdaSpace{}%
\AgdaBound{Δ}\AgdaSymbol{\}\{}\AgdaBound{u}\AgdaSymbol{\}}\AgdaSpace{} \AgdaSymbol{→}\\
\>[0]\AgdaSpace{}%
\AgdaDatatype{SN*}\AgdaSpace{}%
\AgdaOperator{\AgdaFunction{⟦\AgdaUnderscore{}⟧}}\AgdaSpace{}%
\AgdaBound{u}\AgdaSpace{}%
\AgdaSymbol{→}\AgdaSpace{}%
\AgdaDatatype{SN*}\AgdaSpace{}%
\AgdaOperator{\AgdaFunction{⟦\AgdaUnderscore{}⟧}}\AgdaSpace{}%
\AgdaSymbol{(}\AgdaOperator{\AgdaFunction{⟪}}\AgdaSpace{}%
\AgdaBound{u}\AgdaSpace{}%
\AgdaOperator{\AgdaFunction{•}}\AgdaSpace{}%
\AgdaSymbol{(}\AgdaFunction{ids}\AgdaSpace{}%
\AgdaOperator{\AgdaFunction{∘}}\AgdaSpace{}%
\AgdaBound{ρ}\AgdaSymbol{)}\AgdaSpace{}%
\AgdaOperator{\AgdaFunction{⟫}}\AgdaSpace{}%
\AgdaBound{t}\AgdaSymbol{)}\<%
\\
\>[0]\AgdaOperator{\AgdaFunction{⟦}}\AgdaSpace{}%
\AgdaOperator{\AgdaInductiveConstructor{⟨}}\AgdaSpace{}%
\AgdaBound{a}\AgdaSpace{}%
\AgdaOperator{\AgdaInductiveConstructor{,}}\AgdaSpace{}%
\AgdaBound{b}\AgdaSpace{}%
\AgdaOperator{\AgdaInductiveConstructor{⟩}}\AgdaSpace{}%
\AgdaOperator{\AgdaFunction{⟧}}\AgdaSpace{}%
\>[1]\AgdaSymbol{=}\AgdaSpace{}%
\AgdaDatatype{SN*}\AgdaSpace{}%
\AgdaOperator{\AgdaFunction{⟦\AgdaUnderscore{}⟧}}\AgdaSpace{}%
\AgdaBound{a}\AgdaSpace{}%
\AgdaOperator{\AgdaFunction{⊗}}\AgdaSpace{}%
\AgdaDatatype{SN*}\AgdaSpace{}%
\AgdaOperator{\AgdaFunction{⟦\AgdaUnderscore{}⟧}}\AgdaSpace{}%
\AgdaBound{b}\<%
\\
\>[0]\AgdaCatchallClause{\AgdaOperator{\AgdaFunction{⟦}}}\AgdaSpace{}%
\AgdaCatchallClause{\AgdaBound{t}}\AgdaSpace{}%
\AgdaCatchallClause{\AgdaOperator{\AgdaFunction{⟧}}}\AgdaSpace{}%
\>[1]\AgdaSymbol{=}\AgdaSpace{}%
\AgdaRecord{Top}\<%
\end{code}

\noindent where \AgdaOperator{\AgdaFunction{⊗}} is the renaming of the operator
\AgdaOperator{\AgdaFunction{×}} of the module Data.Product,
and \AgdaRecord{Top} the renaming of \ttop of Data.Unit.
The definition of interpretation of terms can be extended to substitutions by defining a
predicate that states that a substitution \AgdaBound{σ} is adequate in a context \AgdaBound{Γ},
written as \AgdaBound{Γ} \AgdaSymbol{⊨} \AgdaBound{σ},
when all terms that result from the application of \AgdaBound{σ} to any variable are strongly
normalizing.

\begin{code}%
\>[0]\AgdaOperator{\AgdaFunction{\AgdaUnderscore{}⊨\AgdaUnderscore{}}}\AgdaSpace{}%
\AgdaSymbol{:}\AgdaSpace{}%
\AgdaSymbol{∀\{}\AgdaBound{Δ}\AgdaSymbol{\}}\AgdaSpace{}%
\AgdaSymbol{→}\AgdaSpace{}%
\AgdaSymbol{(}\AgdaBound{Γ}\AgdaSpace{}%
\AgdaSymbol{:}\AgdaSpace{}%
\AgdaDatatype{Context}\AgdaSymbol{)}\AgdaSpace{}%
\AgdaSymbol{→}\AgdaSpace{}%
\AgdaSymbol{(}\AgdaBound{σ}\AgdaSpace{}%
\AgdaSymbol{:}\AgdaSpace{}%
\AgdaFunction{Subst}\AgdaSpace{}%
\AgdaBound{Γ}\AgdaSpace{}%
\AgdaBound{Δ}\AgdaSymbol{)}\AgdaSpace{}%
\AgdaSymbol{→}\AgdaSpace{}%
\AgdaPrimitive{Set}\<%
\\
\>[0]\AgdaBound{Γ}\AgdaSpace{}%
\AgdaOperator{\AgdaFunction{⊨}}\AgdaSpace{}%
\AgdaBound{σ}\AgdaSpace{}%
\AgdaSymbol{=}\AgdaSpace{}%
\AgdaSymbol{∀\{}\AgdaBound{A}\AgdaSymbol{\}}\AgdaSpace{}%
\AgdaSymbol{(}\AgdaBound{v}\AgdaSpace{}%
\AgdaSymbol{:}\AgdaSpace{}%
\AgdaBound{Γ}\AgdaSpace{}%
\AgdaOperator{\AgdaDatatype{∋}}\AgdaSpace{}%
\AgdaBound{A}\AgdaSymbol{)}\AgdaSpace{}%
\AgdaSymbol{→}\AgdaSpace{}%
\AgdaDatatype{SN*}\AgdaSpace{}%
\AgdaOperator{\AgdaFunction{⟦\AgdaUnderscore{}⟧}}\AgdaSpace{}%
\AgdaSymbol{(}\AgdaBound{σ}\AgdaSpace{}%
\AgdaSymbol{\{}\AgdaBound{A}\AgdaSymbol{\}}\AgdaSpace{}%
\AgdaBound{v}\AgdaSymbol{)}\<%
\end{code}

In particular, the identity substitution \AgdaBound{ids} is an adequate substitution:

\begin{code}%
\>[0]\AgdaFunction{⊨ids}\AgdaSpace{}%
\AgdaSymbol{:}\AgdaSpace{}%
\AgdaSymbol{∀\{}\AgdaBound{Γ}\AgdaSymbol{\}}\AgdaSpace{}%
\AgdaSymbol{→}\AgdaSpace{}%
\AgdaBound{Γ}\AgdaSpace{}%
\AgdaOperator{\AgdaFunction{⊨}}\AgdaSpace{}%
\AgdaFunction{ids}\<%
\\
\>[0]\AgdaFunction{⊨ids}\AgdaSpace{}%
\AgdaSymbol{\AgdaUnderscore{}}\AgdaSpace{}%
\AgdaSymbol{=} \AgdaSpace{}%
\AgdaFunction{SN*-rename}\AgdaSpace{}%
\AgdaOperator{\AgdaInductiveConstructor{S\AgdaUnderscore{}}}\AgdaSpace{}%
\AgdaSymbol{(}\AgdaBound{σ}\AgdaSpace{}%
\AgdaBound{v}\AgdaSymbol{)}\<%
\end{code}

\noindent The complexity of the normalization proof lies in the proof of a fundamental theorem
named \func{adequacy}, which states that for any term \AgdaBound{t} and adequate substitution \AgdaBound{σ}, \snstar\parens{\substi{\bound{σ}}{\bound{t}}} holds.
The proof of this theorem is extensive and requires proving some extra lemmas.
In the next subsection we present some of the most relevant cases of the proof;
here we just give the type of the theorem:

\begin{code}
\>[0]\AgdaFunction{adequacy}\AgdaSpace{}%
\AgdaSymbol{:}\AgdaSpace{}%
\AgdaSymbol{∀}\AgdaSpace{}%
\AgdaSymbol{\{}\AgdaBound{Γ}\AgdaSpace{}%
\AgdaBound{Δ}\AgdaSpace{}%
\AgdaBound{A}\AgdaSymbol{\}}\AgdaSpace{}%
\AgdaSymbol{\{}\AgdaBound{σ}\AgdaSpace{}%
\AgdaSymbol{:}\AgdaSpace{}%
\AgdaFunction{Subst}\AgdaSpace{}%
\AgdaBound{Γ}\AgdaSpace{}%
\AgdaBound{Δ}\AgdaSymbol{\}}\AgdaSpace{}%
\AgdaSymbol{→}\AgdaSpace{}%
\AgdaSymbol{(}\AgdaBound{t}\AgdaSpace{}%
\AgdaSymbol{:}\AgdaSpace{}%
\AgdaBound{Γ}\AgdaSpace{}%
\AgdaOperator{\AgdaDatatype{⊢}}\AgdaSpace{}%
\AgdaBound{A}\AgdaSymbol{)}\AgdaSpace{}%
\AgdaSymbol{→}\AgdaSpace{} \\
\>[0]\AgdaBound{Γ}\AgdaSpace{}%
\AgdaOperator{\AgdaFunction{⊨}}\AgdaSpace{}%
\AgdaBound{σ}\AgdaSpace{}%
\AgdaSymbol{→}\AgdaSpace{}%
\AgdaDatatype{SN*}\AgdaSpace{}%
\AgdaOperator{\AgdaFunction{⟦\AgdaUnderscore{}⟧}}\AgdaSpace{}%
\AgdaSymbol{(}\AgdaOperator{\AgdaFunction{⟪}}\AgdaSpace{}%
\AgdaBound{σ}\AgdaSpace{}%
\AgdaOperator{\AgdaFunction{⟫}}\AgdaSpace{}%
\AgdaBound{t}\AgdaSymbol{)}\<
\end{code}

Finally, the strong normalization property is proved by instantiating \func{adequacy} with the identity substitution:

\begin{code}%
\>[0]\AgdaFunction{strong-norm}\AgdaSpace{}%
\AgdaSymbol{:}\AgdaSpace{}%
\AgdaSymbol{∀}\AgdaSpace{}%
\AgdaSymbol{\{}\AgdaBound{Γ}\AgdaSpace{}%
\AgdaBound{A}\AgdaSymbol{\}}\AgdaSpace{}%
\AgdaSymbol{(}\AgdaBound{t}\AgdaSpace{}%
\AgdaSymbol{:}\AgdaSpace{}%
\AgdaBound{Γ}\AgdaSpace{}%
\AgdaOperator{\AgdaDatatype{⊢}}\AgdaSpace{}%
\AgdaBound{A}\AgdaSymbol{)}\AgdaSpace{}%
\AgdaSymbol{→}\AgdaSpace{}%
\AgdaDatatype{SN}\AgdaSpace{}%
\AgdaBound{t}\<%
\\
\>[0]\AgdaFunction{strong-norm}\AgdaSpace{}%
\AgdaBound{t}\AgdaSpace{}%
\AgdaSymbol{=}\AgdaSpace{}%
\AgdaFunction{transport}\AgdaSpace{}%
\AgdaDatatype{SN}\AgdaSpace{}%
\AgdaFunction{sub-id}\AgdaSpace{}%
\AgdaSymbol{(}\AgdaFunction{SN*-SN} \\ 
\>[0]\AgdaSymbol{(}\AgdaFunction{adequacy}\AgdaSpace{}%
\AgdaBound{t}\AgdaSpace{}%
\AgdaFunction{⊨ids}\AgdaSymbol{))}\<%
\end{code}

\noindent where the lemma \func{sub-id}
eliminates the application of the identity substitution to the term, and the function
\func{transport} is a renaming of \func{subst}, defined in the module
Relation.Binary.
\noindent PropositionalEquality.

\subsection{The adequacy function}
We prove the adequacy theorem by induction on the term \bound{t}.
In each case of the proof we use auxiliary lemmas.
To simplify the explanation, we present the proof by cases.

\vspace{0.5em}
{\bf Case Variable}
\vspace{0.3em}

When the term \bound{t} is a variable, we have to prove that the substitution \bound{σ}
applied to the variable satisfies \snstar.
Since we have as a hypothesis that \bound{σ} is an adequate substitution, it suffices to
apply \bound{σ} to the variable:

\begin{code}%
			\>[0]\AgdaFunction{adequacy}\AgdaSpace{}%
			\AgdaSymbol{(}\AgdaOperator{\AgdaInductiveConstructor{`}}\AgdaSpace{}%
			\AgdaBound{v}\AgdaSymbol{)}%
			\>[23]\AgdaBound{⊨σ}\AgdaSpace{}%
			\AgdaSymbol{=}\AgdaSpace{}%
			\AgdaBound{⊨σ}\AgdaSpace{}%
			\AgdaBound{v}\<%
		\end{code}

\vspace{0.5em}
{\bf Case Top}
\vspace{0.3em}

This is the simplest case, since there is no possible reduction step from \const{$\star$}:

\begin{code}%
			\>[0]\AgdaFunction{adequacy}\AgdaSpace{}%
			\AgdaInductiveConstructor{⋆}%
			\>[23]\AgdaSymbol{\AgdaUnderscore{}}\AgdaSpace{}\AgdaSpace{}%
			\AgdaSymbol{=}\AgdaSpace{}%
			\AgdaInductiveConstructor{sn*}\AgdaSpace{}%
            \AgdaInductiveConstructor{tt}\AgdaSpace{}%
            \AgdaSymbol{(λ}\AgdaSpace{}%
			\AgdaSymbol{())}\<%
		\end{code}

\vspace{0.5em}
{\bf Case Pair}
\vspace{0.3em}

When the term \bound{t} is a pair, the following lemma is defined to allow us to conclude
\snstar \bound{t}, if \snstar holds for the subterms of the pair \bound{t}.

\begin{code}%
	\>[0]\AgdaFunction{lemma-⟨,⟩}\AgdaSpace{}%
	\AgdaSymbol{:}\AgdaSpace{}%
	\AgdaSymbol{∀}\AgdaSpace{}%
	\AgdaSymbol{\{}\AgdaBound{Γ}\AgdaSpace{}%
	\AgdaBound{A}\AgdaSpace{}%
	\AgdaBound{B}\AgdaSymbol{\}}\AgdaSpace{}%
	\AgdaSymbol{→}\AgdaSpace{}%
	\AgdaSymbol{\{}\AgdaBound{a}\AgdaSpace{}%
	\AgdaSymbol{:}\AgdaSpace{}%
	\AgdaBound{Γ}\AgdaSpace{}%
	\AgdaOperator{\AgdaDatatype{⊢}}\AgdaSpace{}%
	\AgdaBound{A}\AgdaSymbol{\}}\AgdaSpace{}%
	\AgdaSymbol{\{}\AgdaBound{b}\AgdaSpace{}%
	\AgdaSymbol{:}\AgdaSpace{}%
	\AgdaBound{Γ}\AgdaSpace{}%
	\AgdaOperator{\AgdaDatatype{⊢}}\AgdaSpace{}%
	\AgdaBound{B}\AgdaSymbol{\}}\AgdaSpace{}%
	\AgdaSymbol{→}\<%
	\\
	\>[0][@{}l@{\AgdaIndent{0}}]%
	\>[2]\AgdaDatatype{SN*}\AgdaSpace{}%
	\AgdaOperator{\AgdaFunction{⟦\AgdaUnderscore{}⟧}}\AgdaSpace{}%
	\AgdaBound{a}\AgdaSpace{}%
	\AgdaSymbol{→}\AgdaSpace{}%
	\AgdaDatatype{SN*}\AgdaSpace{}%
	\AgdaOperator{\AgdaFunction{⟦\AgdaUnderscore{}⟧}}\AgdaSpace{}%
	\AgdaBound{b}\AgdaSpace{}%
	\AgdaSymbol{→}\AgdaSpace{}%
	\AgdaDatatype{SN*}\AgdaSpace{}%
	\AgdaOperator{\AgdaFunction{⟦\AgdaUnderscore{}⟧}}\AgdaSpace{}%
	\AgdaSymbol{(}\AgdaOperator{\AgdaInductiveConstructor{⟨}}\AgdaSpace{}%
	\AgdaBound{a}\AgdaSpace{}%
	\AgdaOperator{\AgdaInductiveConstructor{,}}\AgdaSpace{}%
	\AgdaBound{b}\AgdaSpace{}%
	\AgdaOperator{\AgdaInductiveConstructor{⟩}}\AgdaSymbol{)}\<%
\end{code}

\noindent The proof of this lemma is solved by case analysis on the reduction step taken by
\snstar \bound{t}, which can be a left or right congruence
of the relations \reduces and \isoterm; in each case we use the inductive hypotheses on the subterms.

\noindent Then, we complete the proof of \func{adequacy} for this case as follows:

\begin{code}%
			\>[0]\AgdaFunction{adequacy}\AgdaSpace{}%
			\AgdaOperator{\AgdaInductiveConstructor{⟨}}\AgdaSpace{}%
			\AgdaBound{a}\AgdaSpace{}%
			\AgdaOperator{\AgdaInductiveConstructor{,}}\AgdaSpace{}%
			\AgdaBound{b}\AgdaSpace{}%
			\AgdaOperator{\AgdaInductiveConstructor{⟩}}%
			\>[23]\AgdaBound{⊨σ}\AgdaSpace{}%
			\AgdaSymbol{=} \\
			\>[0][@{}l@{\AgdaIndent{0}}]
			\>[2]\AgdaFunction{lemma-⟨,⟩}\AgdaSpace{}%
			\AgdaSymbol{(}\AgdaFunction{adequacy}\AgdaSpace{}%
			\AgdaBound{a}\AgdaSpace{}%
			\AgdaBound{⊨σ}\AgdaSymbol{)}\AgdaSpace{}%
			\AgdaSymbol{(}\AgdaFunction{adequacy}\AgdaSpace{}%
			\AgdaBound{b}\AgdaSpace{}%
			\AgdaBound{⊨σ}\AgdaSymbol{)}\<%
		\end{code}

\vspace{0.5em}
{\bf Case Application}
\vspace{0.3em}

A similar lemma is needed for the case in which \bound{t} is an application:

\begin{code}%
\>[0]\AgdaFunction{lemma-·}\AgdaSpace{}%
\AgdaSymbol{:}\AgdaSpace{}%
\AgdaSymbol{∀}\AgdaSpace{}%
\AgdaSymbol{\{}\AgdaBound{Γ}\AgdaSpace{}%
\AgdaBound{A}\AgdaSpace{}%
\AgdaBound{B}\AgdaSymbol{\}}\AgdaSpace{}%
\AgdaSymbol{→}\AgdaSpace{}%
\AgdaSymbol{\{}\AgdaBound{a}\AgdaSpace{}%
\AgdaSymbol{:}\AgdaSpace{}%
\AgdaBound{Γ}\AgdaSpace{}%
\AgdaOperator{\AgdaDatatype{⊢}}\AgdaSpace{}%
\AgdaBound{A}\AgdaSpace{}%
\AgdaOperator{\AgdaInductiveConstructor{⇒}}\AgdaSpace{}%
\AgdaBound{B}\AgdaSymbol{\}}\AgdaSpace{}%
\AgdaSymbol{\{}\AgdaBound{b}\AgdaSpace{}%
\AgdaSymbol{:}\AgdaSpace{}%
\AgdaBound{Γ}\AgdaSpace{}%
\AgdaOperator{\AgdaDatatype{⊢}}\AgdaSpace{}%
\AgdaBound{A}\AgdaSymbol{\}}\AgdaSpace{}%
\AgdaSymbol{→}\<%
\\
\>[0][@{}l@{\AgdaIndent{0}}]%
\>[2]\AgdaDatatype{SN*}\AgdaSpace{}%
\AgdaOperator{\AgdaFunction{⟦\AgdaUnderscore{}⟧}}\AgdaSpace{}%
\AgdaBound{a}\AgdaSpace{}%
\AgdaSymbol{→}\AgdaSpace{}%
\AgdaDatatype{SN*}\AgdaSpace{}%
\AgdaOperator{\AgdaFunction{⟦\AgdaUnderscore{}⟧}}\AgdaSpace{}%
\AgdaBound{b}\AgdaSpace{}%
\AgdaSymbol{→}\AgdaSpace{}%
\AgdaDatatype{SN*}\AgdaSpace{}%
\AgdaOperator{\AgdaFunction{⟦\AgdaUnderscore{}⟧}}\AgdaSpace{}%
\AgdaSymbol{(}\AgdaBound{a}\AgdaSpace{}%
\AgdaOperator{\AgdaInductiveConstructor{·}}\AgdaSpace{}%
\AgdaBound{b}\AgdaSymbol{)}\<%
\end{code}

\noindent In this lemma we proceed similarly as in the previous lemma,
with the exception that now the step can also be a $\beta$-reduction. This case is solved by applying
the interpretation of the left term of
the application (which must be an abstraction) to the inductive hypothesis of the right term.

\noindent The adequacy lemma for this case is solved as:

\begin{code}%
			\>[0]\AgdaFunction{adequacy}\AgdaSpace{}%
			\AgdaSymbol{(}\AgdaBound{a}\AgdaSpace{}%
			\AgdaOperator{\AgdaInductiveConstructor{·}}\AgdaSpace{}%
			\AgdaBound{b}\AgdaSymbol{)}%
			\>[23]\AgdaBound{⊨σ}\AgdaSpace{}%
			\AgdaSymbol{=} \\ 
			\>[0][@{}l@{\AgdaIndent{0}}]
			\>[2]
			\AgdaFunction{lemma-·}\AgdaSpace{}%
			\AgdaSymbol{(}\AgdaFunction{adequacy}\AgdaSpace{}%
			\AgdaBound{a}\AgdaSpace{}%
			\AgdaBound{⊨σ}\AgdaSymbol{)}\AgdaSpace{}%
			\AgdaSymbol{(}\AgdaFunction{adequacy}\AgdaSpace{}%
			\AgdaBound{b}\AgdaSpace{}%
			\AgdaBound{⊨σ}\AgdaSymbol{)}\<%
		\end{code}

\vspace{0.5em}
{\bf Case Projection}
\vspace{0.3em}

The case where \bound{t} is a projection requires the following lemma, which asserts that
\snstar \bound{t} holds if \snstar holds for the
term being projected:

\begin{code}%
	\>[0]\AgdaFunction{lemma-π}\AgdaSpace{}%
	\AgdaSymbol{:}\AgdaSpace{}%
	\AgdaSymbol{∀}\AgdaSpace{}%
	\AgdaSymbol{\{}\AgdaBound{Γ}\AgdaSpace{}%
	\AgdaBound{A}\AgdaSpace{}%
	\AgdaBound{B}\AgdaSpace{}%
	\AgdaBound{C}\AgdaSpace{}%
	\AgdaBound{p}\AgdaSymbol{\}}\AgdaSpace{}%
	\AgdaSymbol{→}\AgdaSpace{}%
	\AgdaSymbol{\{}\AgdaBound{a}\AgdaSpace{}%
	\AgdaSymbol{:}\AgdaSpace{}%
	\AgdaBound{Γ}\AgdaSpace{}%
	\AgdaOperator{\AgdaDatatype{⊢}}\AgdaSpace{}%
	\AgdaBound{A}\AgdaSpace{}%
	\AgdaOperator{\AgdaInductiveConstructor{×}}\AgdaSpace{}%
	\AgdaBound{B}\AgdaSymbol{\}}\AgdaSpace{} \AgdaSymbol{→}\AgdaSpace{}%
	\\
	\>[0][@{}l@{\AgdaIndent{0}}]%
	\>[2]\AgdaDatatype{SN*}\AgdaSpace{}%
	\AgdaOperator{\AgdaFunction{⟦\AgdaUnderscore{}⟧}}\AgdaSpace{}%
	\AgdaBound{a}\AgdaSpace{}%
	\AgdaSymbol{→}\AgdaSpace{}%
	\AgdaDatatype{SN*}\AgdaSpace{}%
	\AgdaOperator{\AgdaFunction{⟦\AgdaUnderscore{}⟧}}\AgdaSpace{}%
	\AgdaSymbol{(}\AgdaInductiveConstructor{π}\AgdaSpace{}%
	\AgdaBound{C}\AgdaSpace{}%
	\AgdaSymbol{\{}\AgdaBound{p}\AgdaSymbol{\}}\AgdaSpace{}%
	\AgdaBound{a}\AgdaSymbol{)}\<%
\end{code}

\noindent The proof of this lemma is similar to the proof of \AgdaFunction{lemma-·} in the sense that when the step is a reduction (\const{β-π₁} or \const{β-π₂}), the lemma is solved using the
interpretation of the subterm, which in this case is the interpretation of a pair \ppair{a}{b} that
has the form \snstar\bound{a} \func{⊗} \snstar\bound{b}.

\noindent The adequacy lemma for this case is solved as:

\begin{code}%
			\>[0]\AgdaFunction{adequacy}\AgdaSpace{}%
			\AgdaSymbol{(}\AgdaInductiveConstructor{π}\AgdaSpace{}%
			\AgdaSymbol{\AgdaUnderscore{}}\AgdaSpace{}%
			\AgdaBound{x}\AgdaSymbol{)}%
			\>[23]\AgdaBound{⊨σ}\AgdaSpace{}%
			\AgdaSymbol{=}\AgdaSpace{}%
			\AgdaFunction{lemma-π}\AgdaSpace{}%
			\AgdaSymbol{(}\AgdaFunction{adequacy}\AgdaSpace{}%
			\AgdaBound{x}\AgdaSpace{}%
			\AgdaBound{⊨σ}\AgdaSymbol{)}\<%
		\end{code}

\vspace{0.5em}
{\bf Case Abstraction}
\vspace{0.3em}

The case where \bound{t} is an abstraction is more difficult than the previous ones.
As in these cases, we need to prove \snstar (\const{ƛ} \bound{t}) from \snstar \bound{t}, given
the interpretation of (\const{ƛ} \bound{t}):

\begin{code}%
\>[0]\AgdaFunction{lemma-ƛ}\AgdaSpace{}%
\AgdaSymbol{:}\AgdaSpace{}%
\AgdaSymbol{∀}\AgdaSpace{}%
\AgdaSymbol{\{}\AgdaBound{Γ}\AgdaSpace{}%
\AgdaBound{A}\AgdaSpace{}%
\AgdaBound{B}\AgdaSymbol{\}}\AgdaSpace{}%
\AgdaSymbol{→}\AgdaSpace{}%
\AgdaSymbol{\{}\AgdaBound{t}\AgdaSpace{}%
\AgdaSymbol{:}\AgdaSpace{}%
\AgdaBound{Γ}\AgdaSpace{}%
\AgdaOperator{\AgdaInductiveConstructor{,}}\AgdaSpace{}%
\AgdaBound{B}\AgdaSpace{}%
\AgdaOperator{\AgdaDatatype{⊢}}\AgdaSpace{}%
\AgdaBound{A}\AgdaSymbol{\}}\AgdaSpace{}%
\AgdaSymbol{→}\AgdaSpace{} \\
\>[0][@{}l@{\AgdaIndent{0}}]%
\>[2] \AgdaOperator{\AgdaFunction{⟦}}\AgdaSpace{}%
\AgdaOperator{\AgdaInductiveConstructor{ƛ}}\AgdaSpace{}%
\AgdaBound{t}\AgdaSpace{}%
\AgdaOperator{\AgdaFunction{⟧}}\AgdaSpace{}%
\AgdaSymbol{→}\AgdaSpace{}%
\AgdaDatatype{SN*}\AgdaSpace{}%
\AgdaOperator{\AgdaFunction{⟦\AgdaUnderscore{}⟧}}\AgdaSpace{}%
\AgdaBound{t}\AgdaSpace{}%
\AgdaSymbol{→}\AgdaSpace{}%
\AgdaDatatype{SN*}\AgdaSpace{}%
\AgdaOperator{\AgdaFunction{⟦\AgdaUnderscore{}⟧}}\AgdaSpace{}%
\AgdaSymbol{(}\AgdaOperator{\AgdaInductiveConstructor{ƛ}}\AgdaSpace{}%
\AgdaBound{t}\AgdaSymbol{)}\<%
\end{code}

\noindent This proof required several lemmas, for example
it was necessary to prove that for any terms \bound{u} and \bound{v}, and substitution \AgdaBound{σ}:

\begin{enumerate}
 \item if \bound{u} \reduces \bound{v} then, \AgdaOperator{\AgdaFunction{⟪}}\AgdaSpace{}\AgdaBound{σ}\AgdaSpace{}\AgdaOperator{\AgdaFunction{⟫}}\AgdaSpace{} \bound{u} \reduces \AgdaOperator{\AgdaFunction{⟪}}\AgdaSpace{}
\AgdaBound{σ} \AgdaOperator{\AgdaFunction{⟫}}\AgdaSpace{} \bound{v}
\item if \bound{u} \isoterm \bound{v} then, \AgdaOperator{\AgdaFunction{⟪}}\AgdaSpace{}\AgdaBound{σ}\AgdaSpace{}\AgdaOperator{\AgdaFunction{⟫}}\AgdaSpace{} \bound{u} \isoterm \AgdaOperator{\AgdaFunction{⟪}}\AgdaSpace{}
\AgdaBound{σ} \AgdaOperator{\AgdaFunction{⟫}}\AgdaSpace{} \bound{v}
\item if \bound{u} is normalizing and \AgdaBound{σ} is adequate, then \cons{\AgdaBound{u}}{\AgdaBound{σ}} is adequate.
\item if \bound{u} \reduces \bound{v} then \snstar \bound{u} \reduces \snstar \bound{v}
\item if \bound{u} \isoterm \bound{v} then \snstar \bound{u} \isoterm \snstar \bound{v}

\end{enumerate}

\noindent In addition, we need to prove that an adequate substitution composed with a rename
is also an adequate substitution:

\begin{code}%
\>[0]\AgdaFunction{⊨rename}\AgdaSpace{}%
\AgdaSymbol{:}\AgdaSpace{}%
\AgdaSymbol{∀\{}\AgdaBound{Γ}\AgdaSpace{}%
\AgdaBound{Δ}\AgdaSpace{}%
\AgdaBound{Δ₁}\AgdaSymbol{\}}\AgdaSpace{}%
\AgdaSymbol{\{}\AgdaBound{σ}\AgdaSpace{}%
\AgdaSymbol{:}\AgdaSpace{}%
\AgdaFunction{Subst}\AgdaSpace{}%
\AgdaBound{Γ}\AgdaSpace{}%
\AgdaBound{Δ}\AgdaSymbol{\}}\AgdaSpace{}%
\AgdaSymbol{→}\<%
\\
\>[0][@{}l@{\AgdaIndent{0}}]%
\>[2]\AgdaBound{Γ}\AgdaSpace{}%
\AgdaOperator{\AgdaFunction{⊨}}\AgdaSpace{}%
\AgdaBound{σ}\AgdaSpace{}%
\AgdaSymbol{→}\AgdaSpace{}%
\AgdaSymbol{(}\AgdaBound{ρ}\AgdaSpace{}%
\AgdaSymbol{:}\AgdaSpace{}%
\AgdaFunction{Rename}\AgdaSpace{}%
\AgdaBound{Δ}\AgdaSpace{}%
\AgdaBound{Δ₁}\AgdaSymbol{)}\AgdaSpace{}%
\AgdaSymbol{→}\AgdaSpace{}%
\AgdaBound{Γ}\AgdaSpace{}%
\AgdaOperator{\AgdaFunction{⊨}}\AgdaSpace{}%
\AgdaSymbol{(}\AgdaOperator{\AgdaFunction{⟪}}\AgdaSpace{}%
\AgdaFunction{ids}\AgdaSpace{}%
\AgdaOperator{\AgdaFunction{∘}}\AgdaSpace{}%
\AgdaBound{ρ}\AgdaSpace{}%
\AgdaOperator{\AgdaFunction{⟫}}\AgdaSpace{}%
\AgdaOperator{\AgdaFunction{∘}}\AgdaSpace{}%
\AgdaBound{σ}\AgdaSymbol{)}\<%
\end{code}

\noindent and that an extension of an adequate substitution is also adequate:

\begin{code}%
\>[0]\AgdaFunction{⊨exts}\AgdaSpace{}%
\AgdaSymbol{:}\AgdaSpace{}%
\AgdaSymbol{∀\{}\AgdaBound{Γ}\AgdaSpace{}%
\AgdaBound{Δ}\AgdaSpace{}%
\AgdaBound{A}\AgdaSymbol{\}}\AgdaSpace{}%
\AgdaSymbol{\{}\AgdaBound{σ}\AgdaSpace{}%
\AgdaSymbol{:}\AgdaSpace{}%
\AgdaFunction{Subst}\AgdaSpace{}%
\AgdaBound{Γ}\AgdaSpace{}%
\AgdaBound{Δ}\AgdaSymbol{\}} \AgdaSymbol{→}\AgdaSpace{} \\ 
\>[0][@{}l@{\AgdaIndent{0}}]%
\>[2] \AgdaBound{Γ}\AgdaSpace{}%
\AgdaOperator{\AgdaFunction{⊨}}\AgdaSpace{}%
\AgdaBound{σ}\AgdaSpace{}%
\AgdaSymbol{→}\AgdaSpace{}%
\AgdaSymbol{(}\AgdaBound{Γ}\AgdaSpace{}%
\AgdaOperator{\AgdaInductiveConstructor{,}}\AgdaSpace{}%
\AgdaBound{A}\AgdaSymbol{)}\AgdaSpace{}%
\AgdaOperator{\AgdaFunction{⊨}}\AgdaSpace{}%
\AgdaSymbol{(}\AgdaFunction{exts}\AgdaSpace{}%
\AgdaBound{σ}\AgdaSymbol{)}\<%
\end{code}

\noindent Now, it is possible to define the case of \func{adequacy} for abstraction as follows:

\begin{samepage}
	\begin{code}%
			\>[0]\AgdaFunction{adequacy}\AgdaSpace{}%
			\AgdaSymbol{\{}\AgdaArgument{σ}\AgdaSpace{}%
			\AgdaSymbol{=}\AgdaSpace{}%
			\AgdaBound{σ}\AgdaSymbol{\}}\AgdaSpace{}%
			\AgdaSymbol{(}\AgdaOperator{\AgdaInductiveConstructor{ƛ}}\AgdaSpace{}%
			\AgdaBound{t}\AgdaSymbol{)}\AgdaSpace{}%
			\AgdaBound{⊨σ}\AgdaSpace{}%
			\AgdaSymbol{=}\<%
			\\
			\>[0][@{}l@{\AgdaIndent{0}}]%
			\>[2]\AgdaFunction{lemma-ƛ}\<%
			\\
			\>[2][@{}l@{\AgdaIndent{0}}]%
			\>[4]\AgdaSymbol{(λ}\AgdaSpace{}%
			\AgdaSymbol{\{}\AgdaSpace{}%
			\AgdaSymbol{\{}\AgdaArgument{ρ}\AgdaSpace{}%
			\AgdaSymbol{=}\AgdaSpace{}%
			\AgdaBound{ρ}\AgdaSymbol{\}\{}\AgdaArgument{u}\AgdaSpace{}%
			\AgdaSymbol{=}\AgdaSpace{}%
			\AgdaBound{u}\AgdaSymbol{\}}\AgdaSpace{}%
			\AgdaBound{SNu}\AgdaSpace{}%
			\AgdaSymbol{→}\<%
			\\
			\>[4][@{}l@{\AgdaIndent{0}}]%
			\>[6]\AgdaFunction{transport}\AgdaSpace{}%
			\AgdaSymbol{(}\AgdaDatatype{SN*}\AgdaSpace{}%
			\AgdaOperator{\AgdaFunction{⟦\AgdaUnderscore{}⟧}}\AgdaSymbol{)}\<%
			\\
			\>[6][@{}l@{\AgdaIndent{0}}]%
			\>[8]\AgdaSymbol{(}\AgdaFunction{subst-split}\AgdaSpace{}%
			\AgdaSymbol{\{}\AgdaArgument{t}\AgdaSpace{}%
			\AgdaSymbol{=}\AgdaSpace{}%
			\AgdaBound{t}\AgdaSymbol{\})}\<%
			\\
			\>[8]\AgdaSymbol{(}\AgdaFunction{adequacy}\AgdaSpace{}%
			\AgdaBound{t}\AgdaSpace{}%
			\AgdaSymbol{(}\AgdaOperator{\AgdaFunction{⊨}}\AgdaSpace{}%
			\AgdaBound{SNu}\AgdaSpace{}%
			\AgdaOperator{\AgdaFunction{•}}\AgdaSpace{}%
			\AgdaSymbol{(}\AgdaFunction{⊨rename}\AgdaSpace{}%
			\AgdaBound{⊨σ}\AgdaSpace{}%
			\AgdaBound{ρ}\AgdaSymbol{)))\})}\AgdaSpace{}\<%
			\\
			\>[4]\AgdaSymbol{(}\AgdaFunction{adequacy}\AgdaSpace{}%
			\AgdaBound{t}\AgdaSpace{}%
			\AgdaSymbol{(}\AgdaFunction{⊨exts}\AgdaSpace{}%
			\AgdaBound{⊨σ}\AgdaSymbol{))}\<%
		\end{code}
\end{samepage}

\noindent The function \func{subst-split} is used to combine the substitutions \bound{σ} and
\cons{\bound{u}}{\parens{\comp{\ids}{\bound{$\rho$}}}} into a single substitution.

\vspace{0.5em}
{\bf Case Iso}
\vspace{0.3em}

Finally, the case where t is the constructor of isomorphism is the most difficult case,
because we must resolve the cases for each equivalence of terms.
As in the other cases, we have a principal lemma:

\begin{code}%
			\>[0]\AgdaFunction{lemma-≡}\AgdaSpace{}%
			\AgdaSymbol{:}\AgdaSpace{}%
			\AgdaSymbol{∀}\AgdaSpace{}%
			\AgdaSymbol{\{}\AgdaBound{Γ}\AgdaSpace{}%
			\AgdaBound{A}\AgdaSpace{}%
			\AgdaBound{B}\AgdaSpace{}%
			\AgdaBound{iso}\AgdaSymbol{\}}\AgdaSpace{}%
			\AgdaSymbol{→}\AgdaSpace{}%
			\AgdaSymbol{\{}\AgdaBound{t}\AgdaSpace{}%
			\AgdaSymbol{:}\AgdaSpace{}%
			\AgdaBound{Γ}\AgdaSpace{}%
			\AgdaOperator{\AgdaDatatype{⊢}}\AgdaSpace{}%
			\AgdaBound{A}\AgdaSymbol{\}}\AgdaSpace{}%
			\AgdaSymbol{→} \\ 
			\>[8][@{}l@{\AgdaIndent{0}}]%
			\>[10]
			\AgdaDatatype{SN*}\AgdaSpace{}%
			\AgdaOperator{\AgdaFunction{⟦\AgdaUnderscore{}⟧}}\AgdaSpace{}%
			\AgdaBound{t}\AgdaSpace{}%
			\AgdaSymbol{→}\AgdaSpace{}%
			\AgdaDatatype{SN*}\AgdaSpace{}%
			\AgdaSymbol{\{}\AgdaArgument{A}\AgdaSpace{}%
			\AgdaSymbol{=}\AgdaSpace{}%
			\AgdaBound{B}\AgdaSymbol{\}}\AgdaSpace{}%
			\AgdaOperator{\AgdaFunction{⟦\AgdaUnderscore{}⟧}}\AgdaSpace{}%
			\AgdaSymbol{(}\AgdaOperator{\AgdaInductiveConstructor{[}}\AgdaSpace{}%
			\AgdaBound{iso}\AgdaSpace{}%
			\AgdaOperator{\AgdaInductiveConstructor{]≡}}\AgdaSpace{}%
			\AgdaBound{t}\AgdaSymbol{)}\<%
			\end{code}

\noindent This function is defined in terms of this auxiliary lemma:

\begin{code}
\>[0]\AgdaFunction{aux}\AgdaSpace{}
\AgdaSymbol{:}\AgdaSpace{}
\AgdaSymbol{∀}\AgdaSpace{}
\AgdaSymbol{\{}\AgdaBound{Γ}\AgdaSpace{}
\AgdaBound{A}\AgdaSpace{}
\AgdaBound{iso}\AgdaSpace{}
\AgdaBound{t'}\AgdaSymbol{\}}\AgdaSpace{}
\AgdaSymbol{→}\AgdaSpace{}
\AgdaSymbol{\{}\AgdaBound{t}\AgdaSpace{}
\AgdaSymbol{:}\AgdaSpace{}
\AgdaBound{Γ}\AgdaSpace{}
\AgdaOperator{\AgdaDatatype{⊢}}\AgdaSpace{}
\AgdaBound{A}\AgdaSymbol{\}}\AgdaSpace{}
\AgdaSymbol{→} \\ 
\>[0][@{}l@{\AgdaIndent{0}}]
\>[2] \AgdaDatatype{SN*}\AgdaSpace{}
\AgdaOperator{\AgdaFunction{⟦\AgdaUnderscore{}⟧}}\AgdaSpace{}
\AgdaBound{t}\AgdaSpace{}
\AgdaSymbol{→}\AgdaSpace{}
\AgdaSymbol{(}\AgdaOperator{\AgdaInductiveConstructor{[}}\AgdaSpace{}
\AgdaBound{iso}\AgdaSpace{}
\AgdaOperator{\AgdaInductiveConstructor{]≡}}\AgdaSpace{}
\AgdaBound{t}\AgdaSymbol{)}\AgdaSpace{}
\AgdaOperator{\AgdaDatatype{⇝}}\AgdaSpace{}
\AgdaBound{t'}\AgdaSpace{}
\AgdaSymbol{→}\AgdaSpace{}
\AgdaDatatype{SN*}\AgdaSpace{}
\AgdaOperator{\AgdaFunction{⟦\AgdaUnderscore{}⟧}}\AgdaSpace{}
\AgdaBound{t'}
\end{code}

\noindent which is defined by case analysis on the second argument (the reduction step).
The goal of this lemma is to derive \snstar \bound{t'} from \snstar \bound{t},
where \bound{t} and \bound{t'} are the terms related by the corresponding isomorphism.
In this proof we use the lemmas defined above for the different term constructors.

We will not present the proof of this theorem here due to its length and complexity,
we briefly outline some of the critical points of the proof and the techniques used:

\begin{enumerate}
 \item The case in which the isomorphism is $\langle \lambda x^A.r, \lambda x^A.s \rangle \rightleftarrows \lambda x^A. \langle r,s \rangle$, presents the difficulty of instantiating the interpretations of the left abstractions for constructing the interpretation of the right abstraction.
To resolve this, we use a lemma that concludes \snstar \bound{t} from \snstar (\const{ƛ} \bound{t}).
The idea behind the lemma is to replace the first variable of the term with index zero, thus obtaining
exactly the same term.
\item In cases involving the curry isomorphism, multiple substitutions must be combined into a single
one that can be obtained through the interpretation of the abstraction.
\end{enumerate}

As a conclusion, we note that the interpretations of the terms allow us to resolve the cases of elimination
(projection and application), but, on the other hand, the cases of introduction (product and abstraction)
become more complex, since it is in these cases that such interpretations are constructed.

\section{Evaluation}
\label{sec:eval}
Once the strong normalization property has been proven, the evaluation function can be defined in Agda since we have proof that it terminates.

First, we define the relation \type{$\_\rightsquigarrow*\_$} as the transitive and reflexive closure of
\type{$\_\rightsquigarrow\_$}:

\begin{code}
\>[0]\AgdaKeyword{data}\AgdaSpace{}%
\AgdaOperator{\AgdaDatatype{\AgdaUnderscore{}⇝*\AgdaUnderscore{}}}\AgdaSpace{}%
\AgdaSymbol{\{}\AgdaBound{Γ}\AgdaSpace{}%
\AgdaBound{A}\AgdaSymbol{\}}\AgdaSpace{}%
\AgdaSymbol{:}\AgdaSpace{}%
\AgdaSymbol{(}\AgdaBound{Γ}\AgdaSpace{}%
\AgdaOperator{\AgdaDatatype{⊢}}\AgdaSpace{}%
\AgdaBound{A}\AgdaSymbol{)}\AgdaSpace{}%
\AgdaSymbol{→}\AgdaSpace{}%
\AgdaSymbol{(}\AgdaBound{Γ}\AgdaSpace{}%
\AgdaOperator{\AgdaDatatype{⊢}}\AgdaSpace{}%
\AgdaBound{A}\AgdaSymbol{)}\AgdaSpace{}%
\AgdaSymbol{→}\AgdaSpace{}%
\AgdaPrimitive{Set}\AgdaSpace{}%
\AgdaKeyword{where}\<%
\\
\>[0][@{}l@{\AgdaIndent{0}}]%
\>[2]\AgdaOperator{\AgdaInductiveConstructor{\AgdaUnderscore{}∎}}%
\>[39I]\AgdaSymbol{:}\AgdaSpace{}%
\AgdaSymbol{(}\AgdaBound{t}\AgdaSpace{}%
\AgdaSymbol{:}\AgdaSpace{}%
\AgdaBound{Γ}\AgdaSpace{}%
\AgdaOperator{\AgdaDatatype{⊢}}\AgdaSpace{}%
\AgdaBound{A}\AgdaSymbol{)} \AgdaSpace{}
\AgdaSymbol{→}\AgdaSpace{}%
\AgdaBound{t}\AgdaSpace{}%
\AgdaOperator{\AgdaDatatype{⇝*}}\AgdaSpace{}%
\AgdaBound{t}\<%
\\
%
\>[2]\AgdaOperator{\AgdaInductiveConstructor{\AgdaUnderscore{}⇄⟨\AgdaUnderscore{}⟩\AgdaUnderscore{}}}\AgdaSpace{}%
\AgdaSymbol{:}\AgdaSpace{}%
\AgdaSymbol{(}\AgdaBound{t}\AgdaSpace{}%
\AgdaSymbol{:}\AgdaSpace{}%
\AgdaBound{Γ}\AgdaSpace{}%
\AgdaOperator{\AgdaDatatype{⊢}}\AgdaSpace{}%
\AgdaBound{A}\AgdaSymbol{)}\AgdaSpace{}%
\AgdaSymbol{\{}\AgdaBound{r}\AgdaSpace{}%
\AgdaBound{t'}\AgdaSpace{}%
\AgdaSymbol{:}\AgdaSpace{}%
\AgdaBound{Γ}\AgdaSpace{}%
\AgdaOperator{\AgdaDatatype{⊢}}\AgdaSpace{}%
\AgdaBound{A}\AgdaSymbol{\}} \AgdaSpace{} \AgdaSymbol{→} 
\\
\>[2][@{}l@{\AgdaIndent{0}}]%
\>[4]
\AgdaBound{t}\AgdaSpace{}%
\AgdaOperator{\AgdaDatatype{⇄}}\AgdaSpace{}%
\AgdaBound{r} 
%
\AgdaSymbol{→} %
\AgdaBound{r}\AgdaSpace{}%
\AgdaOperator{\AgdaDatatype{⇝*}}\AgdaSpace{}%
\AgdaBound{t'}
%
\AgdaSpace{} \AgdaSymbol{→}\AgdaSpace{}%
\AgdaBound{t}\AgdaSpace{}%
\AgdaOperator{\AgdaDatatype{⇝*}}\AgdaSpace{}%
\AgdaBound{t'}\<%
\\
%
%
\>[2]\AgdaOperator{\AgdaInductiveConstructor{\AgdaUnderscore{}↪⟨\AgdaUnderscore{}⟩\AgdaUnderscore{}}}\AgdaSpace{}%
\AgdaSymbol{:}\AgdaSpace{}%
\AgdaSymbol{(}\AgdaBound{t}\AgdaSpace{}%
\AgdaSymbol{:}\AgdaSpace{}%
\AgdaBound{Γ}\AgdaSpace{}%
\AgdaOperator{\AgdaDatatype{⊢}}\AgdaSpace{}%
\AgdaBound{A}\AgdaSymbol{)}\AgdaSpace{}%
\AgdaSymbol{\{}\AgdaBound{r}\AgdaSpace{}%
\AgdaBound{t'}\AgdaSpace{}%
\AgdaSymbol{:}\AgdaSpace{}%
\AgdaBound{Γ}\AgdaSpace{}%
\AgdaOperator{\AgdaDatatype{⊢}}\AgdaSpace{}%
\AgdaBound{A}\AgdaSymbol{\}} \AgdaSpace{} \AgdaSymbol{→}\<%
\\
\>[2][@{}l@{\AgdaIndent{0}}]%
\>[4]%
\AgdaBound{t}\AgdaSpace{}%
\AgdaOperator{\AgdaDatatype{↪}}\AgdaSpace{}%
\AgdaBound{r} \AgdaSpace{} 
%
\AgdaSymbol{→} \AgdaSpace{}%
\AgdaBound{r}\AgdaSpace{}%
\AgdaOperator{\AgdaDatatype{⇝*}}\AgdaSpace{}%
\AgdaBound{t'}
%
\AgdaSpace{} \AgdaSymbol{→}\AgdaSpace{}%
\AgdaBound{t}\AgdaSpace{}%
\AgdaOperator{\AgdaDatatype{⇝*}}\AgdaSpace{}%
\AgdaBound{t'}\<%
\end{code}

Given a closed typed term \bound{t}, the return type of the evaluation function will be the reduction sequence from \bound{t} to a value \bound{t'}, defined as the following data type:

\begin{code}
\>[0]\AgdaKeyword{data}\AgdaSpace{}%
\AgdaDatatype{Steps}\AgdaSpace{}%
\AgdaSymbol{\{}\AgdaBound{A}\AgdaSymbol{\}}\AgdaSpace{}%
\AgdaSymbol{:}\AgdaSpace{}%
\AgdaInductiveConstructor{∅}\AgdaSpace{}%
\AgdaOperator{\AgdaDatatype{⊢}}\AgdaSpace{}%
\AgdaBound{A}\AgdaSpace{}%
\AgdaSymbol{→}\AgdaSpace{}%
\AgdaPrimitive{Set}\AgdaSpace{}%
\AgdaKeyword{where}\<%
\\
\>[0][@{}l@{\AgdaIndent{0}}]%
\>[2]\AgdaInductiveConstructor{steps}\AgdaSpace{}%
\AgdaSymbol{:}\AgdaSpace{}%
\AgdaSymbol{\{}\AgdaBound{t}\AgdaSpace{}%
\AgdaBound{t'}\AgdaSpace{}%
\AgdaSymbol{:}\AgdaSpace{}%
\AgdaInductiveConstructor{∅}\AgdaSpace{}%
\AgdaOperator{\AgdaDatatype{⊢}}\AgdaSpace{}%
\AgdaBound{A}\AgdaSymbol{\}} \AgdaSpace{} \AgdaSymbol{→} \<%
\\
\>[2][@{}l@{\AgdaIndent{0}}]%
\>[4]%
\AgdaBound{t}\AgdaSpace{}%
\AgdaOperator{\AgdaDatatype{⇝*}}\AgdaSpace{}%
\AgdaBound{t'} \AgdaSpace{} 
%
\AgdaSymbol{→} \AgdaSpace{}
\AgdaDatatype{Value}\AgdaSpace{}%
\AgdaBound{t'} 
%
\AgdaSymbol{→}\AgdaSpace{}%
\AgdaDatatype{Steps}\AgdaSpace{}%
\AgdaBound{t}\<%
\end{code}

Then, the evaluation function is defined in terms of this function:

\begin{code}
\>[0]\AgdaFunction{eval´}\AgdaSpace{}%
\AgdaSymbol{:}\AgdaSpace{}%
\AgdaSymbol{∀}\AgdaSpace{}%
\AgdaSymbol{\{}\AgdaBound{A}\AgdaSymbol{\}}
\AgdaSpace{}\AgdaSymbol{→}\AgdaSpace{}%
\AgdaSymbol{(}\AgdaBound{t}\AgdaSpace{}%
\AgdaSymbol{:}\AgdaSpace{}%
\AgdaInductiveConstructor{∅}\AgdaSpace{}%
\AgdaOperator{\AgdaDatatype{⊢}}\AgdaSpace{}%
\AgdaBound{A}\AgdaSymbol{)}
\AgdaSpace{}\AgdaSymbol{→}\AgdaSpace{}%
\AgdaDatatype{SN}\AgdaSpace{}%
\AgdaBound{t}

\AgdaSpace{}\AgdaSymbol{→}\AgdaSpace{}%
\AgdaDatatype{Steps}\AgdaSpace{}%
\AgdaBound{t}\<%
\\
\>[0]\AgdaFunction{eval´}\AgdaSpace{}%
\AgdaBound{t}\AgdaSpace{}%
\AgdaSymbol{\AgdaUnderscore{}}\AgdaSpace{}%
\AgdaKeyword{with}\AgdaSpace{}%
\AgdaFunction{progress}\AgdaSpace{}%
\AgdaBound{t}\<%
\\
\>[0]\AgdaFunction{eval´}\AgdaSpace{}%
\AgdaBound{t}\AgdaSpace{}%
\AgdaSymbol{\AgdaUnderscore{}}%
\>[15]\AgdaSymbol{$\mid$}\AgdaSpace{}%
\AgdaInductiveConstructor{done}\AgdaSpace{}%
\AgdaBound{⇑t}%
\>[35]\AgdaSymbol{=}%
\>[38]\AgdaInductiveConstructor{steps}\AgdaSpace{}%
\AgdaSymbol{(}\AgdaBound{t}\AgdaSpace{}%
\AgdaOperator{\AgdaInductiveConstructor{∎}}\AgdaSymbol{)}\AgdaSpace{}%
\AgdaSymbol{(}\AgdaFunction{closed⇑→Value}\AgdaSpace{}%
\AgdaBound{⇑t}\AgdaSymbol{)}
\end{code}

\begin{code}
\>[0] \AgdaFunction{eval´}\AgdaSpace{}%
\AgdaBound{t}\AgdaSpace{}%
\AgdaSymbol{(}\AgdaInductiveConstructor{sn}\AgdaSpace{}%
\AgdaBound{f}\AgdaSymbol{)}\AgdaSpace{}%
\\
\>[0] \AgdaSpace{} \AgdaSpace{} \AgdaSpace{} \AgdaSymbol{$\mid$}\AgdaSpace{}%
\AgdaInductiveConstructor{step⇄}\AgdaSpace{}%
\AgdaSymbol{\{}\AgdaBound{r}\AgdaSymbol{\}}\AgdaSpace{}%
\AgdaBound{t⇄r}\AgdaSpace{}%
\AgdaKeyword{with}\AgdaSpace{}%
\AgdaFunction{eval´}\AgdaSpace{}%
\AgdaBound{r}\AgdaSpace{}%
\AgdaSymbol{(}\AgdaBound{f}\AgdaSpace{}%
\AgdaSymbol{(}\AgdaInductiveConstructor{inj₂}\AgdaSpace{}%
\AgdaBound{t⇄r}\AgdaSymbol{))} 
\end{code}

\begin{code}
\>[0] \AgdaSymbol{...}%
\AgdaSymbol{$\mid$}\AgdaSpace{}%
\AgdaInductiveConstructor{steps}\AgdaSpace{}%
\AgdaBound{r⇝t'}\AgdaSpace{}%
\AgdaBound{fin}%
\AgdaSymbol{=}%
\AgdaInductiveConstructor{steps}\AgdaSpace{}%
\AgdaSymbol{(}\AgdaBound{t}\AgdaSpace{}%
\AgdaOperator{\AgdaInductiveConstructor{⇄⟨}}\AgdaSpace{}%
\AgdaBound{t⇄r}\AgdaSpace{}%
\AgdaOperator{\AgdaInductiveConstructor{⟩}}\AgdaSpace{}%
\AgdaBound{r⇝t'}\AgdaSymbol{)}\AgdaSpace{}%
\AgdaBound{fin}\<%
\\
\>[0] \AgdaFunction{eval´}\AgdaSpace{}%
\AgdaBound{t}\AgdaSpace{}%
\AgdaSymbol{(}\AgdaInductiveConstructor{sn}\AgdaSpace{}%
\AgdaBound{f}\AgdaSymbol{)}\AgdaSpace{}%
\\
\>[0] \AgdaSpace{}\AgdaSpace{}\AgdaSpace{} \AgdaSpace{} \AgdaSymbol{$\mid$}\AgdaSpace{}%
\AgdaInductiveConstructor{step↪}\AgdaSpace{}%
\AgdaSymbol{\{}\AgdaBound{r}\AgdaSymbol{\}}\AgdaSpace{}%
\AgdaBound{t↪r}\AgdaSpace{}%
\AgdaKeyword{with}\AgdaSpace{}%
\AgdaFunction{eval´}\AgdaSpace{}%
\AgdaBound{r}\AgdaSpace{}%
\AgdaSymbol{(}\AgdaBound{f}\AgdaSpace{}%
\AgdaSymbol{(}\AgdaInductiveConstructor{inj₁}\AgdaSpace{}%
\AgdaBound{t↪r}\AgdaSymbol{))}\<%
\end{code}

\begin{code}
\>[0]
\AgdaSymbol{...}%
\AgdaSymbol{$\mid$}\AgdaSpace{}%
\AgdaInductiveConstructor{steps}\AgdaSpace{}%
\AgdaBound{r⇝t'}\AgdaSpace{}%
\AgdaBound{fin}%
\AgdaSymbol{=}%
\AgdaInductiveConstructor{steps}\AgdaSpace{}%
\AgdaSymbol{(}\AgdaBound{t}\AgdaSpace{}%
\AgdaOperator{\AgdaInductiveConstructor{↪⟨}}\AgdaSpace{}%
\AgdaBound{t↪r}\AgdaSpace{}%
\AgdaOperator{\AgdaInductiveConstructor{⟩}}\AgdaSpace{}%
\AgdaBound{r⇝t'}\AgdaSymbol{)}\AgdaSpace{}%
\AgdaBound{fin} 
\end{code}

This function applies progress to the typed term. There are three possibilities:
\begin{enumerate}
 \item If the term \bound{t} is in normal form, the reduction sequence is trivial:
\bound{t} \AgdaOperator{\AgdaDatatype{⇝*}} \bound{t} and \bound{t} is also a value.
\item If \bound{t} \isoterm \bound{r}, we recursively call \bound{eval} with the proof that
\bound{r} is strong normalizing. The result is the reduction sequence obtained
by adding the step \bound{t} \isoterm \bound{r} to the result of the recursion.
\item If \bound{t} \reduces \bound{r}, we proceed similarly.
 \end{enumerate}

We note that the argument that provides the evidence that \bound{t} is strong normalizing is
necessary to pass Agda's termination checker. In each recursive step,  a constructor \bound{sn}
is removed from this argument, making it structurally smaller.

Finally, we define \func{eval} using the strong normalization theorem:

\begin{code}%
%
\>[0] \AgdaFunction{eval}\AgdaSpace{}%
\AgdaSymbol{:}\AgdaSpace{}%
\AgdaSymbol{∀}\AgdaSpace{}%
\AgdaSymbol{\{}\AgdaBound{A}\AgdaSymbol{\}}\AgdaSpace{}%
\AgdaSymbol{→}\AgdaSpace{}%
\AgdaSymbol{(}\AgdaBound{t}\AgdaSpace{}%
\AgdaSymbol{:}\AgdaSpace{}%
\AgdaInductiveConstructor{∅}\AgdaSpace{}%
\AgdaOperator{\AgdaDatatype{⊢}}\AgdaSpace{}%
\AgdaBound{A}\AgdaSymbol{)}\AgdaSpace{}%
\AgdaSymbol{→}\AgdaSpace{}%
\AgdaDatatype{Steps}\AgdaSpace{}%
\AgdaBound{t}
\\
\>[0] \AgdaFunction{eval}\AgdaSpace{}%
\AgdaBound{t}\AgdaSpace{}%
\AgdaSymbol{=}\AgdaSpace{}%
\AgdaFunction{eval´}\AgdaSpace{}%
\AgdaBound{t}\AgdaSpace{}%
\AgdaSymbol{(}\AgdaFunction{strong-norm}\AgdaSpace{}%
\AgdaBound{t}\AgdaSymbol{)}\<%
\end{code}

\paragraph{Example}
Consider the evaluation of the term $\Omega$, that can be defined in Agda as follows:

\begin{code}%
	\>[0]\AgdaFunction{Ω}\AgdaSpace{}%
	\AgdaSymbol{:}\AgdaSpace{}%
	\AgdaInductiveConstructor{∅}\AgdaSpace{}%
	\AgdaOperator{\AgdaDatatype{⊢}}\AgdaSpace{}%
	\AgdaInductiveConstructor{⊤}\<%
	\\
	\>[0]\AgdaFunction{Ω}\AgdaSpace{}%
	\AgdaSymbol{=}\AgdaSpace{}%
	\AgdaSymbol{(}\AgdaOperator{\AgdaInductiveConstructor{ƛ}}\AgdaSpace{}%
	\AgdaOperator{\AgdaInductiveConstructor{[}}\AgdaSpace{}%
	\AgdaInductiveConstructor{sym}\AgdaSpace{}%
	\AgdaInductiveConstructor{abs}\AgdaSpace{}%
	\AgdaOperator{\AgdaInductiveConstructor{]≡}}\AgdaSpace{}%
	\AgdaSymbol{(}\AgdaOperator{\AgdaInductiveConstructor{`}}\AgdaSpace{}%
	\AgdaInductiveConstructor{Z}\AgdaSymbol{)}\AgdaSpace{}%
	\AgdaOperator{\AgdaInductiveConstructor{·}}\AgdaSpace{}%
	\AgdaOperator{\AgdaInductiveConstructor{`}}\AgdaSpace{}%
	\AgdaInductiveConstructor{Z}\AgdaSymbol{)}\AgdaSpace{}%
	\AgdaOperator{\AgdaInductiveConstructor{·}} \\ 
	\>[0] \AgdaSpace{}\AgdaSpace{} \AgdaSpace{} \AgdaSpace{} \AgdaSpace{}
	\AgdaSymbol{(}\AgdaOperator{\AgdaInductiveConstructor{[}}\AgdaSpace{}%
	\AgdaInductiveConstructor{abs}\AgdaSpace{}%
	\AgdaOperator{\AgdaInductiveConstructor{]≡}}\AgdaSpace{}%
	\AgdaSymbol{(}\AgdaOperator{\AgdaInductiveConstructor{ƛ}}\AgdaSpace{}%
	\AgdaOperator{\AgdaInductiveConstructor{[}}\AgdaSpace{}%
	\AgdaInductiveConstructor{sym}\AgdaSpace{}%
	\AgdaInductiveConstructor{abs}\AgdaSpace{}%
	\AgdaOperator{\AgdaInductiveConstructor{]≡}}\AgdaSpace{}%
	\AgdaSymbol{(}\AgdaOperator{\AgdaInductiveConstructor{`}}\AgdaSpace{}%
	\AgdaInductiveConstructor{Z}\AgdaSymbol{)}\AgdaSpace{}%
	\AgdaOperator{\AgdaInductiveConstructor{·}}\AgdaSpace{}%
	\AgdaOperator{\AgdaInductiveConstructor{`}}\AgdaSpace{}%
	\AgdaInductiveConstructor{Z}\AgdaSymbol{))}\<%
\end{code}

\noindent This definition corresponds to the formalization of the typing derivation given in Figure \ref{Omega}. The evaluation of this term is as follows:

\begin{code}%
\>[0]\AgdaFunction{Ω⇝⋆}\AgdaSpace{}%
\AgdaSymbol{:}\AgdaSpace{}%
\AgdaFunction{Ω}\AgdaSpace{}%
\AgdaOperator{\AgdaDatatype{⇝*}}\AgdaSpace{}%
\AgdaInductiveConstructor{⋆}\<%
\\
\>[0]\AgdaFunction{Ω⇝⋆}%
\>[745I]\AgdaSymbol{=}\<%
\\
\>[.][@{}l@{}]\<[745I]%
\>[5]\AgdaFunction{Ω}\<%
\\
\>[5][@{}l@{\AgdaIndent{0}}]%
\>[7]\AgdaOperator{\AgdaInductiveConstructor{⇄⟨}}\AgdaSpace{}%
\AgdaInductiveConstructor{ξ-·₁}\AgdaSpace{}%
\AgdaSymbol{(}\AgdaInductiveConstructor{ζ}\AgdaSpace{}%
\AgdaSymbol{(}\AgdaInductiveConstructor{ξ-·₁}\AgdaSpace{}%
\AgdaInductiveConstructor{sym-abs}\AgdaSymbol{))}\AgdaSpace{}%
\AgdaOperator{\AgdaInductiveConstructor{⟩}}\<%
\\
\>[5]\AgdaSymbol{(}\AgdaOperator{\AgdaInductiveConstructor{ƛ}}\AgdaSpace{}%
\AgdaSymbol{(}\AgdaOperator{\AgdaInductiveConstructor{ƛ}}\AgdaSpace{}%
\AgdaOperator{\AgdaInductiveConstructor{`}}\AgdaSpace{}%
\AgdaSymbol{(}\AgdaOperator{\AgdaInductiveConstructor{S}}\AgdaSpace{}%
\AgdaInductiveConstructor{Z}\AgdaSymbol{))}\AgdaSpace{}%
\AgdaOperator{\AgdaInductiveConstructor{·}}\AgdaSpace{}%
\AgdaOperator{\AgdaInductiveConstructor{`}}\AgdaSpace{}%
\AgdaInductiveConstructor{Z}\AgdaSymbol{)}\AgdaSpace{}%
\AgdaOperator{\AgdaInductiveConstructor{·}} \\
\>[5] \AgdaSymbol{(}\AgdaOperator{\AgdaInductiveConstructor{[}}\AgdaSpace{}%
\AgdaInductiveConstructor{abs}\AgdaSpace{}%
\AgdaOperator{\AgdaInductiveConstructor{]≡}}\AgdaSpace{}%
\AgdaSymbol{(}\AgdaOperator{\AgdaInductiveConstructor{ƛ}}\AgdaSpace{}%
\AgdaSymbol{(}\AgdaOperator{\AgdaInductiveConstructor{[}}\AgdaSpace{}%
\AgdaInductiveConstructor{sym}\AgdaSpace{}%
\AgdaInductiveConstructor{abs}\AgdaSpace{}%
\AgdaOperator{\AgdaInductiveConstructor{]≡}}\AgdaSpace{}%
\AgdaSymbol{(}\AgdaOperator{\AgdaInductiveConstructor{`}}\AgdaSpace{}%
\AgdaInductiveConstructor{Z}\AgdaSymbol{))}\AgdaSpace{}%
\AgdaOperator{\AgdaInductiveConstructor{·}}\AgdaSpace{}%
\AgdaOperator{\AgdaInductiveConstructor{`}}\AgdaSpace{}%
\AgdaInductiveConstructor{Z}\AgdaSymbol{))}\<%
\\
\>[5]\AgdaComment{--\ (λx.\ (λy.\ x)\ ·\ x)\ ·\ (λx.\ x\ ·\ x)}\<%
\\
\>[5][@{}l@{\AgdaIndent{0}}]%
\>[7]\AgdaOperator{\AgdaInductiveConstructor{↪⟨}}\AgdaSpace{}%
\AgdaInductiveConstructor{ξ-·₁}\AgdaSpace{}%
\AgdaSymbol{(}\AgdaInductiveConstructor{ζ}\AgdaSpace{}%
\AgdaInductiveConstructor{β-ƛ}\AgdaSymbol{)}\AgdaSpace{}%
\AgdaOperator{\AgdaInductiveConstructor{⟩}}\<%
\\
\>[5]\AgdaSymbol{(}\AgdaOperator{\AgdaInductiveConstructor{ƛ}}\AgdaSpace{}%
\AgdaOperator{\AgdaInductiveConstructor{`}}\AgdaSpace{}%
\AgdaInductiveConstructor{Z}\AgdaSymbol{)}\AgdaSpace{}%
\AgdaOperator{\AgdaInductiveConstructor{·}}\AgdaSpace{}%
\AgdaSymbol{(}\AgdaOperator{\AgdaInductiveConstructor{[}}\AgdaSpace{}%
\AgdaInductiveConstructor{abs}\AgdaSpace{}%
\AgdaOperator{\AgdaInductiveConstructor{]≡}}\AgdaSpace{}%
\AgdaSymbol{(}\AgdaOperator{\AgdaInductiveConstructor{ƛ}}\AgdaSpace{}%
\AgdaSymbol{(}\AgdaOperator{\AgdaInductiveConstructor{[}}\AgdaSpace{}%
\AgdaInductiveConstructor{sym}\AgdaSpace{}%
\AgdaInductiveConstructor{abs}\AgdaSpace{}%
\AgdaOperator{\AgdaInductiveConstructor{]≡}}\AgdaSpace{}%
\AgdaSymbol{(}\AgdaOperator{\AgdaInductiveConstructor{`}}\AgdaSpace{}%
\AgdaInductiveConstructor{Z}\AgdaSymbol{))}\AgdaSpace{}%
\AgdaOperator{\AgdaInductiveConstructor{·}}\AgdaSpace{}%
\AgdaOperator{\AgdaInductiveConstructor{`}}\AgdaSpace{}%
\AgdaInductiveConstructor{Z}\AgdaSymbol{))}\<%
\\
\>[5]\AgdaComment{--\ (λx.\ x)\ ·\ (λx.\ x\ ·\ x)}\<%
\\
\>[5][@{}l@{\AgdaIndent{0}}]%
\>[7]\AgdaOperator{\AgdaInductiveConstructor{⇄⟨}}\AgdaSpace{}%
\AgdaInductiveConstructor{ξ-·₂}\AgdaSpace{}%
\AgdaSymbol{(}\AgdaInductiveConstructor{ξ-≡}\AgdaSpace{}%
\AgdaSymbol{(}\AgdaInductiveConstructor{ζ}\AgdaSpace{}%
\AgdaSymbol{(}\AgdaInductiveConstructor{ξ-·₁}\AgdaSpace{}%
\AgdaInductiveConstructor{sym-abs}\AgdaSymbol{)))}\AgdaSpace{}%
\AgdaOperator{\AgdaInductiveConstructor{⟩}}\<%
\\
\>[5]\AgdaSymbol{(}\AgdaOperator{\AgdaInductiveConstructor{ƛ}}\AgdaSpace{}%
\AgdaOperator{\AgdaInductiveConstructor{`}}\AgdaSpace{}%
\AgdaInductiveConstructor{Z}\AgdaSymbol{)}\AgdaSpace{}%
\AgdaOperator{\AgdaInductiveConstructor{·}}\AgdaSpace{}%
\AgdaSymbol{(}\AgdaOperator{\AgdaInductiveConstructor{[}}\AgdaSpace{}%
\AgdaInductiveConstructor{abs}\AgdaSpace{}%
\AgdaOperator{\AgdaInductiveConstructor{]≡}}\AgdaSpace{}%
\AgdaSymbol{(}\AgdaOperator{\AgdaInductiveConstructor{ƛ}}\AgdaSpace{}%
\AgdaSymbol{(}\AgdaOperator{\AgdaInductiveConstructor{ƛ}}\AgdaSpace{}%
\AgdaOperator{\AgdaInductiveConstructor{`}}\AgdaSpace{}%
\AgdaSymbol{(}\AgdaOperator{\AgdaInductiveConstructor{S}}\AgdaSpace{}%
\AgdaInductiveConstructor{Z}\AgdaSymbol{))}\AgdaSpace{}%
\AgdaOperator{\AgdaInductiveConstructor{·}}\AgdaSpace{}%
\AgdaOperator{\AgdaInductiveConstructor{`}}\AgdaSpace{}%
\AgdaInductiveConstructor{Z}\AgdaSymbol{))}\<%
\\
\>[5]\AgdaComment{--\ (λx.\ x)\ ·\ (λx.\ (λy.\ x)\ ·\ x)}\<%
\\
\>[5][@{}l@{\AgdaIndent{0}}]%
\>[7]\AgdaOperator{\AgdaInductiveConstructor{↪⟨}}\AgdaSpace{}%
\AgdaInductiveConstructor{ξ-·₂}\AgdaSpace{}%
\AgdaSymbol{(}\AgdaInductiveConstructor{ξ-≡}\AgdaSpace{}%
\AgdaSymbol{(}\AgdaInductiveConstructor{ζ}\AgdaSpace{}%
\AgdaInductiveConstructor{β-ƛ}\AgdaSymbol{))}\AgdaSpace{}%
\AgdaOperator{\AgdaInductiveConstructor{⟩}}\<%
\\
\>[5]\AgdaSymbol{(}\AgdaOperator{\AgdaInductiveConstructor{ƛ}}\AgdaSpace{}%
\AgdaOperator{\AgdaInductiveConstructor{`}}\AgdaSpace{}%
\AgdaInductiveConstructor{Z}\AgdaSymbol{)}\AgdaSpace{}%
\AgdaOperator{\AgdaInductiveConstructor{·}}\AgdaSpace{}%
\AgdaSymbol{(}\AgdaOperator{\AgdaInductiveConstructor{[}}\AgdaSpace{}%
\AgdaInductiveConstructor{abs}\AgdaSpace{}%
\AgdaOperator{\AgdaInductiveConstructor{]≡}}\AgdaSpace{}%
\AgdaSymbol{(}\AgdaOperator{\AgdaInductiveConstructor{ƛ}}\AgdaSpace{}%
\AgdaOperator{\AgdaInductiveConstructor{`}}\AgdaSpace{}%
\AgdaInductiveConstructor{Z}\AgdaSymbol{))}\<%
\\
\>[5]\AgdaComment{--\ (λx.\ x)\ ·\ (λx.\ x)}\<%
\\
\>[5][@{}l@{\AgdaIndent{0}}]%
\>[7]\AgdaOperator{\AgdaInductiveConstructor{⇄⟨}}\AgdaSpace{}%
\AgdaInductiveConstructor{ξ-·₂}\AgdaSpace{}%
\AgdaInductiveConstructor{abs}\AgdaSpace{}%
\AgdaOperator{\AgdaInductiveConstructor{⟩}}\<%
\\
\>[5]\AgdaSymbol{(}\AgdaOperator{\AgdaInductiveConstructor{ƛ}}\AgdaSpace{}%
\AgdaOperator{\AgdaInductiveConstructor{`}}\AgdaSpace{}%
\AgdaInductiveConstructor{Z}\AgdaSymbol{)}\AgdaSpace{}%
\AgdaOperator{\AgdaInductiveConstructor{·}}\AgdaSpace{}%
\AgdaInductiveConstructor{⋆}\<%
\\
\>[5]\AgdaComment{--\ (λx.\ x)\ ·\ ⋆}\<%
\\
\>[5][@{}l@{\AgdaIndent{0}}]%
\>[7]\AgdaOperator{\AgdaInductiveConstructor{↪⟨}}\AgdaSpace{}%
\AgdaInductiveConstructor{β-ƛ}\AgdaSpace{}%
\AgdaOperator{\AgdaInductiveConstructor{⟩}}\<%
\\
\>[5]\AgdaInductiveConstructor{⋆}\<%
\\
\>[0][@{}l@{\AgdaIndent{0}}]%
\>[2]\AgdaOperator{\AgdaInductiveConstructor{∎}}\<%
\\
\\[\AgdaEmptyExtraSkip]%
\>[0]\AgdaFunction{\AgdaUnderscore{}}\AgdaSpace{}%
\AgdaSymbol{:}\AgdaSpace{}%
\AgdaFunction{eval}\AgdaSpace{}%
\AgdaFunction{Ω}\AgdaSpace{}%
\AgdaOperator{\AgdaDatatype{≋}}\AgdaSpace{}%
\AgdaInductiveConstructor{steps}\AgdaSpace{}%
\AgdaFunction{Ω⇝⋆}\AgdaSpace{}%
\AgdaInductiveConstructor{V-⋆}\<%
\\
\>[0]\AgdaSymbol{\AgdaUnderscore{}}\AgdaSpace{}%
\AgdaSymbol{=}\AgdaSpace{}%
\AgdaInductiveConstructor{refl}\<%
\end{code}

We can see that in order to apply the $\beta$-reduction, the term on the left must be an
abstraction. Therefore, to reduce a term of the form
(\const{[} \bound{iso} \const{]≡} \bound{r}) \const{·} \bound{s}
the constructor \const{[\_]≡\_} must be removed from the left side by applying a term equivalence.

On the other hand, the terms produced by the equivalences \const{sym-abs} and \const{sym-id-⇒}
add a new abstraction that does not capture any variable within the body of the abstraction.
Although it might seem unnecessary, it's actually crucial for enabling self-application of the variable.

\section{Conclusions}
\label{sec:conclusions}

In this work, we formalized System I with the addition of the type Top in Agda, proving the strong normalization and progress theorems.

The strong normalization theorem is important for two main reasons.
From a formalization perspective, the constructive proof allows us to define a total evaluation function for the calculus, by induction on the strong normalization evidence.
From the Curry–Howard perspective, considering the language as a logic, it ensures its consistency.

The formalization given in this work has some interesting features. One of them is
the representation of intrinsically typed terms, where each term of the calculus
is a direct implementation of a typing rule.
As a result of this encoding, the implementation of functions over these typed terms
preserves types, and the application of term isomorphisms is syntax directed.
As a consequence, the relation \isoterm eliminates the type isomorphisms of a term,
in the same way that the relation \reduces eliminates applications
and projections. This point is essential for making reduction sequences finite,
since it prevents some term isomorphisms, such as \pair{r}{s} \isoterm \pair{s}{r},
from being applied an indefinite number of times.

Another interesting feature is that the proof given for strong normalization follows Schäfer's technique, which, while framed within Tait and Girard's reducibility, presents several changes aimed at a smoother formalization process. Within that approach, we chose Kovács' variant, more appropriate for the constructive setting of Agda.
It is important to note that this is the first version we know of for a formalization
of the normalization proof of a lambda calculus modulo isomorphisms.

In this kind of calculus, the number of reduction cases that must be considered in the normalization
proof is very large due to the equivalence relation on terms, so working in a rigorous setting such as that
provided by Agda was very helpful.

\section{Future work}
\label{sec:futurework}

In the future, we would like to adapt the formalization to the setting of other systems modulo isomorphisms, such as Polymorphic System I~\cite{PSI}, \ie System F modulo isomorphisms, and the distributive $\lambda$-calculus~\cite{ADC20}, which is an untyped variant of System I where only the distributive rule is considered.

In a similar direction we want to explore the relation
of the calculus we have formalized with System I$_{\eta}$~\cite{DCD23}, which adds
extensionality to System I. Their approach differs from ours; they introduce
$\eta$-expansion rule as a reduction relation, whereas we combine $\eta$-expansion with some term isomorphism rules.

It would also be interesting to define a variant of the formalization without syntactic witness in the terms. This would produce terms that are not in correspondence with their type derivations, hence a form of inference of type isomorphisms would be needed to determine which term isomorphism could be applied to the term.

\bibliographystyle{plain}
\bibliography{biblio}

\end{document}